\documentclass[12pt]{article}
\pdfoutput=1
\usepackage[utf8]{inputenc}
\usepackage[left=2.55cm, right=2.55cm, top=2.55cm, bottom=2.55cm]{geometry}
\usepackage{tikz}
\usepackage{tikz-feynman}
\usepackage{contour}
\usepackage{amsmath,amssymb,amsbsy}
\usepackage{slashed}
\usepackage{xcolor}
\usepackage{graphicx}
\usepackage{tikz}
\usepackage{tikz-feynman}
\usepackage{contour}
\usepackage{url}
\usepackage{cancel}
\usepackage{cite}
\usepackage[colorlinks=true,allcolors=blue,pdfborder={0 0 0},linktocpage=false,pdfencoding=auto]{hyperref}
\usepackage{tabularx,booktabs}
\usepackage{multicol}
\usepackage{feynmp}
\usepackage{units}
\usepackage{xspace}
\usepackage[labelfont=bf]{caption}
\usepackage[section]{placeins}
\usepackage{subcaption}
\usepackage{soul}
\usepackage{bbold}
\usepackage{parskip}
\usepackage{float}
\usepackage{tabulary}

\definecolor{darkred}{rgb}{0.6,0,0}
\definecolor{darkpurple}{rgb}{0.5,0,0.5}

\newcommand{\beqn}{\begin{eqnarray}}
\newcommand{\eeqn}{\end{eqnarray}}
\def\non{\nonumber\\}

\begin{document}
\author{Wan-Zhe Feng$^a$\footnote{\href{mailto:vicf@tju.edu.cn}{vicf@tju.edu.cn}}, 
Jinzheng Li$^b$\footnote{\href{mailto:li.jinzh@northeastern.edu}{li.jinzh@northeastern.edu}} ~and
Pran Nath$^b$\footnote{\href{mailto:p.nath@northeastern.edu}{p.nath@northeastern.edu}}\\
$^{a}$\textit{\normalsize Center for Joint Quantum Studies and Department of Physics, School of Science,}\\\textit{\normalsize
 Tianjin University, Tianjin 300350, PR. China}\\
$^{b}$\textit{\normalsize Department of Physics, Northeastern University, Boston, MA 02115-5000, USA} }

\title{\vspace{-2cm}\begin{flushright}
\end{flushright}
\vspace{1cm}
\Large \bf
Cosmologically Consistent Analysis of Gravitational Waves
from hidden sectors
 \vspace{0.0cm}
}
 
\date{}
\maketitle
%%%%%%%%%%%%%%%%%%%%%
\begin{abstract}
Production of gravitational waves in the early universe is discussed in a  cosmologically consistent analysis within a first order phase transition involving a hidden sector feebly coupled
 with the visible sector.  Each sector resides in its own heat bath leading to a potential dependent on two temperatures, and on two fields: one a standard model Higgs and the other a scalar arising from a  hidden sector $U(1)$ gauge theory. A synchronous evolution of the hidden and visible sector temperatures
is carried out from the reheat temperature down to the electroweak scale. 
The hydrodynamics of two-field phase transitions, one for the visible and the other for the
hidden is discussed, which leads to separate tunneling temperatures, and different 
sound speeds for the two sectors.  Gravitational waves emerging from the two sectors are computed and their imprint on the measured gravitational wave power spectrum vs frequency
is analyzed in terms of bubble nucleation signature, i.e., detonation, deflagration,
and hybrid. It is shown that the two-field model predicts gravitational waves accessible at several proposed gravitational wave detectors: LISA, DECIGO, BBO, Taiji
 and their discovery would probe specific regions of the  hidden sector parameter
space and may also shed light on the nature of bubble nucleation in the early universe.
The analysis presented here indicates that the
cosmologically preferred models  are those where the tunneling in the visible sector
precedes the tunneling in the hidden sector and the sound speed $c_s$ lies below its
maximum, i.e., $c^2_s<\frac{1}{3}$. It is of interest to investigate if these features
are universal and applicable to a wider class of cosmologically consistent models.

    \end{abstract}
\numberwithin{equation}{section}
\newpage

{  \hrule height 0.4mm \hypersetup{colorlinks=black,linktocpage=true} 
\tableofcontents
\vspace{0.5cm}
 \hrule height 0.4mm} 

%%%%%%%%%%%%%%%%%%%%%%%%%%%%%%%%%%%%%%%%%%%%
\section{Introduction \label{sec:1}}
The observation of gravitational waves in black hole mergers in 2016 
\cite{LIGOScientific:2016aoc} opened up a new avenue to explore
fundamental physics in a broader context using stochastic background of gravitational waves
which arise from a variety of phenomena including those from cosmic phase transitions.
The cosmic phase transitions occur at finite temperatures 
\cite{Kirzhnits:1972iw,Kirzhnits:1972ut,Witten:1980ez,Guth:1981uk,Steinhardt:1981ct}
 and give rise to stochastic gravitational waves \cite{Witten:1984rs,Hogan:1986qda,Kamionkowski:1993fg}.
Several other sources of stochastic gravitational waves exist such as from the decay of the inflaton into 
standard model particles at the end of inflation\cite{Khlebnikov:1997di,Easther:2006vd,GarciaBellido:2007af}.
It is also suggested that phase transitions may be linked to generation of
matter-antimatter asymmetry, and specially to baryogenesis
~\cite{Cohen:1990py,Carena:2000id,Cline:2006ts,White:2016nbo,Cline:2018fuq}.
The study of cosmic phase transitions involves finite temperature field theory 
which has been investigated in several early works\cite{Dolan:1973qd,Weinberg:1974hy}. A significant amount of further work already exists in this area, see e.g., 
~\cite{Adams:1993zs,Parwani:1991gq,Arnold:1992rz,Espinosa:1992kf,Quiros:1992ez,Curtin:2016urg,Espinosa:2007qk,Espinosa:2008kw,Azevedo:2018fmj,Mohamadnejad:2021tke,Biermann:2022meg,Wang:2022akn,Bringmann:2023iuz,Addazi:2016fbj,Aoki:2017aws,Pasechnik:2023hwv,Schwaller:2015tja,Addazi:2017gpt,Fairbairn:2019xog,Co:2021rhi,Borah:2021ocu,Abe:2023zja,Imtiaz:2018dfn,Paul:2020wbz}. For reviews of phase transitions see  \cite{Quiros:1999jp,Morrissey:2012db,Mazumdar:2018dfl}).

In the current analysis we discuss phase transitions and gravitational wave generation from
hidden sectors which arise in supergravity, string and extra dimensional models which improves on some of the previous works in that the analysis is cosmologically consistent.
This implies a number of things which we mention briefly. First the
 gravitational wave models need to satisfy constraints at different temperatures, e.g., at the
 tunneling temperature (10-100) GeV and at the BBN temperature $\sim 1$ MeV which
 requires an extrapolation over four to five orders of magnitude. This is due to the fact that at the tunneling temperature
  the phase transition is controlled in part by the parameter $\alpha=\epsilon/\rho$  where $\epsilon$ is the latent heat in the phase transition and $\rho$ is the total energy density which includes the energy density of the standard model and of the hidden sector. In general the hidden sector and the visible sector are at different temperatures
  and we need to know their precise correlation as a function of temperatures to compute 
  $\alpha$ correctly. 
     Further, as noted we need to extrapolate to BBN time which constraints 
     the extra degrees of freedom $\Delta N_{eff}$  above the standard model prediction,
     which requires we determine the hidden sector temperature at BBN time.
   Often this correlation is done by assuming separate entropy conservation in the visible sector
  and in the hidden sector. In this case the ratio $\xi(T)= T_h/T$, where $T_h$ is the temperature
  in the hidden sector and $T$ is the temperature in the visible sector, is correlated with 
  the ratio $\xi(T_0)$ at temperature $T_0$ so that 
     \begin{align} 
     \frac{h^h_{eff}(\xi(T_0)T_0)}{h^v_{eff}(T_0)}\xi^3(T_0) =    
\frac{h^h_{eff}(\xi(T)T)}{h^v_{eff}(T)}\xi^3(T)
\label{no-coupling}
\end{align}  
where $h^v_{eff}$ and $h^h_{eff}$ are the entropy degrees of freedom 
at their respective temperatures of the visible sector and of the hidden sector.
  However, it was shown in ~\cite{Li:2023nez,Nath:2024mgr} that the separate entropy
  conservation approximation is highly inaccurate and leads to erroneous results for 
  $\Delta N_{\rm eff}$ by up to 500\%. There is another basic problem with relations of 
  the type above for cases where the decoupling in the dark sector occurs below the 
  mass threshold of the dark particles. In this case the assumption of using thermal
  equilibrium to compute the effective degrees of freedom in the hidden sector 
  breaks down as it gives essentially $h^{h}_{eff}(T_0)=0$ requiring $\xi(T_0)$ to blow up.
  Here the accurate analysis used in this work is essential as explained in Appendix E.
  
In the analysis we carry out a synchronous evolution of the temperatures in the
  visible and in the hidden sectors. Central to the analysis is the evolution equation for $\xi(T)$
  which is solved together with the yield equations for the particles in the hidden sector
  and the visible sector with an assumed boundary condition on $\xi(T)$ at the reheat 
  temperature which leads to an accurate prediction for $\xi(T)$ at any temperature.
    There are also other aspects of the analysis which we briefly comment on. In the 
    current analysis we have nucleation arising from two bubble formations, one in the 
    visible sector and the other in the hidden sector and we give a combined treatment of 
    both. This leads to two different critical temperatures and tunnelings arising 
    from the visible sector and from the hidden sector. 
    Further, often in gravitational wave analyses
    a sound speed of $c^2_s=1/3$ is assumed which is the terminal relativistic speed of sound
    waves in a fluid. However, in the presence of true (broken) and false (symmetric) vacua
    for the visible and hidden sectors four different possibilities for the sound speed arise:
    with two possibilities for the visible sector depending on whether  the vacuum is true or false,
    and similarly for the hidden sector. We  discuss these
    possibilities and show that the gravitational wave power spectrum depends sensitively on 
   sound speed. 
    Finally, we have investigated the possibility of identifying the nature of bubble dynamics 
    and nucleation, i.e., detonation, deflagration and hybrid for their possible imprint on the gravitational wave spectrum. While we draw no firm conclusion, we notice that among the candidates models that satisfy all the constraints (i.e., constraints from first order phase transition (FOPT), from  relic density, and from $\Delta N_{eff}$),  the hybrid nucleation modes exhibit the largest gravitational wave power spectrum.
   
%%%%%%%%%%%%%%%%%%%%%%%%%%%%%%%%%%%%%%%%%%%%%%
The outline of the rest of the paper is as follows: 
In section \ref{sec:2} we write the hidden sector model and discuss its temperature 
dependent potential including thermal contributions to the field dependent masses
including the daisy summed multi-loop contribution. Then we define the two-field potential including
the temperature dependent potential for the standard model Higgs.
In this section we also give a brief discussion of synchronous evolution of coupled hidden and visible sectors . 
In  section \ref{sec:3} we discuss nucleation and vacuum decay during phase transition 
for the case of a single field and then for the two-field case.  
In section \ref{sec:4} we discuss the hydrodynamics of bubble formation during phase transition. Here we discuss the  sound velocity in the visible and in the hidden sectors 
for symmetric and broken phases and
 give an analysis of relativistic fluid equations and of bubble dynamics.
 Gravitational wave spectra arising from first order phase transitions from 
   the visible and the hidden sectors are discussed in section
 \ref{sec:5}. A detailed numerical analysis of gravitational wave power spectrum is given in 
 section \ref{sec:6}. 
 Thus in subsection \ref{sec:6.1} we exhibit  the parameter space of models 
 investigated in Monte Carlo simulations and the theoretical and experimental
 constraints placed on the allowed set of models.
 The nucleation temperature and the resulting gravity power spectrum is discussed in  subsection \ref{sec:6.2}.  
 In subsection \ref{sec:6.3} we discuss the effect of sound velocity on the gravitational wave power
 spectrum and in subsection \ref{sec:6.4} we investigate the dependence of  sound velocity
  on the nucleation temperature. An analysis of the $\Delta N_{eff}$ 
 constraint is given in subsection \ref{sec:6.5}.
 In subsection \ref{sec:6.6} we discuss
 the gravity power spectrum for different nucleation modes, i.e., detonation, deflagration and
 hybrid. 
  It is shown that a significant part of the parameter space of the assumed hidden sector model can be accessed by the planned space based gravity experiments such as LISA,
 DECIGO, BBO, Taiji and other experiments. Conclusions are given in section \ref{sec:7}.

 Additional details of the analysis are given in the Appendices A-E. Thus in Appendix A, we give 
 further details of the temperature dependent potential for the hidden sector and 
  and computation of temperature dependent corrections to the bosonic masses for a $U(1)$ gauge theory including the contribution of the daisy resummation.  In Appendix B, we give a 
  summary of the known results on the temperature dependent Higgs potential for the visible sector. In Appendix C we give further details of visible and hidden sector interactions that enter in the
  combined analysis of the two sectors, and in  Appendix D we give the scattering cross sections
 that enter in the yield equations for the dark scalar, the dark fermion and for the dark gauge
 boson. Finally in Appendix E we discuss the energy and pressure densities away from equilibrium as they are relevant for freeze-out and decoupling in the hidden sector. 
 
%%%%%%%%%%%%%%%%%%%
\section{Two-field phase transition involving the standard model and a hidden sector\label{sec:2}}
  As noted in the Introduction,
  cosmological phase transitions have been investigated in a significant number of previous works 
  (for reviews, e.g.,\cite{Sher:1988mj,Morrissey:2012db,Weir:2017wfa,Athron:2023xlk}). 
  Most of the previous works using beyond the standard model (BSM) physics, 
  involve dynamics of only one field. Such an analysis does not fully take into account the
  effect of the standard model on computing the strength of the phase transition $\alpha$
  in tunneling  and the proper imposition of the $\Delta N_{eff}$ constraint at BBN time.
      Thus, as noted earlier a more complete analysis needs to consider an analysis
      involving BSM physics along with the  standard model which in our case implies 
    a two-field analysis including the Higgs field of the standard model along with the Higgs field of the hidden sector.
  Further,
  since the visible sector and the hidden sector would normally be in different heat baths,
  the thermal potential governing the phase transition will depend on two temperatures,
  one of the visible and the other of the hidden sector. In the presence of a coupling between
  the two, as is most likely via a variety of portals, a synchronous evolution of the visible and 
  the hidden sector temperature is essential for reliable predictions of phenomena related to the
  cosmological phase transition and specifically on predictions of the power spectrum of gravity
  waves resulting from the phase transition. This aspect of the cosmological phase transition is
  one of the focus points of the current analysis.

\subsection{The hidden sector model and its temperature dependent potential\label{sec:4}}
We discuss now the case of phase transitions  which involve two scalar
fields one of which is the standard models Higgs and the other is a hidden
sector Higgs scalar. In this case we consider the Lagrangian of the form 
 \begin{equation}
\mathcal{L}=\mathcal{L}_{\rm SM}+\Delta\mathcal{L},
\label{totL}
\end{equation}
where $\mathcal{L}_{\rm SM}$ is the standard model Lagrangian, and  
$\Delta\mathcal{L}$ is the hidden sector Lagrangian given by
\begin{align}
\Delta\mathcal{L} =&-\frac{1}{4}A_{\mu\nu}A^{\mu\nu} -  
|(\partial_\mu-i g_x A_\mu) \Phi|^2- V^{\rm h}_{\rm eff}(\Phi)\non
&-\bar D (\frac{1}{i}\gamma^\mu \partial_\mu+m_D) D -\frac{\delta}{2} A_{\mu\nu} B^{\mu\nu} 
-g_x Q_D \bar D \gamma^\mu D A_\mu,
\label{bsm}
\end{align}
where $A_\mu$ is the gauge field
of the $U(1)_X$  of the hidden sector, $D$ is the dark fermion, $\Phi$ is a
complex scalar field and $B_\mu$ is the gauge field
of the $U(1)_Y$ and $A_{\mu\nu}= \partial_\mu A_\nu -\partial_\nu A_\mu$ and $B_{\mu\nu}=
\partial_\mu B_\nu- \partial_\nu B_\mu$. 
 Thermal contributions to the zero temperature  potential 
 $V^{\rm h}_{\rm eff}(\Phi)$  will allow a first 
 order phase transition and a VEV growth  for the scalar field $\Phi$ generating a mass for 
  the gauge boson $A_\mu$ and the scalar field in the hidden sector. 
Thus the effective  temperature dependent hidden sector potential including loop corrections 
is given by  
\begin{align}
V^{\rm h}_{{\rm eff}}(\rm \Phi,T_h)&=V_{0h}+V_{1h}^{(0)}
+\Delta V_{1h}^{(T_h)} 
+ V_{h}^{\rm{daisy}}(T_h)\,. 
\label{pot-hid}
\end{align}
Here $V_{0h}$ is the zero temperature tree potential, $V_{1h}^{(0)}$ is the one loop
Coleman-Weinberg zero temperature contribution, $\Delta V_{1h}^{(T_h)}$ is the one-loop
thermal contribution, $V_{1h}^{\rm{daisy}}(T_h)$ is the daisy contribution from multi-loop summation and divergences are cancelled off  by counter terms.
Thus we have
\begin{align}
 V_{0h}&= -\mu_h^2\Phi\Phi^* +\lambda_h (\Phi^*\Phi)^2,~~
\Phi= \frac{1}{\sqrt 2} (\chi_c+ \chi+ i G^0_h)
\label{pot-hid}
\end{align}
where $\chi_c$ is the background field which enters in the tree level potential.
Further, $V_{1h}^{(0)}(\chi)$, the one-loop effective potential at $T=0$, is given by
\begin{equation}
V_{1h}^{(0)}(\chi_c)=\sum_{i}\frac{N_{i}(-1)^{2{\rm s}_{i}}}{64\pi^{2}}m_{i}^{4}(\chi_c)\left[\ln\left(\frac{m_{i}^{2}(\chi_c)}{\Lambda_h^{2}}\right)-\mathcal{C}_{i}\right]\,,
\end{equation}
where $N_i$ is the degrees of particle $i$ and
where the field dependent masses of the hidden sector fields 
$A_{\mu}, \chi, G^0_h$ that enter the potential are given by
 \begin{align}
  m_A^2(\chi_c)= g^2_x \chi_c^2,  ~~m^2(\chi_c)= -\mu^2_h + 3 \lambda_h \chi_c^2,
  ~~m_{G^0_h}^2(\chi_c)=-\mu^2_h+  \lambda_h \chi_c^2.
 \end{align}
 For the  one-loop thermal correction we have
 %%%%%%%%%%%%%%%%%%%%%%%%%%%%%%%
  \begin{align}
\Delta V_{1h}^{(T_h)}(\chi_c,T_h)&=  \frac{T_h^4}{2\pi^2}
\Big[ 3J_{B}\left(\frac{m_{A}}{T_h}\right)
+   J_{B}\left(\frac{m_{\chi}}{T_h}\right)
+   J_{B}\left(\frac{m_{G^0_h}}{T_h}\right)\,\Big]
\label{V1T}
\end{align}
where $J_i$ ($i=B,F)$ is defined so that at one loop 
\begin{align}
J_{i}\left(\frac{m_{i}}{T_h}\right)&=
\int_{0}^{\infty}{\rm d}q\,q^{2}\ln\big[1\mp{\rm exp}\big(-\sqrt{q^{2}+m_{i}^{2}/T_h^{2}}\big)\big],~~i=(B,F)\,,
\label{JBF}
\end{align}
where $(B,F)$ stand for bosonic and fermionic cases. 
    The daisy loop contributions are only for the  longitudinal mode of $A$
    and $\chi$ and are given for mode $i=A,\chi$ so that 
    \begin{align}
V^{{\rm daisy}}(i,T_h) &  =-\frac{T_h}{12\pi}\Big\{\big[m_i^{2}+\Pi_i(T_h)\big]^{3/2}-m_i^{3}\Big\}\,,
\label{vdaisy}
\end{align}    
    where $\Pi_i(T_h)$ is thermal contribution to the zero temperature mass $m^2_i$.
    For the longitudinal mode of $A$ and for $\chi$ they are given by
    \begin{align}
    \Pi_A(T_h)&=\frac{2}{3} g_x^2 T_h^2\,, 
    ~~ \Pi_\chi(T_h)= \frac{1}{4} g_x^2 T_h^2 + \frac{1}{3} \lambda_h T_h^2.
    \label{selfmass}
    \end{align} 
   A deduction of Eqs. (\ref{vdaisy}) and (\ref{selfmass}) is given in Appendix A.
We note that the daisy resummation correction
to the effective potential is equivalent to
replacing the particle mass in $J_{B}$ function so that 
\begin{equation}
m_{i}^{2}\to\big[m_{i}^{(T_{h})}\big]^{2}\equiv m_{i}^{2}+\Pi_{i}(T_{h})\,,
\label{replace}
\end{equation}
where $\Pi_{i}(T_h)$ is the self-energy of the bosonic field for particle $i$
 at finite temperature $T_h$, known as ``Debye mass''.
 Making the replacement of Eq. (\ref{replace}), the effective potential of 
 %Eq.(\ref{pot-hid}) 
 Eq.(2.3)
 now takes the 
 form  
 \begin{align}
V_{{\rm eff}}^{{\rm h}}(\chi_{c},T_{h}) & =V_{0h}+V_{1h}^{(0)}(\chi_{c})+V_{1}^{T}(\chi_{c},T_{h})\nonumber\\
 & =\frac{\mu_{h}^{2}}{2}\chi_{h}^{2}+\frac{\lambda_{h}}{4}\chi_{h}^{4}+\sum_{i}\frac{g_{i}(-1)^{2{\rm s}_{i}}}{64\pi^{2}}m_{i}^{4}(\chi_{c})\left[\ln\left(\frac{m_{i}^{2}(\chi_{c})}{\Lambda_{h}^{2}}\right)-\mathcal{C}_{i}\right]\nonumber \\
 & \,\quad+\frac{T_{h}^{4}}{2\pi^{2}}\sum_{B}g_{B}\int_{0}^{\infty}{\rm d}q\,q^{2}\ln\big[1-{\rm exp}\big(-\sqrt{q^{2}+\big[m_{B}^{(T_{h})}\big]^{2}/T_{h}^{2}}\big)\big]\nonumber \\
 & \,\quad-\frac{T_{h}^{4}}{2\pi^{2}}\sum_{F}g_{F}\int_{0}^{\infty}{\rm d}q\,q^{2}\ln\big[1+{\rm exp}\big(-\sqrt{q^{2}+\big[m_{F}^{(T_{h})}\big]^{2}/T_{h}^{2}}\big)\big]\,. 
\end{align}
This is the potential that is used in the analysis here. 
 In this work we analyze a whole range of temperatures
 which encompass the regions 
 $T_h\ll m$, $T_h\gg m$
  and the regions 
 in between. For this reason we do not use high $T_h$ and low $T_h$ expansions
 but rather use the full integral forms for $J_B$ (and also for $J_F$ in the standard model case).
 Further details on the thermal masses for the hidden sector are given in Appendix A and a summary
 of the temperature dependent potential for  the standard model including corrections due to
  thermal masses and daisy contributions is given in Appendix B.  
 %%%%%%%%%%%%%%%%%%%%%
    
    Let us now consider the case of two sectors together but with no
 interactions between the scalar fields so that the scalar potential is simply
 a sum of potentials in the two sectors, i.e., 
 \begin{align}
     V_{\rm eff}(\phi_c,T; \chi_c,T_h)= V^{\rm v}_{\rm eff}(\phi_c,T) + V^{\rm h}_{\rm eff}(\chi_c,T_h).
     \label{veff-phi-chi}
 \end{align}
 where $V^{\rm v}_{\rm eff}(\phi_c,T)$ is the effective temperature dependent Higgs potential in the standard model,
 which is well known but for easy reference it is given in Appendix B. Here the minimization conditions are 
\begin{align}
  V^{\rm v}_{{\rm eff},\phi_c}=0, V^{\rm h}_{{\rm eff},\chi_c}=0, V^{\rm v}_{{\rm eff},\phi_c\phi_c}>0,  V^{\rm h}_{{\rm eff},\chi_c\chi_c}>0,
  \label{2-minima}
\end{align}
which imply that if the minimization conditions are individually satisfied 
in each sector then the minimization  of the potential overall is also satisfied for the combined system of the visible and the hidden sectors. At the minimum of the potential we define $v=\phi_c$ and $v_h=\chi_c$.
We note, however,
that the two potentials are at different temperatures, one at $T$ and the other at
$T_h$, and 
for a synchronous minimization to occur in the two sectors 
$T$ and $T_h$ must be related by 
\begin{align}
    T_h=\xi(T) T \label{eqxi}
\end{align} 
where $\xi(T)$ is determined
by a synchronous evolution of the visible sector and the hidden sector
from the reheating scale to the low temperature scale where 
phase transitions occur with given initial 
condition on $\xi_0$ at the reheat temperature. In the absence of a synchronous evolution 
$\xi$ has been used \cite{Breitbach:2018ddu} as a free parameter. However, such a procedure
 does not allow one to use temperature constraints consistently at different temperatures such as at
  the time of tunnelings which occur at different temperatures for the visible and the hidden sector
  and to correlate them with the $\Delta N_{eff}$ constraint the BBN time.
In this work, we will solve $\xi(T)$ as a function of $T$ which gives more reliable results. Further, 
as noted earlier we can reliably extrapolate the data to BBN time to include 
  the constraint from $\Delta N_{eff}$ \cite{Aboubrahim:2022gjb,Li:2023nez,Nath:2024mgr}  
  and  from the relic density of dark matter. 
%%%%%%%%%%%%%%%%%%%%%%%\\

\subsection{Synchronous  evolution of coupled hidden and visible sectors
\label{sec:2.2}}

We discuss below an analysis for the evolution of $\xi(T)$ which in general allows for any type of  thermal contact between the visible and the hidden sectors.
Since the standard model explains quite accurately  
a large amount of data at the electroweak scale, the couplings between the hidden and the 
visible sectors need to be extra-weak\cite{Feldman:2006wd} or 
feeble. 
Such couplings could arise via a Higgs portal\cite{Patt:2006fw},
kinetic mixing\cite{Holdom:1985ag} or Stueckelberg mass mixing\cite{Kors:2004dx}
or both\cite{Feldman:2007wj}, as well as other possible combinations such
as Stueckelberg-Higgs portal \cite{Du:2022fqv}, or some higher dimensional operator connecting the two sectors.  
Synchronous thermal evolution between the visible and one hidden sector was discussed  in\cite{Aboubrahim:2020lnr}, for the case with two hidden sectors was discussed in \cite{Aboubrahim:2021ohe} and for multiple hidden sectors in 
 \cite{Aboubrahim:2022bzk}. 
  Here we give a brief review of synchronous evolution
 central to the analysis of this work. 
 
 Thus the energy densities for the visible and the hidden
sectors obey the following coupled Boltzmann equations in  an expanding universe
\begin{align}
&\frac{d\rho_v}{dt} + 3 H(\rho_v+p_v)=j_v,\non
&\frac{d\rho_h}{dt} + 3 H(\rho_h+p_h)=j_h.
\label{2.1} 
\end{align}
Here $\rho_v$ and $p_v$ are the energy and momentum densities
for the visible  sector, 
  and where $(j_v,j_h)$ encode in them all the possible processes exchanging energy between these sectors. They are defined in Appendices C and D. 
The total energy density $\rho=\rho_v+\rho_h$ satisfies the equation
\begin{align} 
\frac{d\rho}{dt}+ 3H(\rho+p)=0,
\end{align}
where $p=p_v+p_h$ is the total pressure density.  We introduce 
 the functions  $\sigma_i= \frac{3}{4}(1+\frac{p_i}{\rho_i})$, where $\sigma_1=\sigma_v, \sigma_2=\sigma_h$ where $\sigma_i=\frac{3}{4}$ for matter dominance and $\sigma_i=1$
for radiation dominance. Similarly we define $\sigma=  \frac{3}{4}(1+\frac{p}{\rho})$.   We note that
 $\sigma_v$, $\sigma_h$, $\sigma$ are temperature dependent  
  and this dependence is taken into account in the evolution equations.
  Using $\sigma's$ the $\rho_i, \rho$  the  evolution equations read
\begin{align} 
\frac{d\rho_i}{dt}+ 4H\sigma_i\rho_i=j_{i}, (i=v, h),
~~\frac{d\rho}{dt}+ 4H\sigma\rho=0.
\label{1.27}
\end{align}
We will use temperature in stead of time and  temperature of the 
visible sector $T$ as the clock.  In this case we can write the evolution 
equations in terms of $T$ using the relation
\begin{align} 
\frac{dT}{dt} = -{4H\sigma\rho}(\frac{d\rho}{dT})^{-1},
\end{align}
and  $d\rho_i/dt=(d\rho_i/dT)(dT/dt)$. Further, 
we can deduce the following evolution equation for $\xi(T)$ which governs the
temperature evolution of the hidden sector relative to that of  the visible sector
\begin{align}
\frac{d\xi}{dT}= \left[ -\xi \frac{d\rho_h}{dT_h} +
\frac{4H\sigma_h\rho_h-j_h}{4H\sigma\rho-4H\sigma_h\rho_h+ j_h} \frac{d\rho_v}{dT}\right] (T \frac{d\rho_h}{dT_h})^{-1}.
\label{DE1}
\end{align}
where $j_h$ is defined in Eq.(\ref{y7}). 
The above analysis is general allowing for any type of 
thermal contact via any type of portal.
In the analysis here we assume a kinetic mixing and do not consider Stueckelberg
mass mixing as it would lead to milli-charges for dark matter\cite{Kors:2004dx,Cheung:2007ut,Kors:2004hz,Aboubrahim:2021ycj}.
Thus we include in the Lagrangian a term 
$-\frac{\delta}{2} A^{\mu\nu}B_{\mu\nu}$ where $B_{\mu\nu}$ is the 
field strength of $U(1)_Y$ hypercharge field $B_\mu$. Further details of the interactions 
between the visible and the hidden sector in the canonically diagonalized basis are given 
in Appendix C.

The evolution equation for $\xi(T)$, Eq.(\ref{DE1}) involves $j_h$ which depends on the 
 yields of the hidden sector
   $Y_D,Y_{\gamma'}$ and $Y_{\chi}$ (see Appendix D). We discuss the Boltzmann equations for
   the yields below
\begin{align}
        \frac{dY_D}{dT}  =& -\frac{\bold{\mathbb{s}}}{H}\left( \frac{d\rho_v/dT}{{4\sigma\rho-4\sigma_h\rho_h}+j_h/H}\right) \Big[\frac{1}{2}\left<\sigma v\right>_{D\bar{D} \rightarrow i\bar{i} }(T)(Y_D^{eq}(T)^2-Y_D^2) \non
   &  -\frac{1}{2}\left<\sigma v\right>_{D\bar{D} \rightarrow \gamma'{\gamma'} }(T_h)\left(Y_D^2-Y_D^{eq}(T_h)^2\frac{Y_{\gamma'}^2}{Y_{\gamma'}^{eq}(T_h)^2}\right)
   \Big]\,,\label{DE2}
  \end{align}
  \begin{align}
    \frac{dY_{\gamma'}}{dT}  =& -\frac{\bold{\mathbb{s}}}{H}\left( \frac{d\rho_v/dT}{{4\sigma\rho-4\sigma_h\rho_h}+j_h/H}\right)  \left[ \frac{1}{2}\left<\sigma v\right>_{D\bar{D} \rightarrow \gamma'{\gamma'} }(T_h)\left(Y_D^2-Y_D^{eq}(T_h)^2\frac{Y_{\gamma'}^2}{Y_{\gamma'}^{eq}(T_h)^2}\right) +\right.\non
    &\frac{1}{2}\left<\sigma v\right>_{\chi\bar{\chi} \rightarrow \gamma'{\gamma'} }(T_h)\left(Y_\chi^2-Y_\chi^{eq}(T_h)^2\frac{Y_{\gamma'}^2}{Y_{\gamma'}^{eq}(T_h)^2}\right)+\left<\sigma v\right>_{i\bar{i}\rightarrow \gamma' }(T)Y_{i}^{eq}(T)^2\non
    &\left.-\frac{1}{\bold{\mathbb{s}}}\left<\Gamma_{\gamma'\rightarrow i\bar{i}}(T_h)\right>Y_{\gamma'}+\left<\Gamma_{\chi\rightarrow \gamma'\gamma'}(T_h)\right>\left(Y_{\chi}-Y^{eq}_\chi(T_h)\frac{Y_{\gamma'}^2}{Y^{eq}_{\gamma'}(T_h)^2}\right)\right]\,,
    \label{DE3}
    \end{align}
    \begin{align}
      \frac{dY_\chi}{dT}  =& -\frac{\bold{\mathbb{s}}}{H}\left( \frac{d\rho_v/dT}{{4\sigma\rho-4\sigma_h\rho_h}+j_h/H}\right)  \big[-\frac{1}{2}\left<\sigma v\right>_{\chi\bar{\chi} \rightarrow \gamma'{\gamma'} }(T_h)\left(Y_\chi^2-Y_\chi^{eq}(T_h)^2\frac{Y_{\gamma'}^2}{Y_{\gamma'}^{eq}(T_h)^2}\right) \non
    &-\frac{1}{\bold{\mathbb{s}}}\left<\Gamma_{\chi\rightarrow \gamma'\gamma'}(T_h)\right>\left(Y_{\chi}-Y^{eq}_\chi(T_h)\frac{Y_{\gamma'}^2}{Y^{eq}_{\gamma'}(T_h)^2}\right) \big]\,.\label{DE4}
\end{align}
where $\bold{\mathbb{s}}$ is the entropy density and yield for particle $i$ is defined by 
$Y_i=n_i/{\bold{\mathbb{s}}}$.
In the analysis here we take account of the hidden sector energy density and pressure density, $\rho_h$ and $p_h$, not only through thermal equilibrium analysis but also by accounting for the contribution of relic abundance. A further discussion of it is provided in Appendix E. For the computation of the visible sector density and pressure we
use the pre-calculated values of  $g^v_{\rm eff}$ and $h^v_{\rm eff}$ which are tabulated results from micrOMEGAs \cite{Belanger:2018ccd}. 
We discuss next the bubble nucleation for the case of the single field first and then for the case of two fields.
%%%%%%%%%%%%%%
\section{Nucleation and vacuum decay \label{sec:3}}
\subsection{Single field nucleation}
  Before proceeding to a discussion of nucleation for the two-field case, we
  first summarize the 
  first-order phase transition involving the decay of the false vacuum
  into the true vacuum involving bubble nucleation of a generic scalar field $\phi$.
   We define the temperature when bubbles start to nucleate as $T_n$. 
   Here at  zero temperature the decay probability per unit time and per unit
 volume is given by $\Gamma= K e^{-S_{4}}$, where $S_{4}$ is the 
 Euclidean action in four dimensions  and $K$ is typically order the fourth
 power of the energy involved in the phase transition\cite{Adams:1993zs}. 
 At finite temperature the  decay probability per unit time and per unit
 volume takes the form $\Gamma= K(T) e^{-S_{3}/T}$ where
 $T$ is the temperature, and $K(T)\sim T^4$. Thus for the case of a single scalar
 field $S_3(T)$ is given by
\begin{align}
    S_3(T) = \int_0^{\infty}4\pi r^2dr \left[\frac{1}{2}\left(\frac{d\phi}{dr}\right)^2+V_{\rm eff}^v(\phi,T)\right]\,,\label{Eq.S3singlefield}
\end{align}
 with the scalar field satisfying the Euclidean O(3) symmetry equation of motion and the appropriate boundary conditions
\begin{align}
    \frac{d^2\phi}{dr^2} + \frac{2}{r}\frac{d\phi}{dr}=\frac{\partial}{\partial\phi}V_{\rm eff}^v(\phi,T),\quad \lim_{r\rightarrow\infty}\phi = 0,\quad \frac{d\phi}{dr}\left.\right|_{r=0}=0.
\end{align}
We use the Mathematica package FindBounce\cite{Guada:2018jek,Guada:2020xnz} to numerically compute $S_3$. 
Once $S_3(T)$ is determined, the nucleation temperature $T_n$ is defined so that
\begin{align}
\int_0^{T_n} \frac{\Gamma dt}{H^3} = \int_{T_n}^{\infty} 
\frac{dT}{T} (\frac{90}{8\pi^3 g_{eff}})^2 (\frac{M_{Pl}}{T})^4 
e^{-S_{3}(T)/T}\simeq 1. \label{Eq.nucl}
\end{align}
This equation is well-approximated by $\frac{S_3(T_n)}{T_n}\sim 140$.
Then the whole vacuum decay process can be characterized by the following temperatures:
      (1) Critical temperature $T_c$: when the effective potential has two degenerate minima.
      (2) Nucleation temperature $T_n$: when the transition occurs or when one bubble is nucleated in one casual Hubble volume.
      (3) Destabilization temperature $T_0$: when the original vacuum is no longer a minimum or when the potential barriers between the false vacuum and true vacuum disappears.
\subsection{Two-field nucleation}
For the two-field nucleation, the calculation here will become complicated because the overshoot/undershoot implementation by some numerical analysis (like CosmoTransitions) is not reliable anymore. The work of \cite{Wainwright:2011kj} discusses such a problem in detail. Thus here we provided a way that can deal with such a situation with the potential given by Eq.(\ref{veff-phi-chi}).
 For the visible (hidden) sector, there will be corresponding temperatures $T_c,T_n, T_0$ ($T_{h,c}, T_{h,n}, T_{h,0}$) with the following orders
  \begin{align}
      T_0<T_n< T_c \quad(T_{h,0}< T_{h,n}< T_{h,c}).
  \end{align}
Since we have $\xi(T)$ to give us the temperature ratio of two sectors at each moment, if we know the temperature of the visible sector, we can then easily find the temperature in the hidden sector, with Eq.(\ref{eqxi}). Correspondingly, we define another function
\begin{align}
    T = \zeta(T_h)T_h 
    \label{eqzeta}
\end{align}
which allows to fix the  temperature in visible sector given the temperature in the hidden sector. 
{We note that  Eqs. (\ref{eqxi}) and (\ref{eqzeta}) are equivalent
so that $\zeta(T_h)=\xi(T)^{-1}$. It is convenient to use Eq.(\ref{eqxi}) [Eq.(\ref{eqzeta})]  when the visible [hidden] sector temperature $T$ [$T_h$] is used as the clock.}
Next we discuss different cases for the nucleation process. \\

Case 1: For this case, we have one of the scalar fields nucleation occurring first and
then the other scalar field nucleation occurring
separately at different time, which means the first scalar fields already reaches its destabilization temperature before the other scalar field reaches its critical temperature, i.e.$(T_0,T_c)\cap (\zeta(T_{h,0})T_{h,0},\zeta(T_{h,c}){T_{h,c}}) = \emptyset$. For this case, the original vacuum  decays first to an intermediate vacuum and then decays into the true vacuum. In this case, we can treat the two-field nucleation
as two single-field vacuum decay problems. Here $T_n$ and ${T_{h,n}}$ can be determined by Eq.(\ref{Eq.S3singlefield} - \ref{Eq.nucl}). 
\begin{figure}[h]
    \centering
    \includegraphics[width=0.4\textwidth]{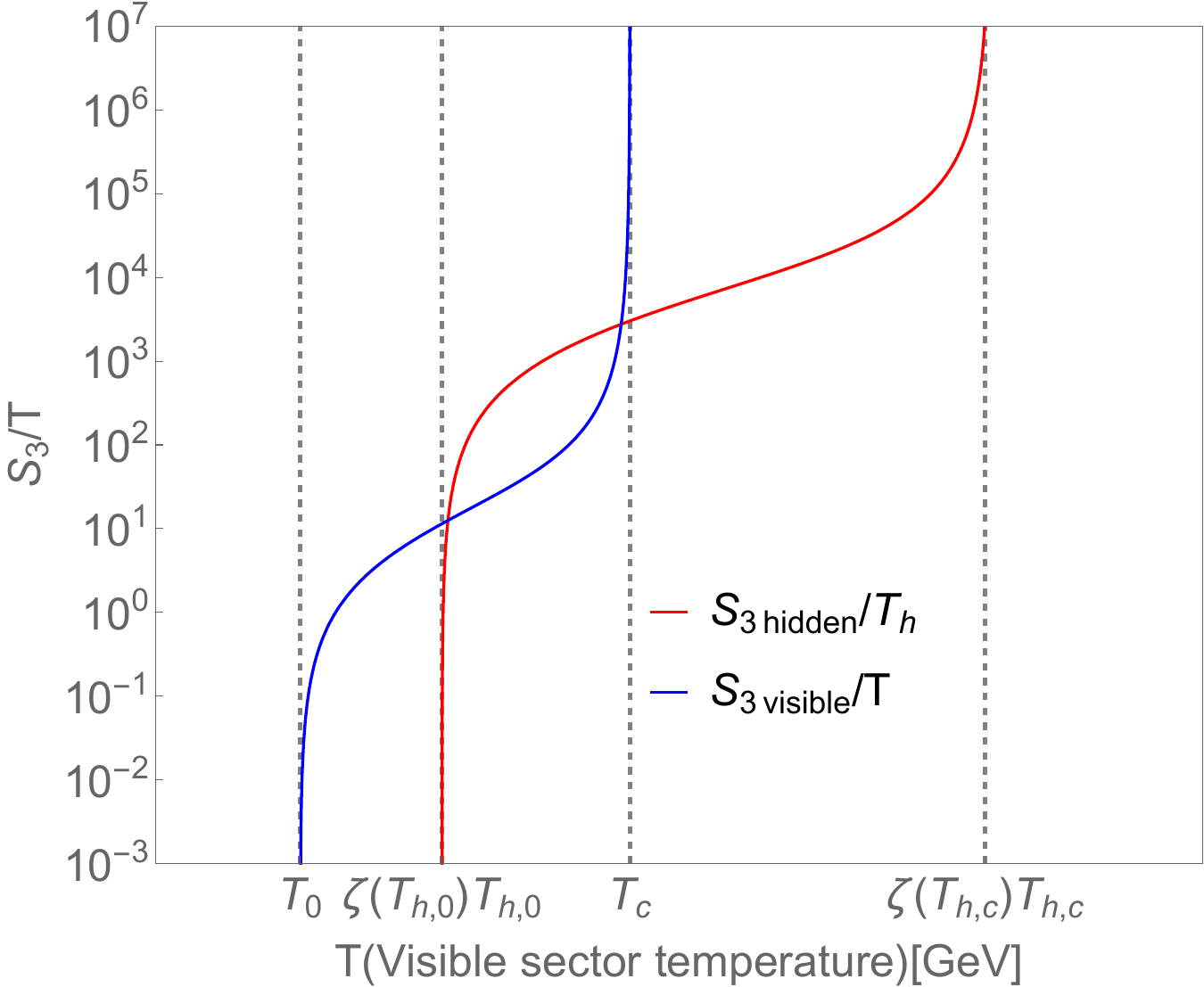}
    \caption{Schematic plot for Case 2. $S_3(T)/T$ v.s. $T$ for $S_{3v}$ and $S_{3h}$.}
    \label{fig:S3schematic}
\end{figure}

Case 2:  For the second case, we have the visible scalar field nucleation and the hidden scalar field nucleation going through the vacuum decay at the same time, which means one of the scalar field reaches its critical temperature before the other scalar field reaches its destabilization temperature, i.e.$(T_0,T_c)\cap (\zeta(T_{h,0})T_{h,0},\zeta(T_{h,c}){T_{h,c}}) \neq \emptyset$. 
For this case, it is possible that the original vacuum decays directly to the final true vacuum and we will have only one transition. Here let us first assume that $T_c < \zeta(T_{h,c}){T_{h,c}}$ so we have
\begin{align}
    T_0<\zeta(T_{h,0}){T_{h,0}}<T_c<\zeta(T_{h,c}){T_{h,c}}
\end{align}
Fig.(\ref{fig:S3schematic}) shows a schematic diagram for such a case. If the first nucleation occurs at $T_c< T< \zeta(T_{h,c}){T_{h,c}}$, then it will be the same as in Case 1 where there will be an intermediate vacuum. If not, then we need consider the possibility that the original vacuum decays directly to the final true vacuum. According to the Eq.(\ref{veff-phi-chi}), there is no interaction between two scalar fields in the potential, i.e there is no terms like $\phi\chi$.  In this case $ S_{3total}(T)$ is given by
\begin{align}
    S_{3total}(T) & = \int_0^{\infty}4\pi r^2dr \left[\frac{1}{2}\left(\frac{d\phi}{dr}\right)^2 + \frac{1}{2}\left(\frac{d\chi}{dr}\right)^2+ V_{\rm eff}(\phi,T; \chi,\xi(T)T)\right] \nonumber\\
    & = \int_0^{\infty}4\pi r^2dr \left[\frac{1}{2}\left(\frac{d\phi}{dr}\right)^2+V^{\rm v}_{\rm eff}(\phi,T)\right] + \int_0^{\infty}4\pi r^2dr \left[\frac{1}{2}\left(\frac{d\chi}{dr}\right)^2+
    V^{\rm h}_{\rm eff}(\chi,\xi(T)T)\right]  \nonumber\\
    & = S_{3v}(T) + S_{3h}(\xi(T)T)
    \label{Eq.s3}
\end{align}
Here the equations of motion are to be solved with Euclidean $O(3)$ symmetry and with appropriate boundary conditions so that 
\begin{align}
    \frac{d^2\phi}{dr^2} + \frac{2}{r}\frac{d\phi}{dr}=
    \frac{\partial}{\partial\phi}V_{\rm eff}(\phi,T; \chi,\xi(T)T)=
    \frac{\partial}{\partial\phi}V^{\rm v}_{\rm eff}(\phi,T),\quad \lim_{r\rightarrow\infty}\phi = 0,\quad \frac{d\phi}{dr}\left.\right|_{r=0}=0.\\
    \frac{d^2\chi}{dr^2} + \frac{2}{r}\frac{d\chi}{dr}=
    \frac{\partial}{\partial\chi}V_{\rm eff}(\phi,T; \chi,\xi(T)T)=
    \frac{\partial}{\partial\chi}V^{\rm h}_{\rm eff}(\chi,T_h),\quad \lim_{r\rightarrow\infty}\chi = 0,\quad \frac{d\chi}{dr}\left.\right|_{r=0}=0.
\end{align}
Since the above two equations can be solved independently, we can just treat each as a single field case. To find $T_n$ for the case original vacuum decays directly to the final true vacuum,
we first assume that such a nucleation happens at $T_{n,total}$ and get
\begin{align}
    \frac{S_{\rm 3toal}(T_{n,total})}{T_{n,total}} =  \frac{S_{\rm 3v}(T_{n,total})}{T_{n,total}} +\frac{S_{\rm 3h}(\xi(T_{n,total})T_{n,total})}{T_{n,total}}\sim 140
\end{align}
which tells us that 
   $\frac{S_{\rm 3v}(T_{n,total})}{T_{n,total}  }< 140$. 
However,  $S_{3v}/T$ is a monotonic increasing function of $T$
which leads to
\begin{align}
 \exists~   T_n>T_{n,total}: \frac{S_{\rm 3v}(T_n)}{T_n  }\sim 140
\end{align}
It tells us that before the original vacuum decays directly
into the final true vacuum, it must decay into an intermediate vacuum first. 
However, it takes some time for the phase transition to complete after the temperature reaches the nucleation temperature. Thus, it is possible that the other sector also reaches its nucleation temperature during this process. It will become more complicated but interesting because now the gravitational wave will be generated by the collision of two types of bubbles from two different sectors.
   {We note, however,  that the interaction between the sectors is too feeble
 to produce any effect.} Further, we assume the phase transition is completed immediately after it reaches the nucleation temperature to avoid this problem {altogether.} 
\begin{figure}
    \centering
    \includegraphics[width=0.6\linewidth]{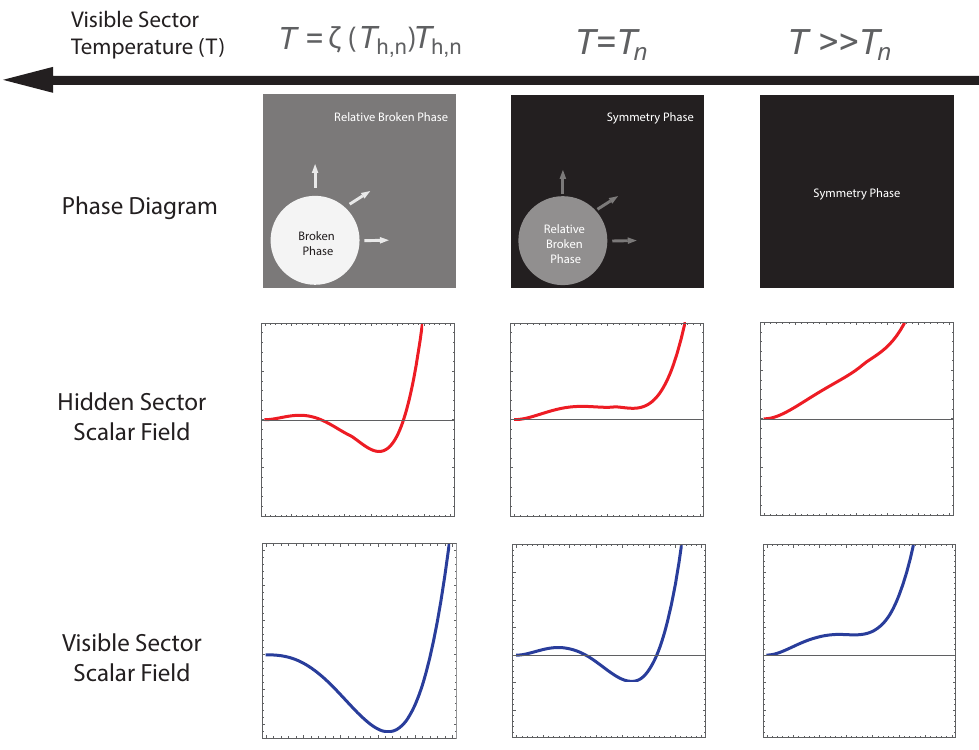}
    \caption{A schematic diagram for the two-step phase transition. The phase of the whole universe transfers from the symmetry phase to the relative broken phase before ultimately reaching the fully broken phase. The plots of $V^{\rm v}_{\rm eff}$ v.s. $\phi$ and $V^{\rm h}_{\rm eff}$ v.s. $\chi$ are shown for different temperatures. }
    \label{fig:schematic}
\end{figure}
Therefore, Case 2 can be treated the same as Case 1 and we have shown that in any case, we can treat the problem as two single-field vacuum decay problems. 
The whole transition will undergo through two-step phase transition at two different nucleation temperatures $T_{n}$ and $T_{h,n}$. For each phase transition, there will be a ``relative symmetry phase" and a ``relative broken phase". A schematic diagram of the entire nucleation process is given by Fig.(\ref{fig:schematic}). 
%%%%%%%%%%%%%
\section{Hydrodynamics of bubble formation during phase transition
\label{sec:4}}
To investigate the gravitational wave generation from cosmological phase transition, we need to first 
study the hydrodynamics of bubble formation during the phase transition. One of the important
 elements in the hydrodynamics of bubbles is the sound velocity in the fluid in the symmetric
 phase and in the broken phase and it is model dependent. We discuss this next.

 \subsection{Sound velocities in the visible and in the hidden sectors \label{sec:4.1}}
  Sound velocity in fluids is known to have
a terminal value so that $c_s^2=1/3$. However, its actual value depends
on whether the phase is unbroken or broken and on the type of the broken phase. We 
start with the thermodynamic quantities: energy density $e$, pressure $p$, and enthalpy density $w$. They are in general given by the following equations.
\begin{align}
    p = -\mathcal{F},\quad e = T\frac{\partial p}{\partial T} - p,\quad w = p+e. 
\end{align}
Here $\mathcal{F}$ is the free energy density where $p$ is given by 
\begin{align}
p(\phi,T;\chi,T_h) &= p_v(\phi,T) + p_h(\chi,T_h) \\
p_v(\phi,T) &=\frac{\pi^2}{90}g^v_{\rm eff} T^4 - V^{\rm v}_{\rm eff}(\phi,T), \label{pv}\\
p_h(\chi,T_h)& =\frac{\pi^2}{90}  g^h_{\rm eff}T_h^4- V^{\rm _h}_{\rm eff}(\chi,T_h)\label{ph}
\end{align}
Correspondingly, we have 
\begin{align}
 e(\phi,T;\chi,T_h)
    &= e_v(\phi,T) + e_h(\chi,T_h) + e_{mix}(\phi,T,\chi,T_h),\\
    e_v(\phi,T)& =T\frac{\partial p_v}{\partial T} - p_v,~~
e_h(\chi,T_h) =T_h\frac{\partial p_h}{\partial T_h} - p_h,\\
e_{mix}(\phi,T;\chi,T_h) &=T^2\frac{\partial p_h}{\partial T_h}\frac{\partial\xi}{\partial T}
\end{align}
and for the sound velocity $c_s^2 = d p/d e$ (total derivative here), we have 
\begin{align}
    c_s^2(\phi,\chi;T,T_h)&
    = \frac{\frac{\partial(p_v)}{\partial T} + \frac{\partial (p_h)}{\partial T_h} \frac{\partial(T_h)}{\partial T }}{\frac{\partial (e_v)}{\partial  T}+ \frac{\partial (e_h)}{\partial T_h}\frac{\partial (T_h)}{\partial T}+\frac{\partial e_{mix}}{\partial T} }  
    \label{Eq.cs2}
\end{align}
where
\begin{align}
    \frac{\partial T_h}{\partial T} = \frac{\partial(\xi T)}{\partial T} = \frac{\partial \xi}{\partial T}T+\xi(T)\,,
\end{align}
and where we used  $\frac{d p_v}{dT} = \frac{\partial p_v}{\partial T},\frac{d e_v}{dT} = \frac{\partial e_v}{\partial T}$ since we are interested in sound velocity in vacuums. 
Further, since explicit integrals for $p_h, e_h$ are known, and numerical tables for
the corresponding visible sector quantities are known, an evaluation of 
$\partial(p_v)/\partial T, \partial(e_v)/\partial T$ and $\partial(p_h)/\partial T_h, \partial(e_h)/\partial T_h$ can be numerically carried out. According to the analysis in the section \ref{sec:3}, there will actually be two sets of symmetry phase velocities and broken phase velocities
possible. For the visible scalar field nucleation, the sound velocity in the symmetric phase and in
the broken phase, i.e. sound velocity outside and inside the bubble of the visible scalar field nucleation, will be labeled $c_{s,+,v}$ and $c_{s,-,v}$. Similarly, for the hidden scalar field nucleation, we have $c_{s,+,h}$ and $c_{s,-,h}$.

We label vacua in the broken phase case for the visible and hidden sectors to be
$\phi_{min}$ and $\chi_{min}$, and they are found numerically. For the case when
nucleation in the visible sector occurs before nucleation in the hidden sector, i.e., 
${T_n}>\zeta(T_{h,n}){T_{h,n}}$, these four velocities are given by
\begin{align}
    c^2_{s,+,v} &=  c_s^2(0,0,T_n,\xi(T_n)T_n)\nonumber\\
    c^2_{s,-,v} &=  c_s^2(\phi_{min},0,T_n,\xi(T_n)T_n)\nonumber\\
    c^2_{s,+,h} &=  c_s^2(\phi_{min},0,\zeta(T_{h,n}){T_{h,n}},T_{h,n})\nonumber\\
    c^2_{s,-,h} &=  c_s^2(\phi_{min},\chi_{min},\zeta(T_{h,n}){T_{h,n}},T_{h,n})\label{csvh}
\end{align}
where the arguments of  $c^2_{s,+,v}$ etc are as defined in Eq. (\ref{Eq.cs2}).
Thus, e.g., $c^2_s(\phi_{min},0,T_n, \xi(T_n)T_n)$ denotes the velocity of the sound wave traveling inside the bubble of the visible phase transition. The visible scalar field is in its broken vacuum while the hidden scalar field is still in its symmetric vacuum. The tunneling temperature of the visible scalar field nucleation is $T_n$ 
and the synchronous temperature of the hidden scalar to it is $\xi(T_n) T_n$.

On the other hand, when ${T_n}<\zeta(T_{h,n}){T_{h,n}}$, these four velocities are given by
\begin{align}
    c^2_{s,+,h} &=  c_s^2(0,0,\zeta(T_{h,n}){T_{h,n}},T_{h,n})\nonumber\\
    c^2_{s,-,h} &=  c_s^2(0,\chi_{min},\zeta(T_{h,n}){T_{h,n}},T_{h,n})\nonumber\\
    c^2_{s,+,v} &=  c_s^2(0,\chi_{min},T_n,\xi(T_n)T_n)\nonumber\\
    c^2_{s,-,v} &=  c_s^2(\phi_{min},\chi_{min},T_n,\xi(T_n)T_n)
   \label{cshv}
\end{align}

\subsection{Relativistic fluid equations and bubble dynamics}
Next, we discuss  hydrodynamics of the bubble expansion\cite{landau,courant,Steinhardt:1981ct,
Kamionkowski:1993fg,Espinosa:2010hh,Wang:2022lyd}. First, we describe the plasma, 
as a relativistic fluid, by its energy-momentum tensor
\begin{align}
    T^{\mu\nu} = w u^{\mu}u^{\nu} + p{g}^{\mu\nu},
\end{align}
Here we are using the metric $g^{\mu\nu}={\rm diag}(-1, 1,1,1)$ where 
$w = e+p$, and $e$ and $p$ are the energy density and pressure as defined 
in section \ref{sec:4.1}, and
 $u^{\mu} = \gamma(v)(1,\vec{v}) (\gamma=1/\sqrt{1-v^2})$ is the four-velocity field.
  The fluid equation of motion is given by the conservation of $T^{\mu\nu}$ 
\begin{align}
    \partial_{\mu}T^{\mu\nu}=\mu^{\nu}\partial_{\mu}(u^{\mu}w)+u^{\mu}w\partial_{\mu}u^{\nu} + \partial^{\nu}p=0.
\end{align}
%%%%%%%%%%%%%%%%%%%%%%%%%%%%%%%%%%%%%%%
The conservation equation can be projected into the 
parallel and perpendicular directions to the flow direction by using $u^{\mu} = \gamma(v)(1,\vec{v})$ and $\bar{u}^{\mu} = \gamma(v)(v,\vec{v}/v)$ such that $\bar{u}_{\mu}u^{\mu} = 0,u_{\nu}\partial_{\mu}u^{\nu} = 0, \bar{u}^2 = 1, u^2 = -1$ which give
\begin{align}
    u_{\nu}\partial_{\mu}T^{\mu\nu}&=\partial_{\mu}(u^{\mu}w)+u^{\mu}\partial_{\mu}p = 0 \\
    \bar{u}_{\nu}\partial_{\mu}T^{\mu\nu}&=\bar{u}_{\nu}u^{\mu}w\partial_{\mu}u^{\nu}+\bar{u}^{\mu}\partial_{\mu}p = 0.
\end{align}
These are the continuity equation and the relativistic Euler equation. 
Further, one assumes a spherically symmetric configuration and since there is no characteristic distance scale in the problem, the solution depends only on a self-similarity coordinate 
$\eta\equiv r/t$, where r is the distance to the bubble center and t is the time since the bubble nucleation. Further, we assume that the bubble reaches a constant terminal velocity after a short expansion time. Thus we can assume that $v_b = \eta_w$. The above two equations 
then take the form
\begin{align}
    (\eta-v)\frac{\partial_{\eta}e}{w} &= 2\frac{v}{\eta}+\gamma^2(1-\eta v)\partial_{\eta}v\,,\\
    (1-\eta v)\frac{\partial_{\eta}p}{w}& = \gamma^2(\eta-v)\partial_{\eta}v\,,
\end{align}
where
$v(\eta)$ is  the fluid velocity at $r = \eta t$ in the frame of the bubble center. Using the definition $c_s^2 = \frac{dp/dT}{de/dT}$,  one gets the following equations
\begin{align}
    2\frac{v}{\eta} = \gamma^2(1-v\eta)(\frac{\mu(\eta,v)^2}{c_s^2}-1)\frac{dv}{d\eta},\label{Eq.dvdxi}\\
    \frac{dw}{d\eta} = w\gamma^2\mu(\eta,v)(\frac{1}{c_s^2}+1)\frac{dv}{d\eta}\label{Eq.dwdxi}
\end{align}
where $\mu(\eta,v) = \frac{\eta-v}{1-\eta v}$.
 In fact, with a steady terminal velocity $\eta_w$, we can use this Lorentz-boost transformation to transform between the bubble wall frame and center of bubble frame by the expression $\mu(\eta_w,v) = \bar{v}$ and 
 $\mu(\eta_w,\bar{v}) = v$.  
   Besides the equations of motion of the plasma given above, we also need junction conditions to connect the symmetry phase and the broken phase. We use subscripts $+$ to denote the symmetric phase and $-$ to denote the broken phase. 
    We note that the junction conditions are to be used infinitely close to the boundary.
   Then assuming the wall is expanding in z-direction, the matching equations are
\begin{align}
    (T_+^{z\nu}-T_-^{z\nu})n_{\nu} = 0, \quad (T_+^{t\nu}-T_-^{t\nu}) n_{\nu} = 0, \quad n_{\mu} = (0,0,0,1)
\end{align}
and we get the continuity equation in the bubble wall frame to be
\begin{align}
    w_+\bar{v}_+\bar{\gamma}_+^2& = w_-\bar{v}_-\bar{\gamma}_-^2 \label{Eq.wboundary}\\
    w_+\bar{v}^2_+\bar{\gamma}_+^2+p_+ &= w_-\bar{v}^2_-\bar{\gamma}_-^2+p_-   
\end{align}
Rearranging it we can get the following equation
\begin{align}
    \bar{v}_+\bar{v}_- = \frac{p_+-p_-}{e_+-e_-}\label{Eq.continuityvv}\\
    \frac{\bar{v}_+}{\bar{v}_-} = \frac{e_-+p_+}{e_+ +p_-}\label{Eq.continuityvoverv}
\end{align}
With boundary condition Eq.(\ref{Eq.continuityvv},\ref{Eq.continuityvoverv}) and the
evolution equation Eq.(\ref{Eq.dvdxi}), we can solve for $v(\eta)$, and there are three different expansion modes: deflagration, hybrid and detonation.  If the wall velocity of the bubble
$v_w$ is subsonic, i.e., $v_w<c_{s,-}$ it gives rise to deflagration where a region of larger density precedes
the bubble wall. For the supersonic case where $v_w>c_{s,-}$, the 
higher density region ahead of the wall does not materialize since the
wall velocity is larger than the sound velocity.  This is the detonation region.
 The region where $v_w\sim c_s$ is a mixture of the two and is referred to 
 as the hybrid region. Once we determine $v({\xi})$ we can apply Eq.(\ref{Eq.dwdxi}) and Eq.(\ref{Eq.wboundary}) to find 
\begin{align}
    w(\eta) = w_0 \exp{\left[\int^{v(\eta)}_{v_0}\left(1+\frac{1}{c_s^2}\right)\gamma^2\mu dv\right]}
\end{align}
The ratio of bulk kinetic energy over the vacuum energy gives the efficiency factor $\kappa$ as
\begin{align}
    \kappa = \frac{3}{\epsilon\eta_w^3}\int w(\eta)v^2\gamma^2\eta^2d\eta\nonumber
\end{align}

In most  analyses of first order phase transition (FOPT), sound velocities are treated approximately often
assuming $c^2_{s,-} = c^2_{s,+} = \frac{1}{3}$ (see, e.g., 
  ~\cite{Espinosa:2010hh} and ~\cite{Wang:2022lyd}). 
 In this case the phase transition strength $\alpha$ is given by
\begin{align}
    \alpha &= \frac{T\frac{d\Delta V_{eff}}{dT} -\Delta V_{eff}}{\rho_{rad}}\label{alphaeq1}
\end{align}
or
\begin{align}
    \alpha' &= \frac{4}{3}\frac{\epsilon_+-\epsilon_-}{w_+} = \frac{\frac{T}{4}\frac{d\Delta V_{eff}}{dT} -\Delta V_{eff}}{\rho_{rad}}\label{alphaeq2}.
\end{align}
However, in this work we  will take into account sound velocity dependence in the analysis as in  \cite{Giese:2020znk} and \cite{Giese:2020rtr}. Here the phase transition strength parameter is given by
\begin{align}
    \alpha_{\bar{\theta}_n} \equiv \frac{D\bar\theta(T_n)}{3w_n}\quad \bar\theta \equiv e-\frac{p}{c_{s,-}^2} \label{alphatheta}\\
    DX(T_n) = X_s(T_n) - X_b(T_n)
\end{align}
with $X = e,p,w$ and the efficiency factor is defined by
\begin{align}
    \kappa = \frac{4}{\alpha_{\bar\theta_n}\eta_w^3}\int d\eta \eta^2v^2\gamma^2\frac{w}{w_n}\label{kappatheta}
\end{align}
In this case $\alpha_{\Bar{\theta}_n}$ and $\kappa$ are both velocity dependent that $\kappa$ depends on $c_{s,+}$, $c_{s,-}$, $\alpha_{\bar\theta_n}$ and $v_w$. We note that for the case $c^2_{s,+} = c^2_{c,b} = 1/3$, it is equivalent to the second definition Eq.(\ref{alphaeq2}). A Python snippet is provided in \cite{Giese:2020znk} and we utilize it in our analysis.
%%%%%%%%%%%%%%%%%%%%%%%%%%%%%%%%%%%%%%%%%%%%
\section{Gravitational wave spectrum with visible and hidden sectors
\label{sec:5}}
The phase transition phenomena is controlled
  by four parameters which are the nucleation temperature $T_n$, the 
  strength of the phase transition $\alpha$, the inverse duration of the 
  transition $\beta$ in comparison with $H_n$ where $H_n$ is the Hubble
  parameter at the time of nucleation and the bubble wall velocity $v_w$.
$T_n$ and $\alpha$ were discussed in section \ref{sec:4}. 
 The time scale
of the phase transition is the inverse of the parameter $\beta$
defined by 
\begin{align}
\beta=\left.-\frac{d(S_3/T)}{dt}\right|_{t=t_n}
 \simeq \left.\frac{1}{\Gamma} 
\frac{d\Gamma}{dt}\right|_{t=t_n} 
\label{2.} 
\end{align}
where $S_3$ is the Euclidean action as already defined. Usually $\beta$ is normalized by $H_n$ and is given by
\begin{align}
    \frac{\beta}{H} = \left.T\frac{d(S_3(T)/T)}{dT}\right|_{T=\{T_n,{T_{h,n}}\}}
    \label{beta}
\end{align}
We note here that a larger  $\alpha$ means a stronger phase transition and a 
 larger value of $\beta$ means a faster phase transition. 
 
Gravitational wave power spectrum has been discussed in a variety of settings
(see, e.g., 
\cite{Kehayias:2009tn,Schwaller:2015tja,Jaeckel:2016jlh,Katz:2016adq,Dev:2019njv,Bian:2019szo,Dror:2019syi,DiBari:2020bvn,Han:2020ekm,Deng:2020dnf,Wang:2022lyd,Wang:2022akn,Jinno:2022fom,Fu:2023mdu,Halverson:2020xpg,Freese:2023fcr}). It is given by
\begin{table}
    \centering
    \begin{tabular}{p{0.1\textwidth}>{\centering\arraybackslash}p{0.25\textwidth}>{\centering\arraybackslash}p{0.25\textwidth}>{\centering\arraybackslash}p{0.25\textwidth}}
    \hline
    \hline
    & Scalar field $\Omega_{\phi}$&Sound waves  $\Omega_{\rm{sw}}$&Turbulence $\Omega_{\rm{turb}}$  \\
    \hline
    $\mathcal{N}$ & 1 &$1.59\times 10^{-1} $ & $2.01\times 10^1$\\    $\kappa_{\phi}$ & $\kappa$ & $\kappa_{\rm{sw}}$ & $\epsilon_{\rm{turb}}\kappa_{\rm{sw}}$\\
    $p$ & 2 & 2& $\frac{3}{2}$\\
    $q$ & 2 & 1& 1\\
    $\Delta$ & $\frac{0.11v_w^3}{0.42+v_w^2}$ & $v_w$ & $v_w$\\
    $f_p$ & $\frac{0.62\beta}{1.8-0.1v_w+v_w^2}$ & $\frac{2\beta}{\sqrt{3}v_w}$ & $\frac{3.5\beta}{2v_w}$\\
    $s(f)$ & $\frac{3.8(f/f_p)^2.8}{1+2.8(f/f_p)^3.8}$ & $(f/f_p)^3 \left(\frac{7}{4+3(f/f_p)^2}\right)^{7/2}$ & $\frac{(f/f_p)^3}{(1+f/f_p)^{11/3}(1+8\pi f/H)}$ \\
    \hline
    \hline
    \end{tabular}

    \caption{Values of the parameters $\mathcal{N}, \kappa, p, q, \Delta,  f_p, s(f)$
    that appear in the gravitational wave power spectrum of Eq.(\ref{gwspec})
   for the three different contributions: $\Omega_{\phi}$ from the scalar field,
  $\Omega_{\rm sw}$ from sound waves,  and  $\Omega_{\rm tb}$  
  from turbulence.} 
    \label{tab:GWcals}
\end{table}
\begin{align}
\Omega_{\rm GW}(f)&=\frac{1}{\rho_c}\frac{d\rho_{\rm GW}}{d\ln f}\simeq \mathcal{N} \Delta \left(\frac{\kappa \alpha}{1+\alpha}\right)^p\left(\frac{H}{\beta}\right)^q s(f),\
~\Omega_{\rm GW}\simeq\Omega_{\phi }+\Omega_{\rm sw}+\Omega_{\rm tb}.
\label{gwspec}
\end{align}
Here $\Omega_{\phi}$ is the contribution to energy density of the 
gravitational wave produced by dynamics of the scalar field,
$\Omega_{\rm sw}$ is the contribution from the sound waves, and
$\Omega_{\rm tb}$ is the contribution by turbulence, $\rho_c$ is the critical density and $f$ is the frequency of the gravitational wave. The rest of the parameters are as discussed in the
text of this section. Further, a detailed discussion of the various contributions can be 
found in \cite{Huber:2008hg,Hindmarsh:2015qta,Caprini:2009yp,Breitbach:2018ddu,Caprini:2015zlo}.
 For the current analysis all the relevant parameters that enter in the computation of $\Omega_{\phi},
\Omega_{\rm sw}, \Omega_{\rm tb}$ which contribute in Eq.(\ref{gwspec}) 
are given in Table.(\ref{tab:GWcals}).
However, we still need to consider the redshift both on the energy density and frequency to deduce the power spectrum $\Omega^0_{GW}(f_0)$ at current temperature $T^0$
from the power spectrum $\Omega_{GW}$ gotten at the tunneling temperature $T_{\rm tun}$. This is accomplished by the following extrapolation 
\cite{Breitbach:2018ddu,Schwaller:2015tja}: 
\begin{align}
   { \Omega^0_{GW}(f_0) = \mathcal{R} \Omega_{GW}\left(\frac{a_0}{a}f_0\right)}
    \label{redshift}
\end{align}
where,
\begin{align}
    {\frac{a_0}{a}} & {= \left(\frac{h_{eff}(T_{\rm tun})}{h^{EQ}_{eff}}\right)^{1/3}\left(\frac{T_{\rm tun}}{T^0}\right)}
  \end{align}
  \begin{align}  
    \mathcal{R} &\equiv \left(\frac{a}{a_0}\right)^4\left(\frac{H}{H_0}\right)^2 \simeq 2.473\times 10^{-5}h^{-2} \left(\frac{h^{EQ}_{eff}}{h_{eff}(T_{\rm tun})}\right)^{4/3}\left(\frac{g_{eff}(T_{\rm tun})}{2}\right)
    \end{align}
    
    \begin{align}
    g_{eff}(T_{\rm tun}) &= g_{eff}^v(T_{\rm tun}) + g_{eff}^h(T_{\rm tun})\xi(T_{\rm tun})^4\\    h_{eff}(T_{\rm tun}) &= h_{eff}^v(T_{\rm tun}) + h_{eff}^h(T_{\rm tun})\xi(T_{\rm tun})^3\\
    h^{EQ}_{eff} &= 3.91 +  h_{eff}^h(T_{eq})\xi(T_{eq})^3.
\end{align}
In the above $T_{\rm tun} = T_n$ for the visible sector nucleation and $T_{\rm tun} = \zeta(T_{h,n})T_{h,n}$ for the hidden sector nucleation. 

It is also necessary to classify whether the bubble wall velocity reaches a terminal velocity. If the bubble wall keeps accelerating, it is called the runaway scenario. If it reaches a terminal velocity, it is called a non-runaway scenario. A detailed discussion can be found in \cite{Breitbach:2018ddu,Caprini:2015zlo,Bodeker:2009qy,Espinosa:2010hh}. To classify these two scenarios, a critical phase transition strength $\alpha_\infty$ is introduced. For the visible sector and hidden sector nucleation, it is given by
\begin{align}
    \alpha_\infty^v &= \frac{(T_n)^2}{\rho_{\rm rad}(T_n)}\left(\sum_{i=bosons}n_i\frac{\Delta m_i^2}{24} + \sum_{i=fermions}n_i\frac{\Delta m_i^2}{48}\right)\\    \alpha_\infty^h &= \frac{(T_{h,n})^2}{\rho_{\rm rad}(\zeta(T_{h,n})T_{h,n})}\left(\sum_{i=bosons}n_i\frac{\Delta m_i^2}{24} + \sum_{i=fermions}n_i\frac{\Delta m_i^2}{48}\right)
\end{align}
When $\alpha_\infty>\alpha$, it is in the non-runaway regime. In this case we have
\begin{align}
    \kappa_\phi = 0,\quad \kappa_{\rm sw} = \kappa(\alpha,c^2_{s,+},c^2_{s,-},v_w) 
\end{align}
When $\alpha_\infty<\alpha$, it is in the runaway regime and we have 
\begin{align}
    \kappa_\phi = 1-\frac{\alpha_\infty}{\alpha},\quad \kappa_{\rm sw} = \frac{\alpha_\infty}{\alpha}\kappa(\alpha_\infty,c^2_{s,+},c^2_{s,-},v_w) 
\end{align}
The bubble wall velocity 
depends on the transition strength $\alpha$ and on the friction between the scalar field and the surrounding particle plasma, described by a friction parameter. Thus {$v_w$} is highly model-dependent.   Since the bubble wall is in the runaway region 
it will keep accelerating, and for reason
we take $v_w\sim 1$. In the non-runaway region, the bubble wall velocity reaches a terminal value, and is model dependent, 
we treat it as a free parameter. It is legal to do so since it is equivalent to introducing additional particles that couple exclusively to the scalar field and affect the friction parameter only {(For recent work on determining wall velocity from initial inputs see, e.g.,
\cite{Bodeker:2017cim,Giese:2020rtr,Giese:2020znk,Ai:2021kak}).}
%%%%%%%%%%%%%%%%%%%%%%%%%%%%% 
 
\section{Simulation of gravitational wave power spectrum\label{sec:6}}
\begin{figure}[H]
    \centering
    \includegraphics[width=0.5\textwidth]{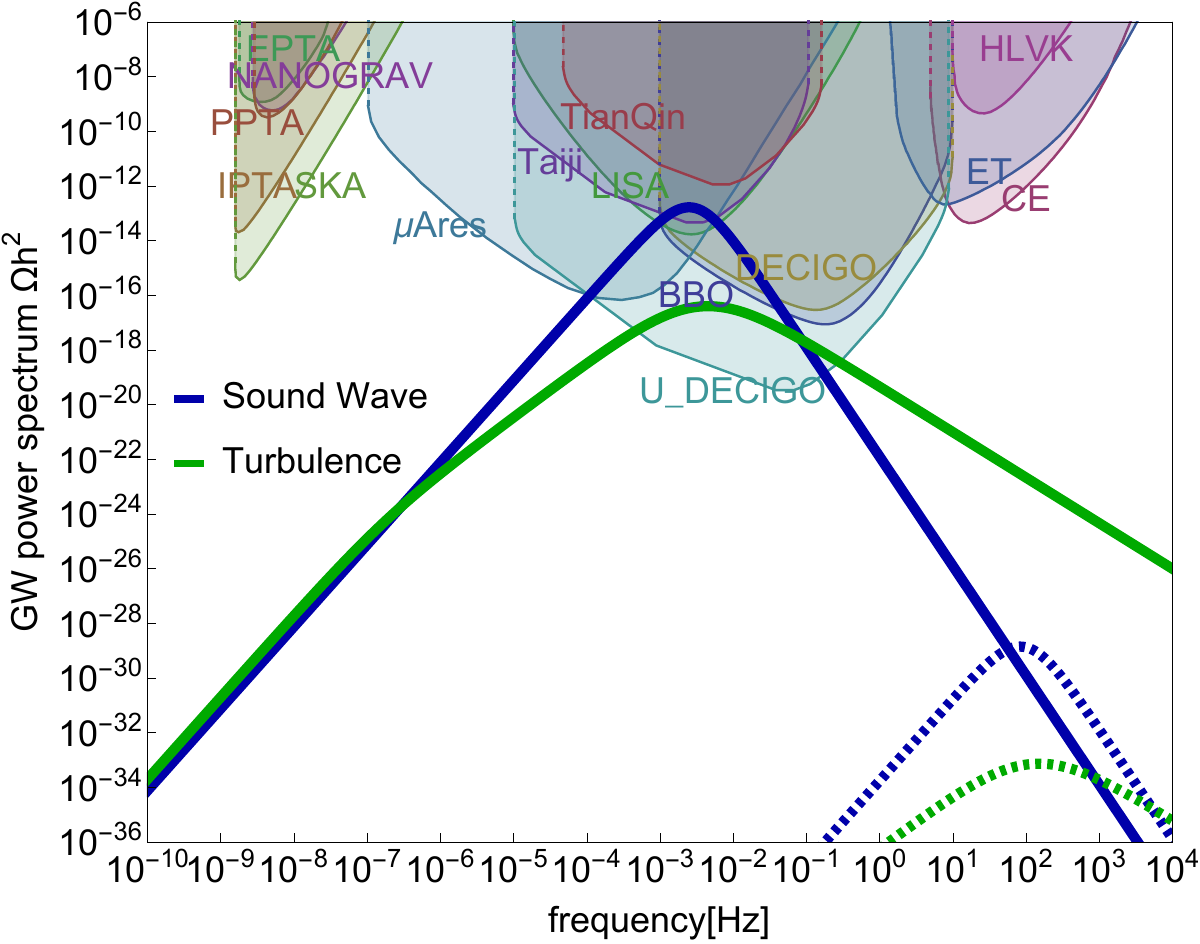}
    \caption{{    
  An exhibition of the gravitational wave power spectrum for model (a) in Table.(\ref{tab:benchmarks}) illustrating the relative contributions from sound wave and turbulence. This is a non-runaway case and $\Omega_{\phi} = 0$. 
    The solid lines are for the hidden sector phase transition while the dashed lines are for the visible sector.  In the analysis we take $\epsilon_{\rm turb}=0.1$  as in \cite{Caprini:2015zlo,Hindmarsh:2015qta}.    
    The regions in color are the power-law integrated sensitivity curves for different experiments, including LISA \cite{LISA:2017pwj,Baker:2019nia,AmaroSeoane:2012km}, EPTA\cite{Kramer:2013kea,Desvignes:2016yex}, 
    HLVK = aLIGO/aVIRGO/KAGRA\cite{Harry:2010zz,VIRGO:2014yos,Somiya:2011np,  TheLIGOScientific:2014jea,TheVirgo:2014hva,Hagihara:2018azu},
    BBO\cite{Grojean:2006bp}, Decigo\cite{Kawamura:2006up}, ET\cite{Punturo:2010zz}, CE\cite{LIGOScientific:2016wof}, Taiji\cite{Ruan:2018tsw}, TianQin\cite{TianQin:2015yph}, $\mu$Ares\cite{Sesana:2019vho},  NANOGrav\cite{NANOGRAV:2018hou,NANOGrav:2023gor}, PPTA\cite{Manchester:2012za}, IPTA\cite{Hobbs:2009yy} and SKA\cite{Weltman:2018zrl}. The data and calculations are from \cite{Schmitz:2020syl}.} 
  }
    \label{fig:GWspectrum0}
\end{figure}

 There are several ongoing gravitational wave experiments and those 
  being proposed that will probe gravitational waves at different frequency regions
  and with different sensitivity.
  These include  LISA \cite{LISA:2017pwj,Baker:2019nia,AmaroSeoane:2012km}, EPTA\cite{Kramer:2013kea,Desvignes:2016yex}, aLIGO/aVIRGO/KAGRA\cite{Harry:2010zz,VIRGO:2014yos,Somiya:2011np,TheLIGOScientific:2014jea,TheVirgo:2014hva,Hagihara:2018azu}, BBO\cite{Grojean:2006bp}, Decigo\cite{Kawamura:2006up}, ET\cite{Punturo:2010zz}, CE\cite{LIGOScientific:2016wof}, Taiji\cite{Ruan:2018tsw}, TianQin\cite{TianQin:2015yph}, $\mu$Ares\cite{Sesana:2019vho}, NANOGrav\cite{NANOGRAV:2018hou,NANOGrav:2023gor}, PPTA\cite{Manchester:2012za}, IPTA\cite{Hobbs:2009yy} and SKA\cite{Weltman:2018zrl}.  
We plot the predictions of the hidden sector model discussed here along
with the expected 
reach of proposed gravitational wave experiments.
Fig.(\ref{fig:GWspectrum0}) provides an example of gravitational wave
 power spectrum  with these experimental constraints for model (a) of 
 Table (2).
  Since the major parameters for SM are already known, we have the visible sector nucleation temperature to be about $161.284$GeV, the phase transition strength $\alpha_v \sim 4\times 10^{-5}$ and the inverse duration of the transition $\beta_v\sim 2.7\times 10^6$. As a result, the direct contribution from the visible sector is very small, which is about $\Omega_{\rm visible}h^2 \sim 10^{-30}$. 

Based on the previous discussion, we calculate the phase transition dynamics and the final gravitational wave power spectrum with different benchmarks on our model. For each model, there are 8 free parameters in total, which are: dark fermion mass $m_D$,  dark photon mass $m_{\gamma'}$, coupling of dark photon and dark fermion $g_x$, kinetic mixing $\delta$, initial temperature ratio $\xi_0$, hidden Higgs field parameter $\mu_h,\lambda_h$, and the bubble wall velocity for hidden sector nucleation $v_{wh}$. Here we provide a table of benchmark models Table.(\ref{tab:benchmarks}) with their outputs given in Table.(\ref{tab:benchmarkresults}).
 \begin{table}[h]
 \centering
     \begin{tabular}{lllllllll}
    \hline
   No. &  $m_D[\text{GeV}]$& $m_{\gamma'}[\text{GeV}]$&$g_X$ & $\delta$(in $10^{-9}$)  & $\xi_0$ & $\mu_h[\text{GeV}]$ & $\lambda_h$ & $v_{wh}$ \\ \hline
(a)&551.7&108.5&0.02059&0.01038&0.671&18.63&0.04973&0.5993\\(b)&204.1&52.52&0.01975&0.01441&0.463&9.922&0.03953&0.5619\\(c)&594.4&221.5&0.002922&0.0281&0.778&41.78&0.08802&0.9599\\(d)&710.&138.6&0.003161&0.03012&0.917&22.03&0.02939&0.6472\\(e)&1111.&113.7&0.02739&0.01174&0.821&18.49&0.03857&0.2674\\(f)&2854.&249.5&0.00821&0.03464&0.795&41.44&0.04183&0.5871\\(g)&530.7&124.7&0.04001&0.02102&0.757&17.14&0.01621&0.6159\\
   \end{tabular}
    \caption{A set of benchmarks covering a range of input parameters 
    used in the computation of tunneling temperature in the hidden sector
    and other relevant outputs in Table.(\ref{tab:benchmarkresults})
    that enter in the computation of the gravitational wave spectrum
    {consistent with all constraints on the dark photon\cite{Aboubrahim:2022qln}}.    
    These benchmarks pass all the constraints mentioned in section \ref{sec:6.1}, 
    and are cosmologically consistent candidate models for the computation of 
    gravitational waves. }
    \label{tab:benchmarks}
\end{table}
 \begin{table}[h]
 \centering
 \resizebox{\textwidth}{!}{
     \begin{tabular}{lllll|llllllllll}
    \hline
   No.&$T_{h,n}$&$\xi(T_n)$&$c^2_{s,+,h}$ & $c^2_{s,-,h}$ & $\alpha_h$ & $\beta_h/H_n$&$\kappa_h$&$\Omega_{DM}h^2$& $f[Hz]$&$\Omega_{GW}h^2$&Mode  \\ \hline
(a)&18.2&0.6724&0.307&0.306&0.0172&294.8&0.223&0.013&0.00257&$1.591\times 10^{-13}$&HYB\\(b)&18.02&0.4635&0.309&0.309&0.00053&1563.&0.0463&0.0267&0.0204&$1.172\times 10^{-18}$&HYB\\
(c)&29.93&0.7794&0.309&0.308&0.043&158.8&0.0523&0.0198&0.0012&$1.524\times 10^{-13}$&DET\\(d)&35.52&0.9181&0.308&0.306&0.0151&871.3&0.0984&0.0385&0.0102&$8.604\times 10^{-15}$&DET\\(e)&19.73&0.8227&0.309&0.308&0.0367&319.7&0.0413&0.0115&0.0055&$9.823\times 10^{-15}$&DEF\\(f)&54.78&0.7956&0.318&0.317&0.0107&845.6&0.187&0.0228&0.0191&$1.42\times 10^{-14}$&HYB\\(g)&28.07&0.758&0.308&0.307&0.0127&565.9&0.175&0.0216&0.00631&$2.843\times 10^{-14}$&HYB\\
   \end{tabular}
   }
    \caption{
    Computation of the nucleation temperature $T_{h,n}$, sound velocities in the symmetric and broken
    phases $c^2_{s,+,h}, c^2_{s,-,h}$, the strength of the phase transition $\alpha_h$, the inverse duration of the transition $\beta_h/H_n$ and  the efficiency factor $\kappa_h$
    all for the hidden sector. Also listed is the dark matter relic density $\Omega_{DM} h^2$, frequency $f(z)$  of the power spectrum at the peak 
    value of the gravitational wave power spectrum, and the gravitational wave power spectrum $\Omega_{GW}
    h^2$ at peak value. The symbols DET, HYB, DEF stand for the nucleation modes 
   detonation, hybrid, deflagration.}
    \label{tab:benchmarkresults}
\end{table}

%%%%%%%%%%%%%%%%%%%%%%%%%%%%%
\subsection{Constraints and Monte Carlo Simulation \label{sec:6.1}}
In the beginning of this section we discussed eight parameters which enter in the analysis of the gravitational wave spectrum. For simulations we take the
following ranges for these parameters
\begin{align}
    m_D\in (10^{-1},10^4)\rm{GeV},&\quad m_{\gamma'}\in (10^{-1},10^4)\rm{GeV},
   \quad g_x\in (10^{-4},10^0), \quad \delta\in (10^{-12},10^{-6}),\non
    \xi_0\in (0,1),&\quad \mu_h\in (10^{-1},10^4)\rm{GeV},
\quad    \lambda_h\in (10^{-5},10^1),\quad v_{wh}\in (0,1).
\label{modelpoints}
\end{align}
In order to investigate the distribution of different nucleation modes, as discussed in section \ref{sec:6.3}, for each event we select  $v_{wh}$ corresponding to three different nucleation modes so that the total number for each type of mode is the same. 
In the Monte Carlo simulation we impose the following constraints
\begin{enumerate}
    \item 
    FOPT constraints: For the first-order phase transition, we require that
    there must be a potential barrier between the false vacuum and the true vacuum. 
     The further constraint is an upper limit of sound velocity so that $c_s^2\leq 1/3$.
    \item $\Delta N_{\rm eff}$ constraint at BBN:
    The  number of effective relativistic degrees of freedom  {$N_{\rm eff}$ at BBN is one of the important constraints on new physics 
    beyond the standard model of particle physics}. 
    The relevant constraint is 
    given by the allowed corridor for the difference between the experimental
    result and the standard model result at the BBN time represented 
    by $\Delta N_{\rm eff}$. For the hidden sector model the extra degrees
    of freedom are given by         
    \begin{align}
        \Delta N_{\rm eff} = \frac{4}{7}g^h_{\rm eff}\left(\frac{11}{4}\right)^{4/3}\xi^4
    \end{align}
     Current experiments observation give us the constraint $\Delta N_{\rm eff} < 0.25$\cite{Planck:2018vyg}.
     \item Relic density constraint. After solving for the yield equations the relic densities for  
     $\chi$ and $D$ can be gotten from their individual yields so that
     \begin{align}
         \Omega_ih^2 = \frac{\bold{s}_0m_iY_i^0h^2}{\rho_c},\quad i\in{(D,\chi)}
     \end{align}
   where $Y_i^0$ is the yield for the $i-$th particle and $\Omega_ih^2$ its relic density
     while the total relic density is the sum of them. In the analysis we use dark matter 
    relic density as an upper limit.  Currently it is given by the Planck experiment\cite{Planck:2018vyg} so that
     \begin{align}
         \Omega_{DM} h^2 = 0.120\pm 0.001
     \end{align}
     Specifically, we impose the constraint $ 0.01< \Omega_{\rm hidden} h^2 < 0.12$.
\end{enumerate}

\begin{figure}[h]
    \centering
    \includegraphics[width=0.3\textwidth]{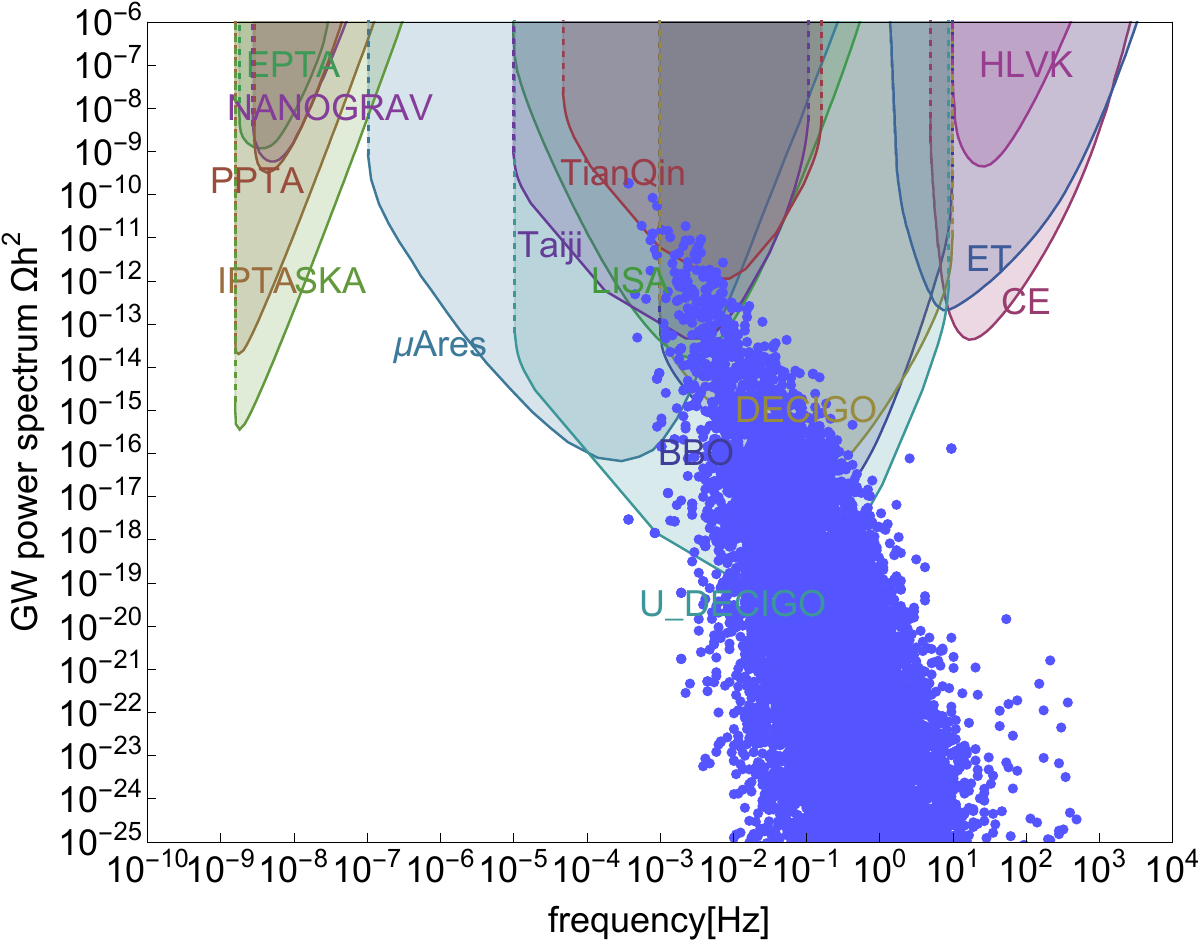}
    \includegraphics[width=0.3\textwidth]{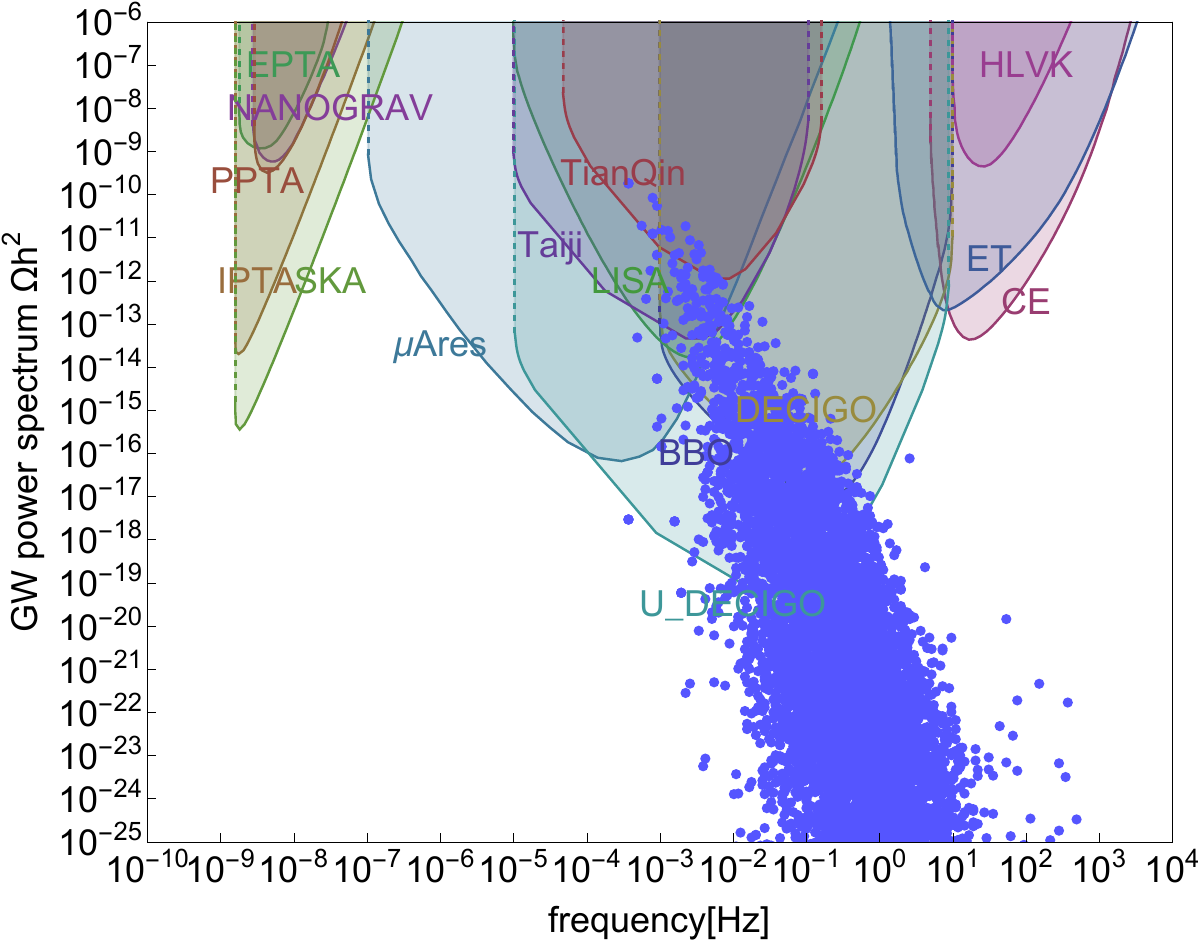}
    \includegraphics[width=0.3\textwidth]{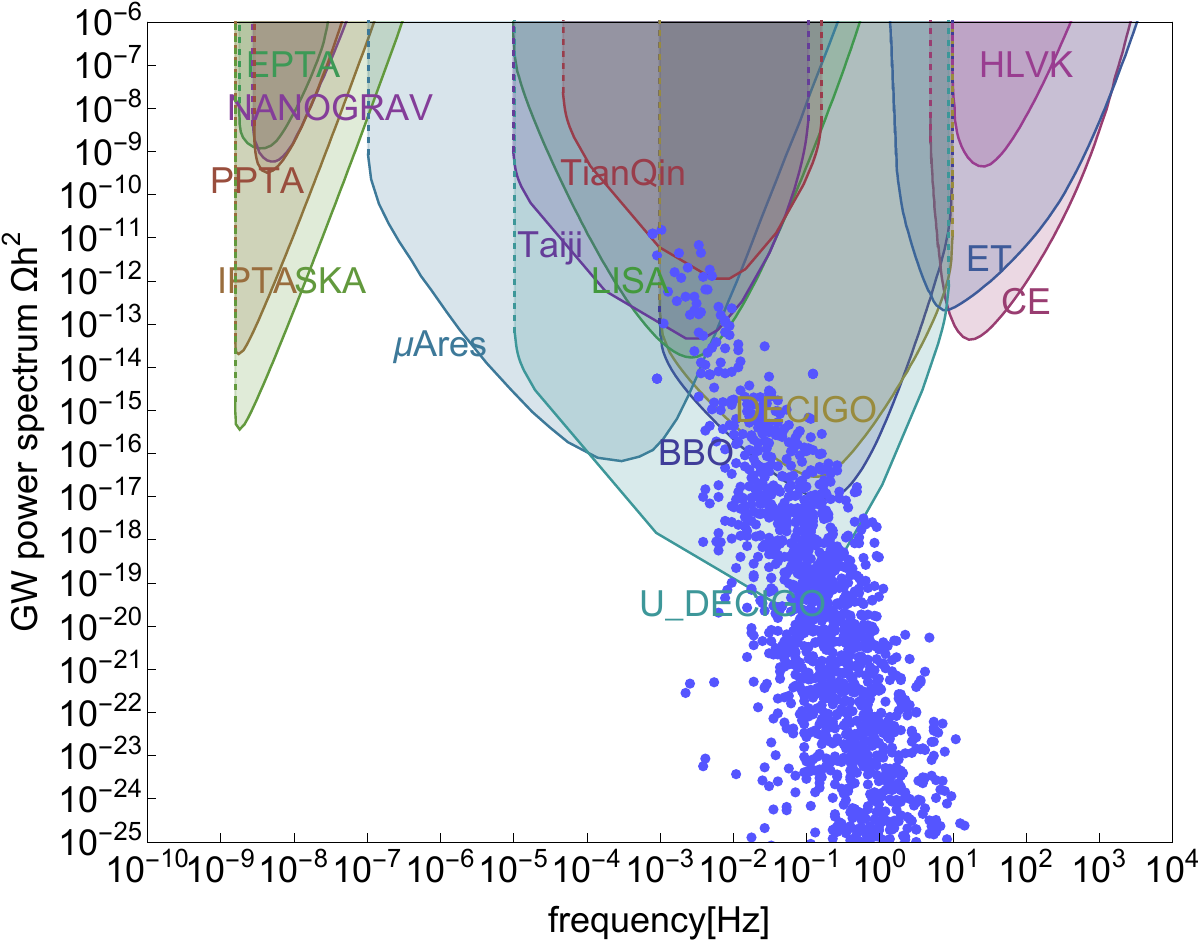}
    \caption{Gravitational wave power spectrum
     resulting from Monte Carlo analysis on 8 free parameters. 
    Left panel:
    scatter plot of the peak value of candidate models at the frequency where the peak value occurs 
    after FOPT constraints are applied. Middle panel: 
 same as the left panel  including  FOPT constraints and the $\Delta N_{\rm eff}$ constraint.
  Right panel: 
 same as the left panel including the FOPT constraints, $\Delta N_{\rm eff}$ constraint and the relic density constraint. }
    \label{GW_MC} 
\end{figure}

For each benchmark model, there will be a corresponding power spectrum curve just like Fig.(\ref{fig:GWspectrum0}). However, plotting the full curve for each model point
would not be illuminating because they would be space filling.  
For that reason we will do a scatter plot on the gravitational wave power spectrum, with each model point represented by the peak of its power spectrum curve at the frequency where 
that peak occurs. An illustration of it 
is given in  Fig.(\ref{GW_MC}).  
%%%%%%%%%%%%%%%%%%%%%%%%%%%%%
\subsection{Nucleation Temperature and GW power spectrum \label{sec:6.2}}
\begin{figure}[H]
    \centering
    \includegraphics[width=0.3\textwidth]{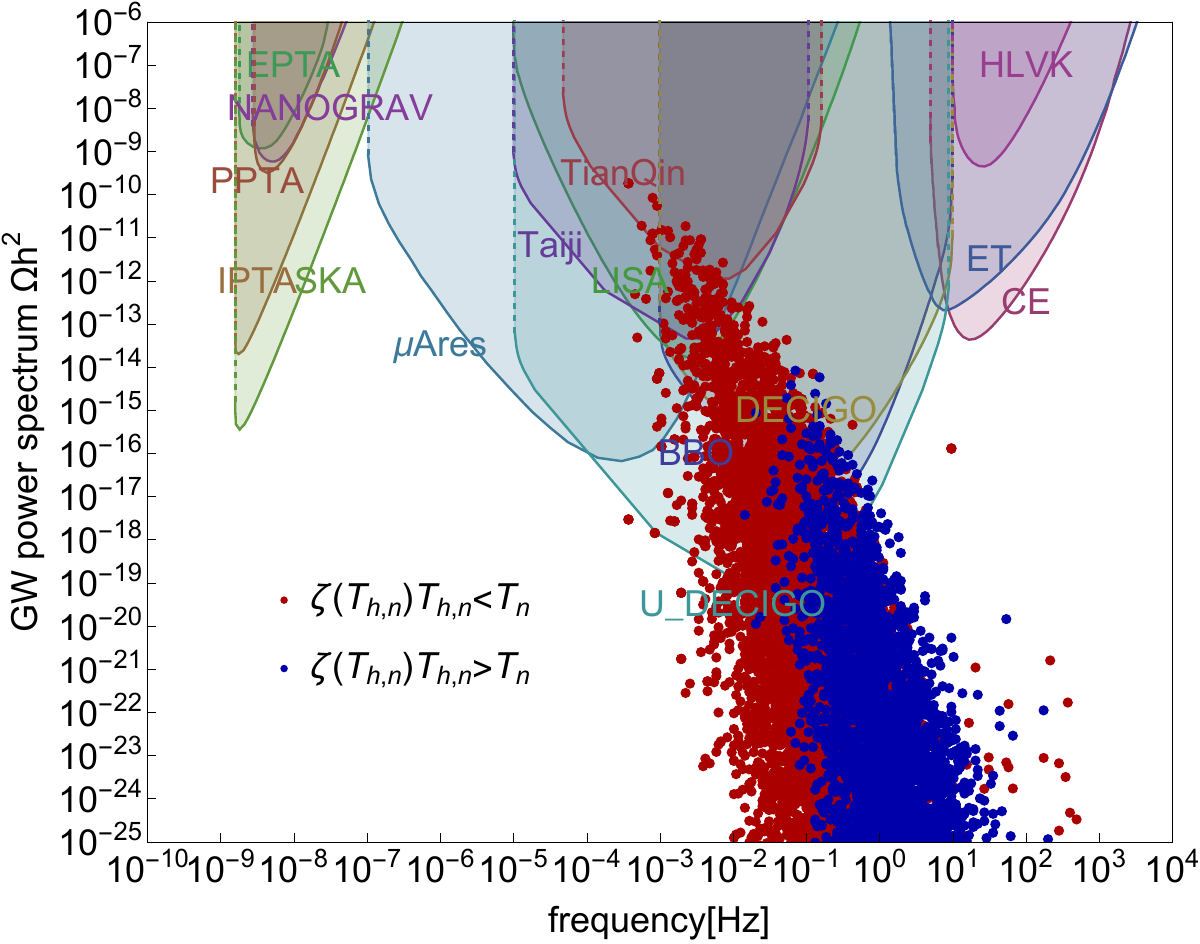}
    \includegraphics[width=0.3\textwidth]{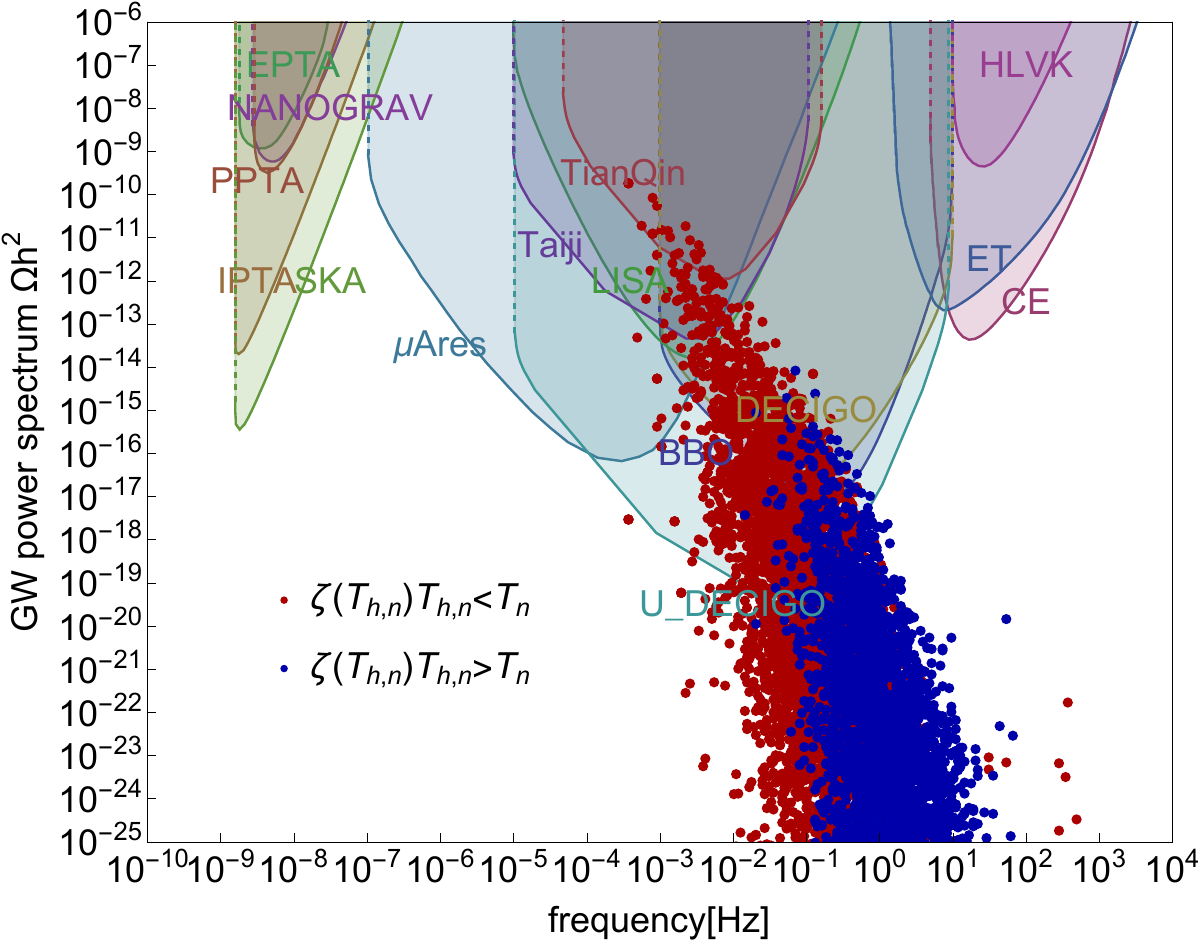}
    \includegraphics[width=0.3\textwidth]{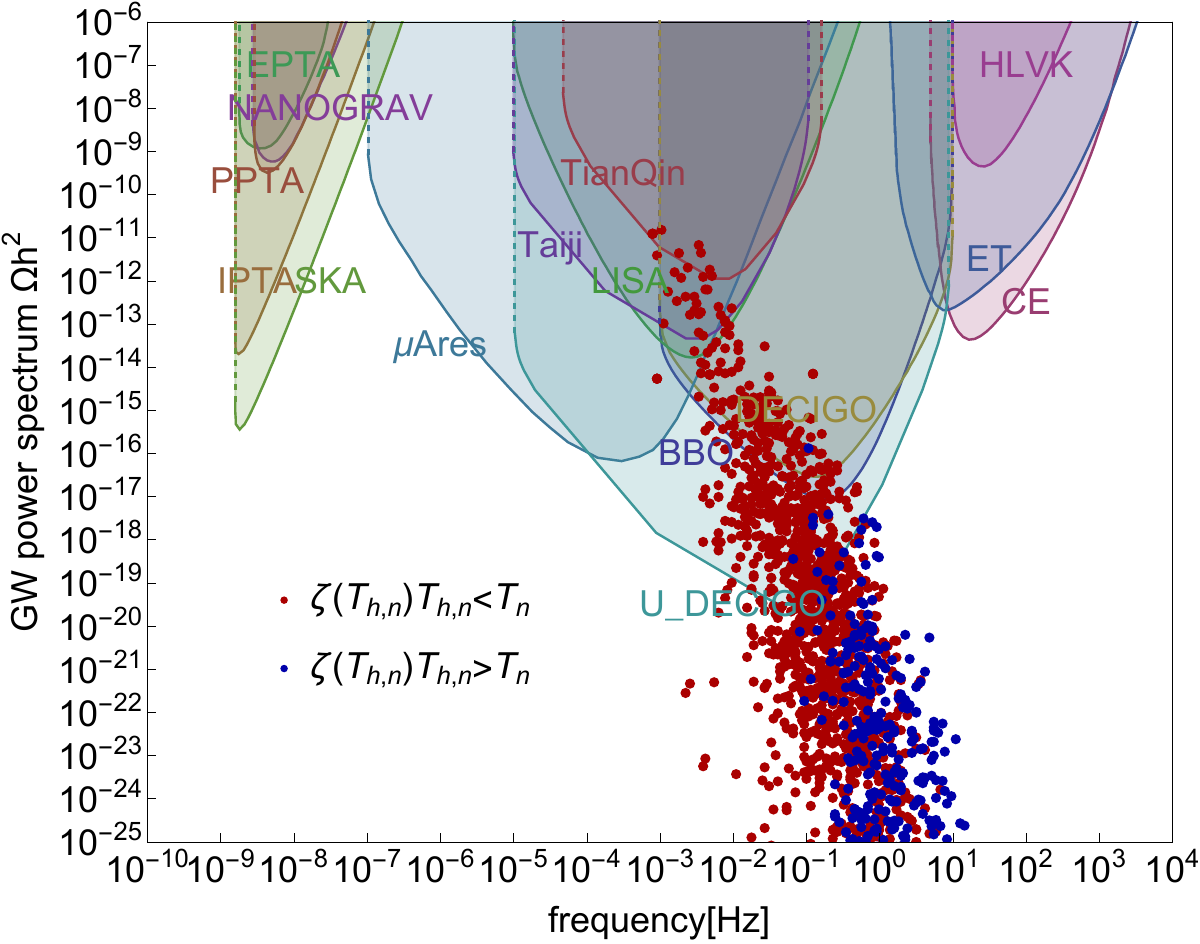}
    \caption{
    Monte Carlo analysis of gravitational  wave
    power spectrum classified by two possible orderings in which nucleation occurs
    in the visible and hidden sector with red model points for 
     $T_n>\zeta(T_{h,n})T_{h,n} $ and blue model points for $\zeta(T_{h,n})T_{h,n} > T_n$. Left
 panel: scatter plot of the candidate models satisfying the    
     FOPT constraints. Middle panel: same as the left panel  satisfying the
      FOPT constraints and  the $\Delta N_{\rm eff}$ constraint. Right panel: same as the
      left panel with models satisfying the
       FOPT constraints, $\Delta N_{\rm eff}$ constraint and the relic density constraint. Here one finds that the residual set of models left after all the constraints are applied are those where the nucleation in the hidden sector happens after nucleation in the visible sector.  
     }
    \label{GW_MC_order}
\end{figure}

Nucleation temperature is one of the key factors in the computation of the
 gravitational wave power spectrum. It affects the spectrum in the following ways: 
\begin{itemize}
    \item[1.]  It enters the phase transition strength $\alpha$ as discussed below
\begin{align}
    \alpha = \frac{\epsilon}{\rho_{rad}}
\end{align}
Although there are several different definitions to the latent heat $\epsilon$ as discussed earlier, the total radiation density of the universe $\rho_{rad}$ is the same and is given by
\begin{align}
    \rho_{rad} &=  \frac{\pi^2}{30}\left(g_{eff}^v(\zeta T_{h,n}) {T_{h,n}}^4\zeta^4+g_{eff}^h(T_{h,n}){T_{h,n}}^4\right)
    \label{rhorad}
\end{align}
It tells us that $\alpha \propto {T_{h,n}}^{-4}$. Thus a smaller $T_{h,n}$ leads to a 
larger $\alpha$ and a larger gravitational wave power spectrum.
\item[2.]  According to Eq.(\ref{redshift}), we have 
$f_{0} \propto T_{h,n}$ 
which implies that a larger power spectrum will arise at lower frequencies.
\end{itemize}
The analysis of Fig.(\ref{GW_MC}) is consistent with the observation above 
that a larger power spectrum will appear at a lower frequency. 
    We also note that for the two-field case, satisfaction of FOPT and other constraints
    is affected by the order in which nucleation in the visible and in the hidden sector occurs.   
Thus we classify all FOPT events into two groups: 1. The {standard model}  Higgs scalar nucleation happens first, i.e. $\zeta(T_{h,n})T_{h,n} < T_n$ (red points)
2. The hidden Higgs scalar nucleation happens first, i.e. $\zeta(T_{h,n})T_{h,n} > T_n$
(blue points).
  The analysis for these two cases  is shown in Fig.(\ref{GW_MC_order}). 
  Here the analysis shows that after the FOPT constraints, $\Delta N_{\rm eff}$ constraint,
   and the relic density constraint are taken into account most of the blue points are eliminated which implies that the hidden sector nucleation occurs after nucleation in 
   the visible sector.
{
%%%%%%%%%%%%%%%%%%%%%%%%%%%%%
\subsection{Sound velocity and GW power spectrum \label{sec:6.3}}
%{
We discussed above the effect of sound velocity on the final power spectrum 
via $\alpha(c^2_{s,-,h})$ according to Eq.(\ref{alphatheta}) and  via $\kappa(\alpha,c^2_{s,+,h},c^2_{s,-,h},v_w) $ according to Eq.(\ref{kappatheta}).
 To demonstrate to what extent sound velocity can change the power spectrum, 
 we investigate the power spectrum for  model (b) from Table.(\ref{tab:benchmarks}) keeping all parameters 
 fixed except for the sound velocity $c^2_{c,b,h}$. 
 The analysis of  Fig.(\ref{cs2changes}) 
  shows that the changes to power spectrums can be as large as a factor of 
$\mathcal{O}(10^3)$.  
  This approach allows us to isolate the effects of sound velocity from other factors, such as the nucleation temperature noted earlier.
\begin{figure}
    \centering
    \includegraphics[width=0.5\textwidth]{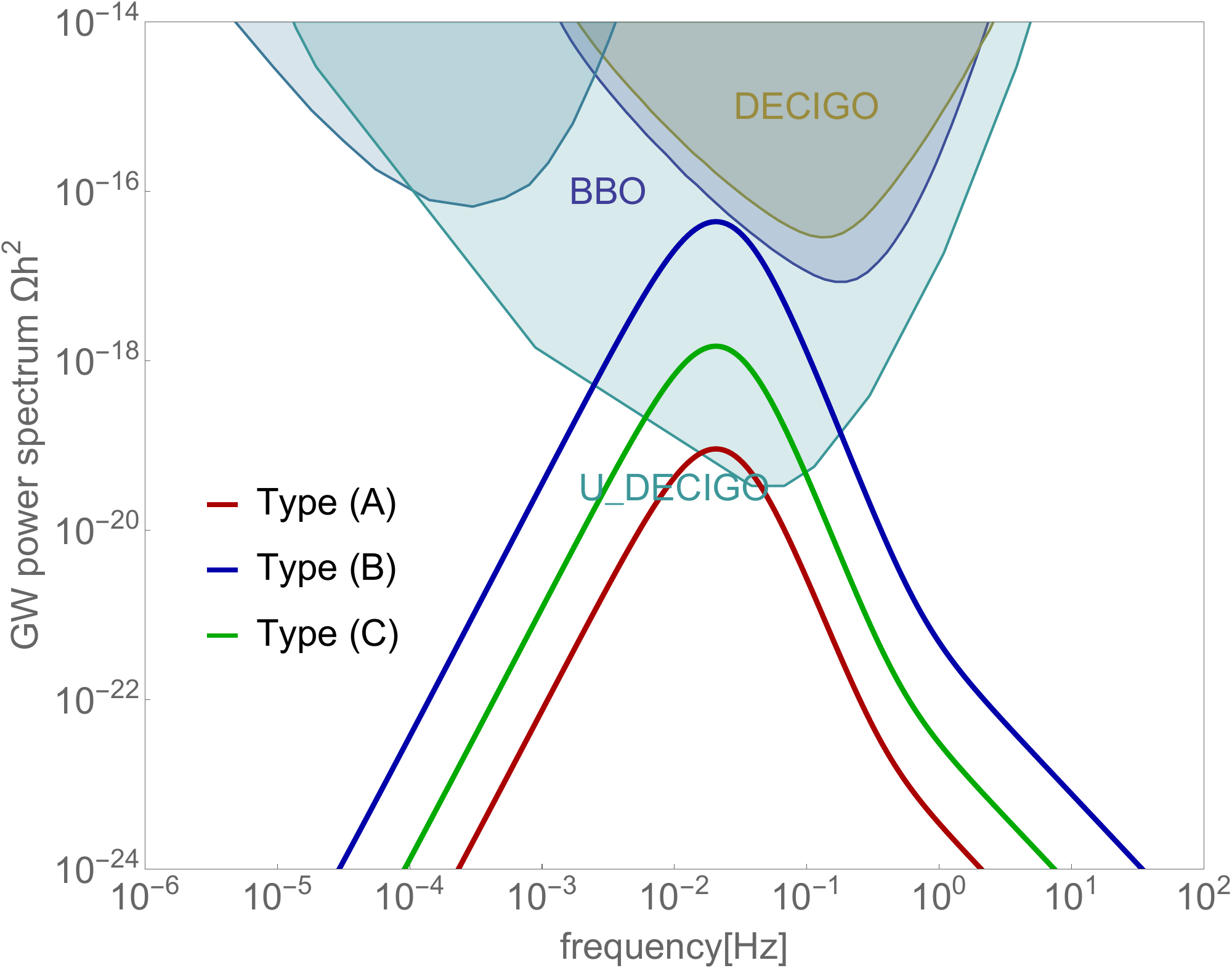}
    \caption{
    An exhibition of the gravitational wave power spectrum for different types of sound velocities
    for a example model. All other parameters, such
    as the nucleation temperature, are kept fixed when $c^2_{s,-,h}$ varies for cases A, B and C discussed in the text.}
    \label{cs2changes}
\end{figure}
The reason we need to discuss this dependence is because there are multiple different analyses on sound velocity among existing works which lead to different results. We classify these as follows
%}
 
   (A). This is the case when one considers just the hidden sector and 
    assumes that the sound velocity takes its maximum value allowed
      in fluids which is $c_s^2=1/3$.  Such an assumption is the one most commonly made,
    see, e.g.,\cite{Breitbach:2018ddu,Wang:2020jrd}.
   
    (B). Here one considers one hidden sector model but including sound velocity dependence. For this class of models 
    $\alpha$ is given by Eq.(\ref{alphatheta}) and $\kappa$ will also be velocity dependent. The sound velocity is given by
    \begin{align}
        c_s^2 = \frac{dp_h/dT_h}{de_h/dT_h}
    \end{align}
    where $p_h$ and $e_h$ are the pressure and the energy density for the hidden sector. 
     Analyses of this type are discussed in, for example\cite{Giese:2020znk,Wang:2021dwl}.
  
    (C). In this work we discuss sound velocity involving two sectors, i.e., the visible and the hidden, 
    and take into account velocity dependence which is
 given by Eq.(\ref{Eq.cs2}). This type of analysis has not been discussed in the existing
  literature to our knowledge.

Applying  the above three types of analyses (A),(B),(C) to model (b), we get sound velocities 
such that $c^2_{s,-,h,\rm{(A)}} = 0.333, c^2_{s,-,h,\rm{(B)}} = 0.234, c^2_{s,-,h,\rm{(C)}} = 0.309$. Correspondingly the gravitational wave power spectrum for the three cases is 
significantly affected due to variations in the sound velocity as illustrated in Fig.(\ref{cs2changes}).

%%%%%%%%%%%%%%%%%%%%%%%%%%%%%
\subsection{Nucleation temperature and sound velocity\label{sec:6.4}}
{In this section, we will analyze how the sound velocity depends on the nucleation temperature. Again, we will focus on $c^2_{s,-,h}$, with sound velocity 
defined as by Eq.(\ref{Eq.cs2}).
 
\begin{figure}[h]
    \centering
    \includegraphics[width=0.3\textwidth]{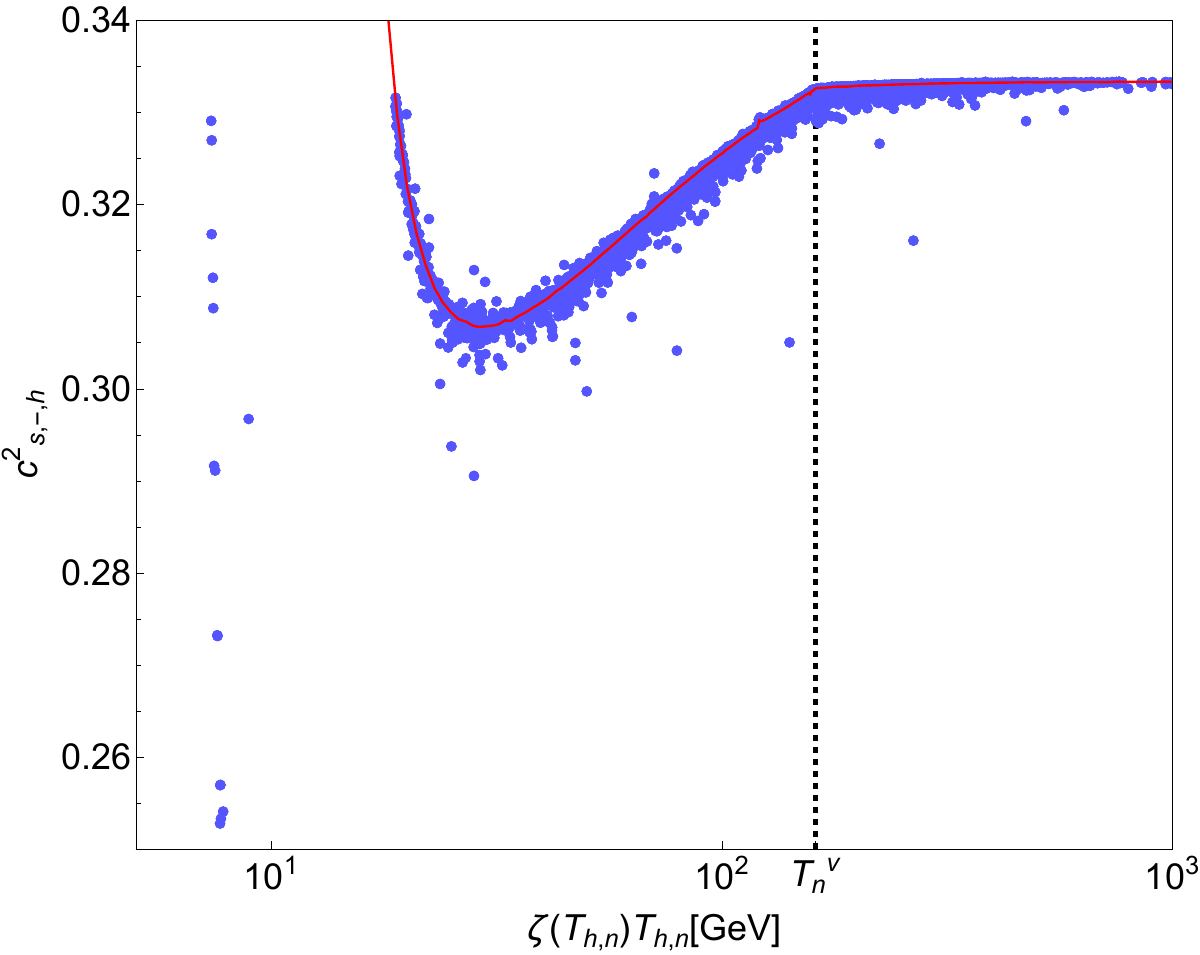}    \includegraphics[width=0.3\textwidth]{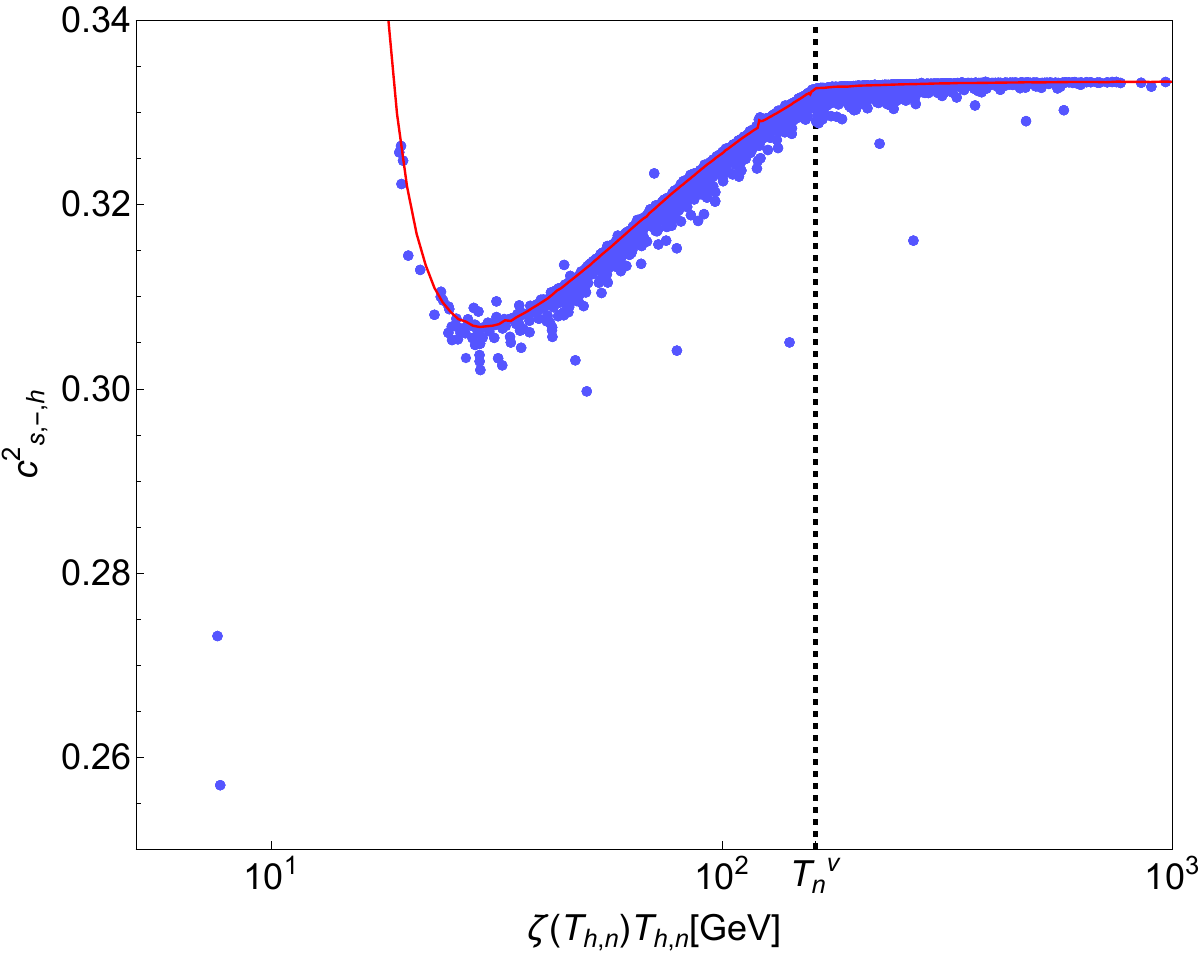}
    \includegraphics[width=0.3\textwidth]{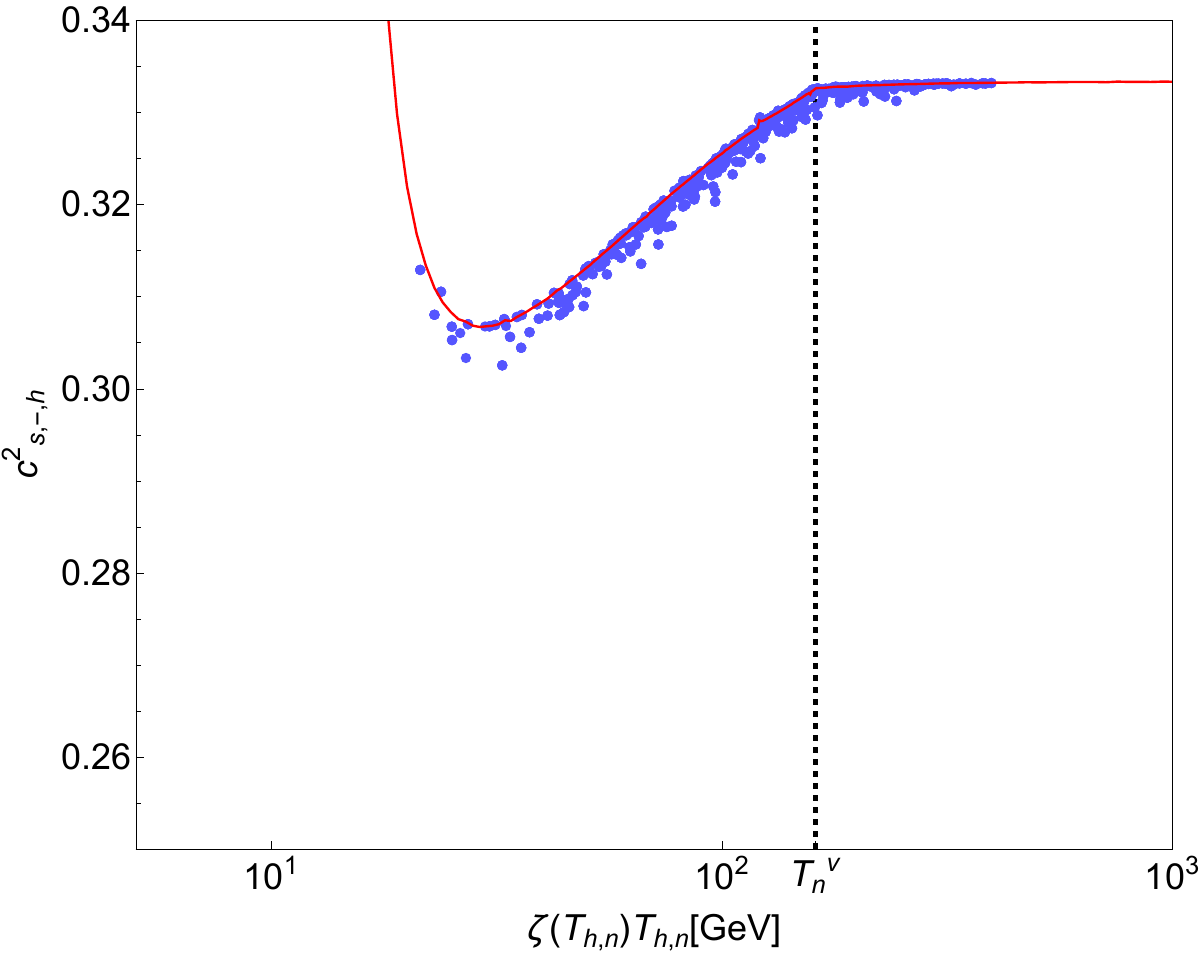}
      \caption{Scatter plots for $c^2_{s,-,h}$ v.s. $\zeta(T_{h,n})T_{h,n}$ for a set of 
     candidate models in the parameter ranges given by
     Eq.(\ref{modelpoints}).
    Here the standard model dominates over the hidden sector in the computation of sound velocity and the sound velocity is close to the one for the visible sector, i.e., 
     $c_s^2 =\frac{d(p_v)/d T}{d(e_v)/d T}$ shown by the red curve. The  vertical black lines 
     give the value of $T_n$.
     Left panel: Scatter plot of $c^2_{s,-,h}$ v.s. $\zeta(T_{h,n})T_{h,n}$ with inclusion of 
      FOPT constraints. Middle panel: same as the left panel including the  
       FOPT constraints and the $\Delta N_{\rm eff}$ constraint. Right panel: same as the left panel
       with FOPT constraints, $\Delta N_{\rm eff}$ constraint, and the relic density constraint.}
    \label{fig:cs2}   
\end{figure}
%%%%%%%%%%%%%
The scatter plot is shown in Fig.(\ref{fig:cs2}). We observe that some of the points in Fig.(\ref{fig:cs2}) are gathered around the curve of $c_s^2 =\frac{d(p_v)/d T}{d(e_v)/d T}$.
This phenomenon happens because the visible sector dominates, i.e. we have the sound velocity so that
\begin{align}
    c_s^2(\phi,\chi,T,T_h) &= \frac{d(p_v+p_h)/d T}{d(e_v+e_h+e_{mix})/d T} \simeq  \frac{d(p_v)/d T}{d(e_v)/d T}
\end{align}
The reason that the visible sector can dominate is because  $p_v\sim \frac{\pi^2}{90} g^v_{eff} T^4$ and $p_h\sim \frac{\pi^2}{90} g^h_{eff} T_h^4$ according to Eq.(\ref{pv},\ref{ph}) and $T>T_h$ when $\xi<1$ and also $g^v_{eff}>g^h_{eff}$.

One may note from Fig.(\ref{fig:cs2}) that for the case 
when the hidden Higgs scalar nucleation occurs first, i.e. $\zeta(T_{h,n})T_{h,n} > T_n$, the red curve stays at $c^2 \sim 1/3$. However,  when the standard model Higgs scalar nucleation
occurs first, i.e. $T_n>\zeta(T_{h,n})T_{h,n} $, we have $c_s^2$ systematically less than $1/3$. 
The different behavior for the two cases arises due to two different constraints, i.e., 
Eq.(\ref{csvh}) and Eq.(\ref{cshv}), for these two different cases. In the analysis of section \ref{sec:6.2} it is found that most of events which pass all the relevant constraints are those where
 $T_n>\zeta(T_{h,n})T_{h,n}$ and where the approximation $c^2_s \sim 1/3$ is typically 
  invalid. In simple terms the cosmologically preferred model points are those where
  $T_n>\zeta(T_{h,n})T_{h,n}$ and $c_s^2< 1/3$. 
%%%%%%%%%%%%%%

{
\subsection{$\Delta N_{eff}$ vs $\xi(T)$. \label{sec:6.5}}
According to the Eq.(\ref{rhorad}),
 the hidden sector nucleation happens at $T_{h,n}$ which lies in the range 
$18-55$ GeV according to Table.(\ref{tab:benchmarkresults}), 
while the BBN temperature is $O(1)$ MeV. This means we need to 
extrapolate the $N_{eff}$ between the two temperatures in a precise way so as to
take account of the $\Delta N_{eff}$ constraint at BBN time which we take to be
$\Delta N_{eff}< 0.25$. In some previous works separate entropy conservation in the 
visible and hidden sectors is used to extrapolate $N_{eff}$ from high temperatures
to low temperatures. Such a procedure is shown to be flawed as it can yield
highly inaccurate estimates on $\Delta N_{eff}$. 
{Thus a more accurate analysis 
is needed as discussed in section \ref{sec:2} and Appendix E. An analysis relevant to the current 
case is given in Fig.(\ref{fig:xi0}). Here we first show that the two sectors decouple at $10^{-2}$GeV and the dark photon also decays out at $10^{-2}$GeV. 
The   left first panel exhibits the decoupling 
more clearly where $n_i\sum_i \left<\sigma v\right>$ for all three hidden sector particles
falls below $H(T)$ at $T\sim 10^{-2}$GeV, which means the complete decoupling of the hidden and visible sectors 
(see Appendix E), and the density of dark relics freezes-out as exhibited in the left second  panel.
The left third panel exhibits $\xi(T)$ vs $T$ which is used to constraint  $\Delta N_{eff}$ at the BBN time as shown in the right panel.
}
 The right panel shows that $\Delta N_{\rm eff}$ drops below the BBN constraint when decoupling happens.
%%%%%%%%%%%%%%%%%%%%%%%%%%%%%
\begin{figure}[H]
    \centering
    \includegraphics[width=0.24\textwidth]{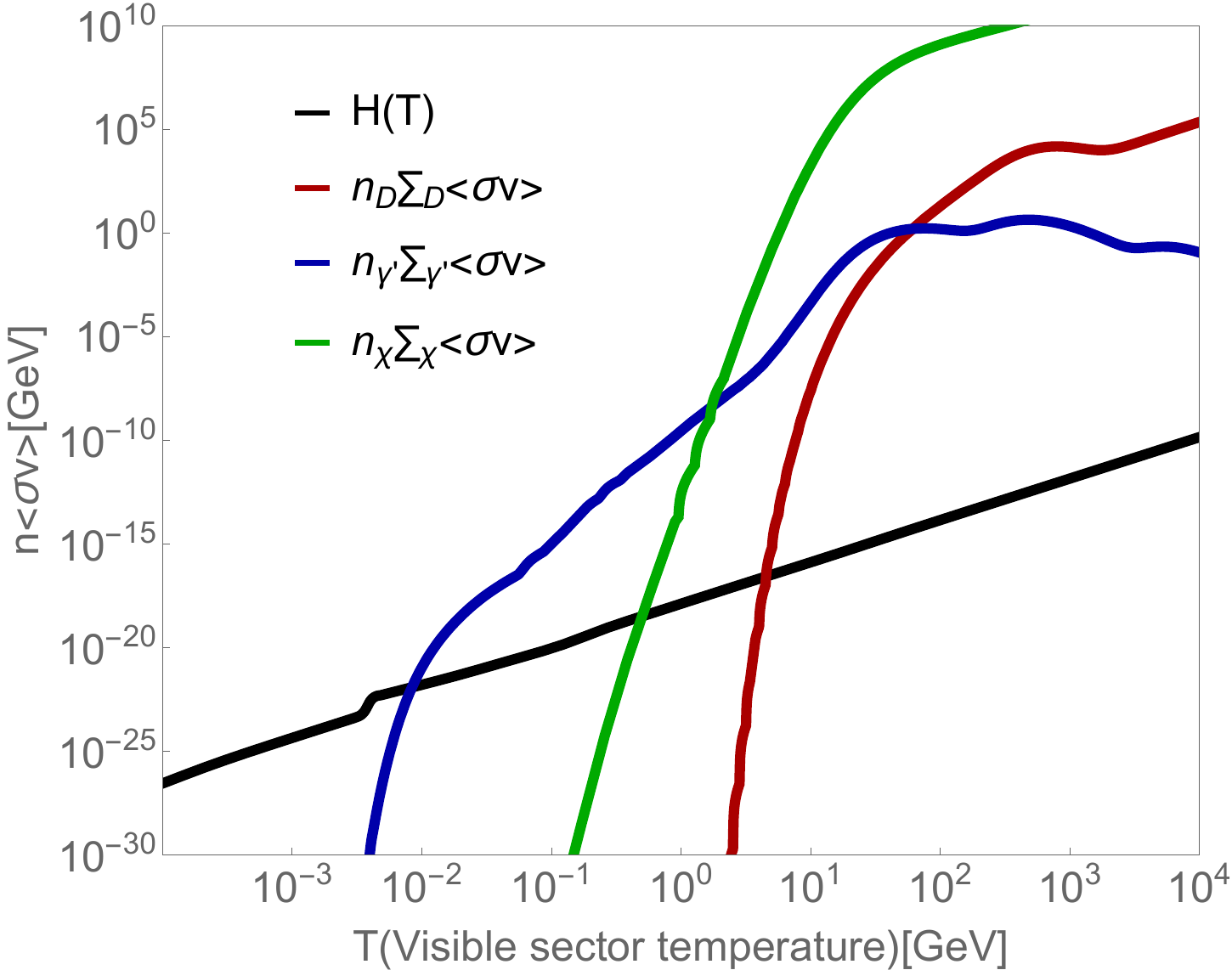} 
    \includegraphics[width=0.24\textwidth]{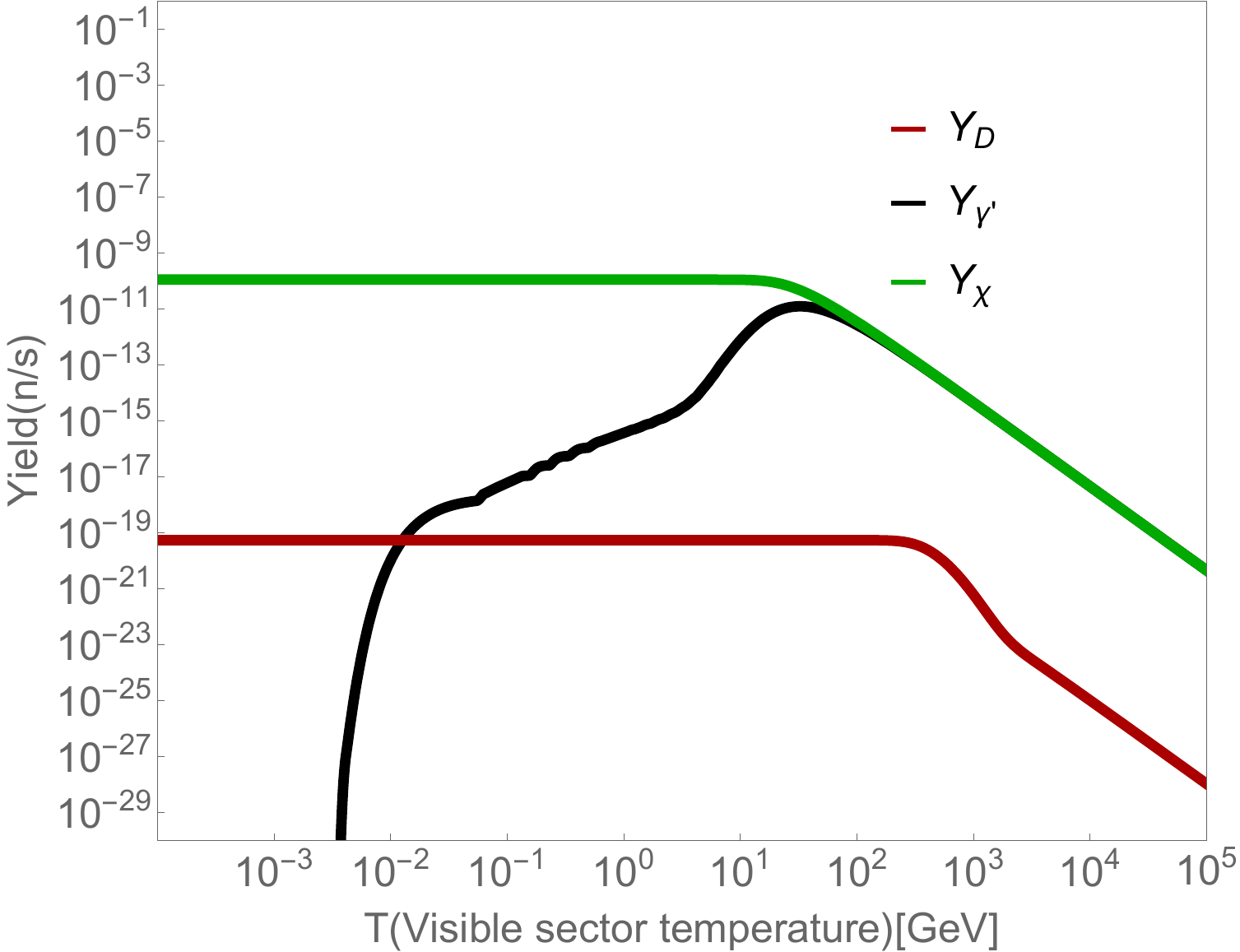}
    \includegraphics[width=0.24\textwidth]{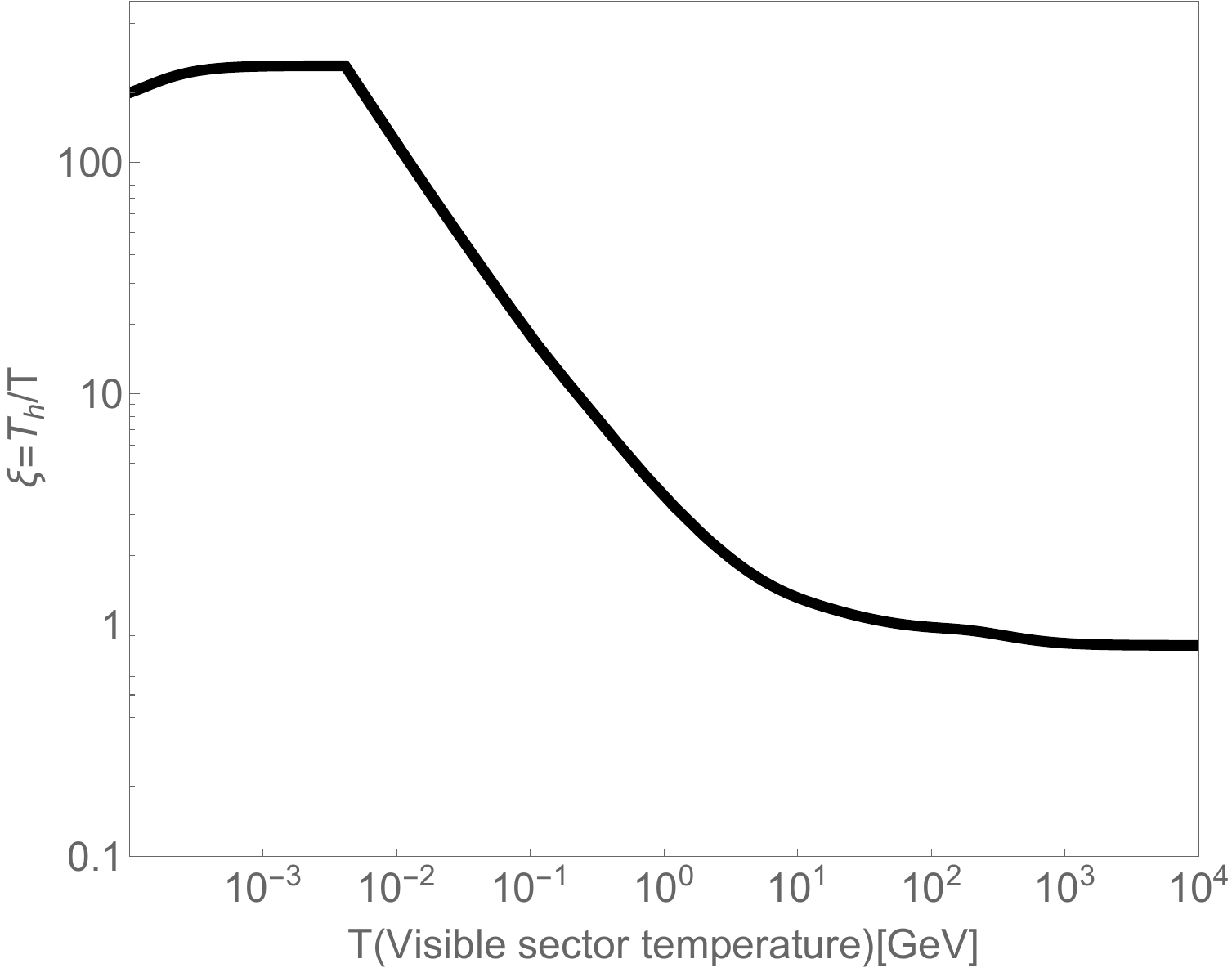} 
    \includegraphics[width=0.24\textwidth]{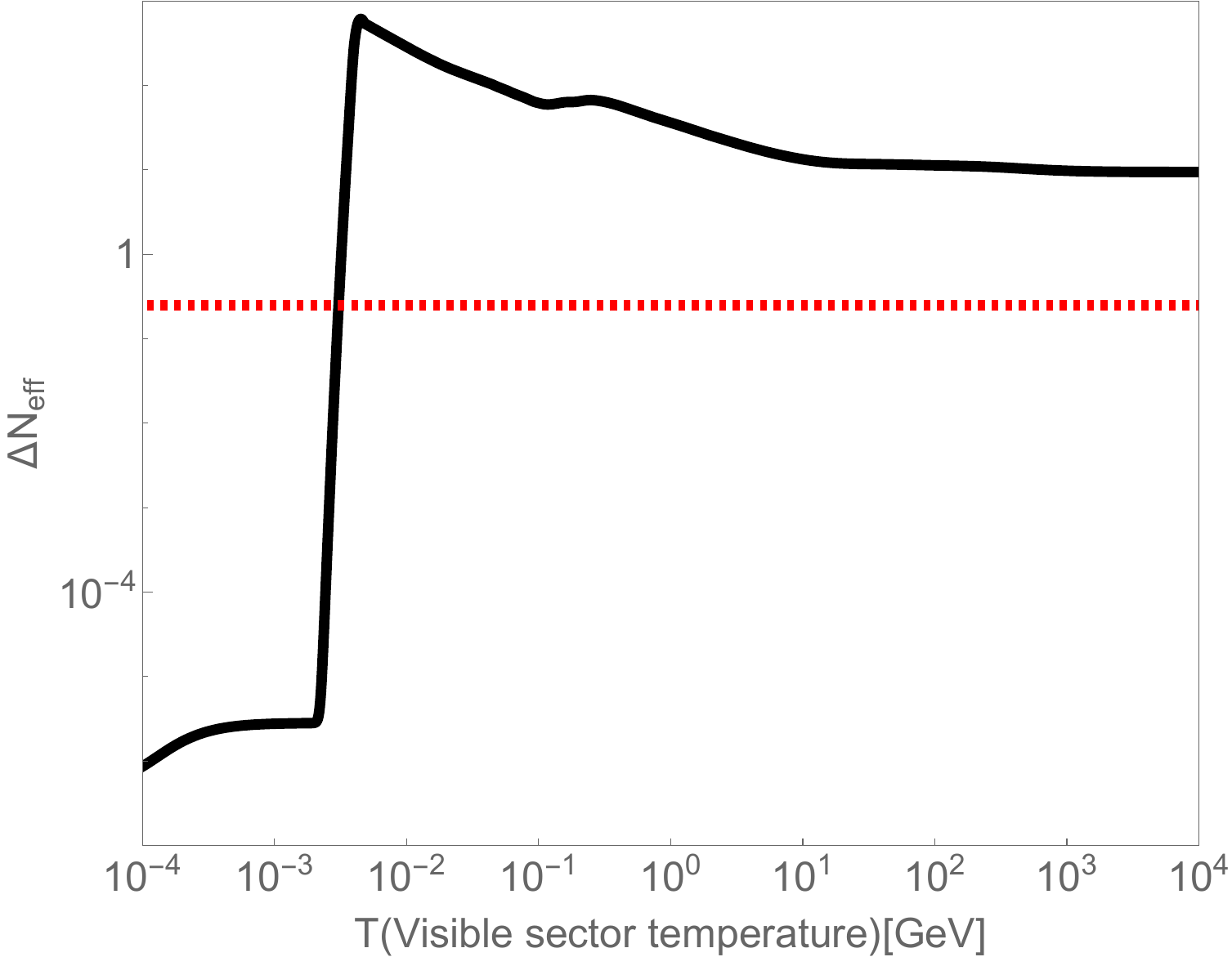} 
    \caption{Left first: An exhibition of the decouplings of hidden sector particles with plots of
          $n_i\sum_i \left<\sigma v\right>$ and $H(T)$ vs $T$ for $i=D,\gamma', \chi$.
           Here we have $n_D\sum_D \left<\sigma v\right> = n_D(\left<\sigma v\right>_{D\bar D\rightarrow i\bar i} + \left<\sigma v\right>_{D\bar D\rightarrow \gamma'{\gamma'}} ),  n_{\gamma'}\sum_{\gamma'} \left<\sigma v\right> = n_{\gamma'}(\left<\sigma v\right>_{D\bar D\rightarrow {\gamma'}{\gamma'}} + \left<\sigma v\right>_{\chi \bar \chi \rightarrow \gamma'{\gamma'}} + \left<\Gamma_{\gamma'\rightarrow i\bar{i}}(T_h)\right>+\left<\Gamma_{\chi\rightarrow \gamma'\gamma'}(T_h)\right> ), n_\chi\sum_\chi \left<\sigma v\right> = n_\chi(\left<\sigma v\right>_{\chi \bar \chi \rightarrow \gamma'{\gamma'}} + \left<\Gamma_{\chi\rightarrow \gamma'\gamma'}(T_h)\right> )$.
           Left second: An exhibition of the decay of dark photon $\gamma'$. The dark photon decays out at about $10^{-3}$GeV. 
           Left third: Evolution of $\xi(T)$ vs $T$ for model (e)
    of Table.(\ref{tab:benchmarks}).Right panel: $\Delta N_{\rm eff}(T)$ v.s. T. The red dashed line is the current limit $\Delta N_{\rm eff} < 0.25$.}
      \label{fig:xi0}
\end{figure}
%%%%%%%%%%%%%%%%%%%%%%%%%%%%%%%
}
\subsection{Gravitational wave power spectrum and the nucleation modes:\\
 detonation, deflagration, hybrid.
\label{sec:6.6}}
Now we discuss the gravitational wave power spectrum for different nucleation modes:
 detonation, deflagration and hybrid. Chapman-Jouguet velocity \cite{landau,courant,Laine:1993ey} 
 is used in part to distinguish different bubble nucleation modes
specifically the detonation and the hybrid modes. It is given by\cite{Giese:2020znk,Giese:2020rtr}
\begin{align}
    v_J = \frac{1+\sqrt{3\alpha_{\bar\theta}(1-c_s^2+3c_s^2\alpha_{\bar\theta})}}{1/c_s+3c_s\alpha_{\bar\theta}}
\end{align}
The bubble nucleation modes are distinguished by the following constraints
\begin{enumerate}
 \item Detonations: $v_w > c_{s,-}$ and $v_w > v_J$
     \item Hybrid: $v_w > c_{s,-}$ and $v_w < v_J$ 
    \item Deflagrations: $v_w < c_{s,-}$
\end{enumerate}
where $v_w$ is the bubble wall velocity.
We apply such classification to all data points in Monte Carlo analysis to produce Fig.(\ref{fig:GW_MC_Mode}). The figure  shows that the hybrid modes are typically the ones with the highest power spectrum.
\begin{figure}[h]
    \centering
    \includegraphics[width=0.3\textwidth]{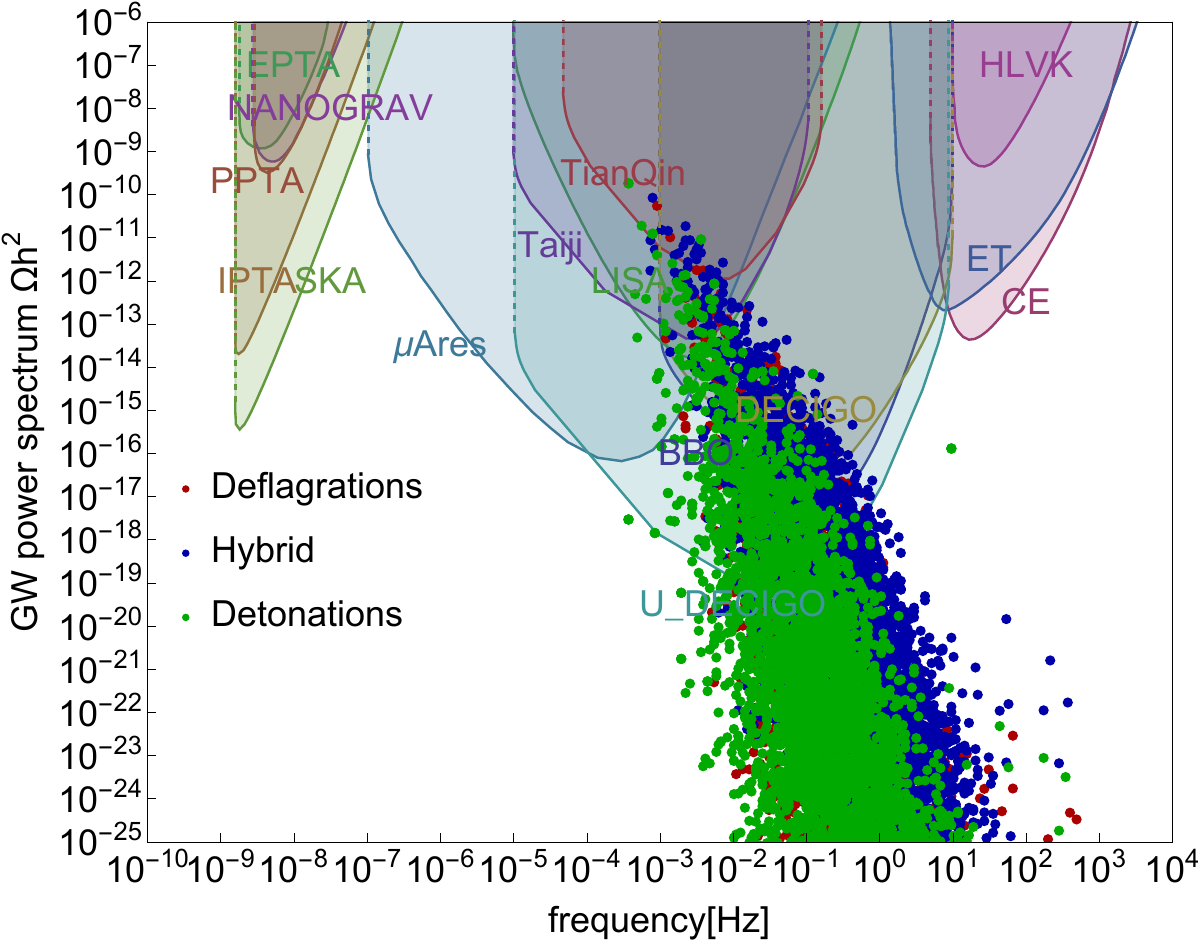}
    \includegraphics[width=0.3\textwidth]{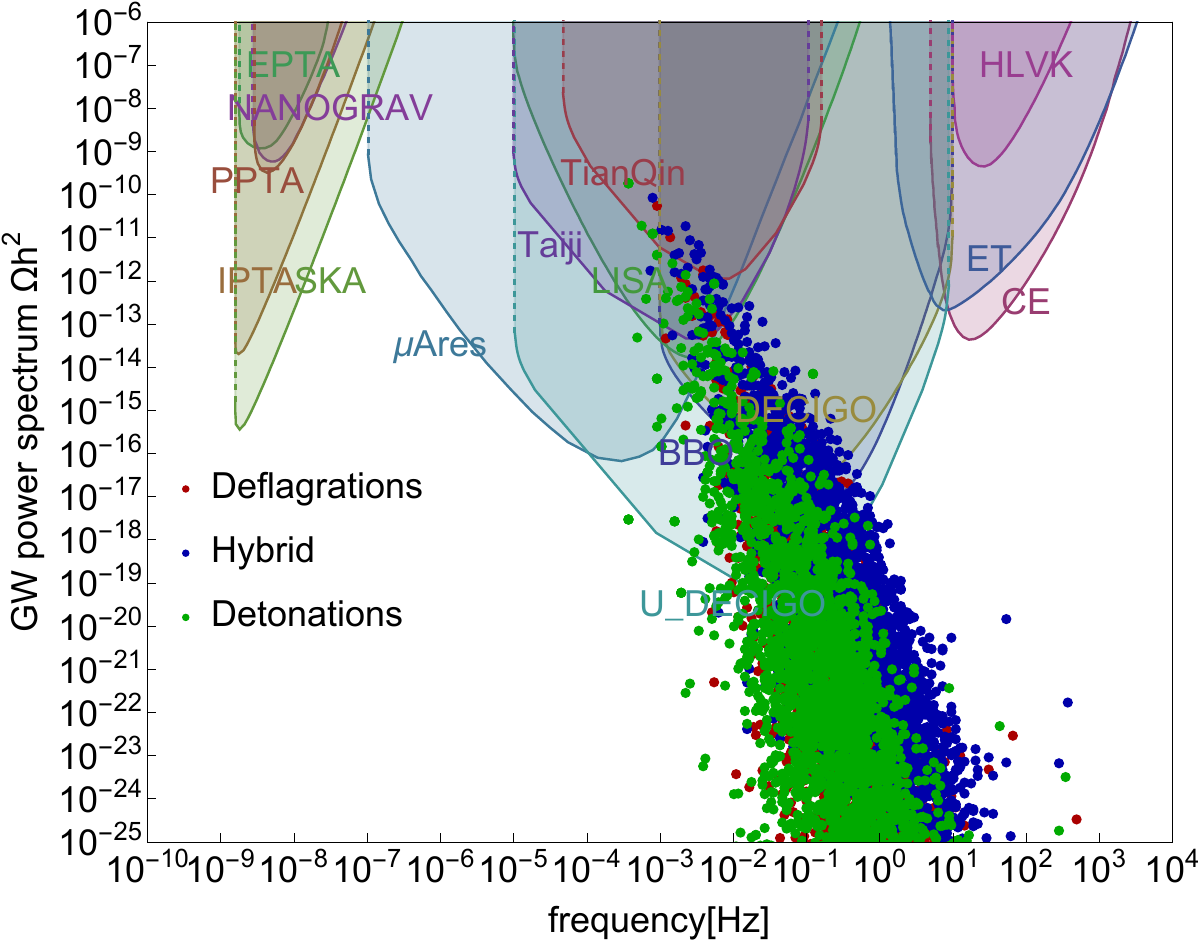}
    \includegraphics[width=0.3\textwidth]{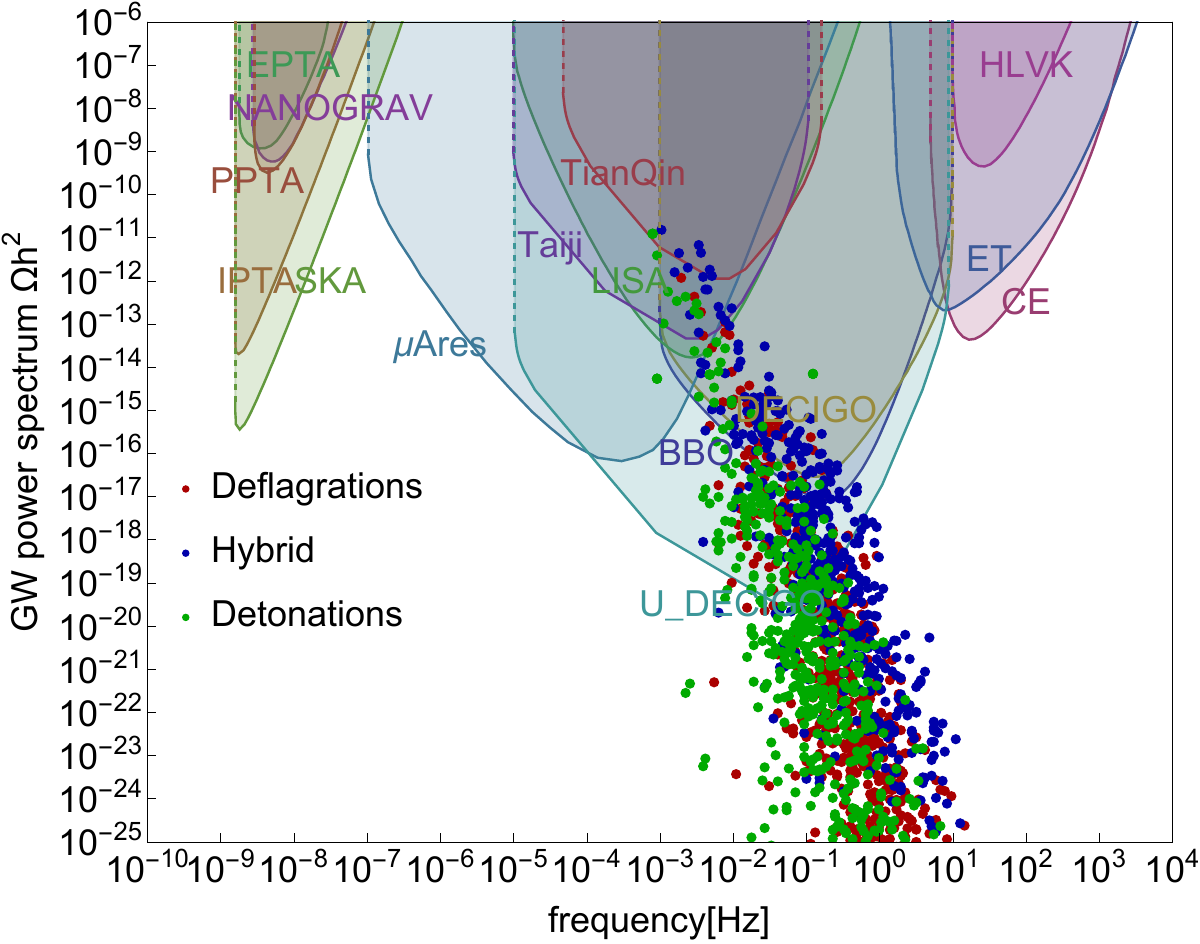}
    \caption{Gravitational wave power spectrum for Monte Carlo analysis classified by different nucleation modes: detonation, deflagration, hybrid. 
   Left panel: allowed set of models satisfying the FOPT constraints.
     Middle panel: same as the left panel with
      models satisfying FOPT constraints  and the $\Delta N_{\rm eff}$ constraint.
       Right: residual set of models satisfying the FOPT constraint,  $\Delta N_{\rm eff}$ constraint and the relic density constraint.}
    \label{fig:GW_MC_Mode}
\end{figure}
%%%%%%%%%%%%%%%%%%%%%%%%%%%%%%%%%%%%%%

\section{Conclusion\label{sec:7}}
In this work we have carried out a cosmologically consistent analysis of gravity
wave power spectrum arising from a first order phase transition involving two sectors: the visible sector and the hidden sector since the two sectors are intrinsically entangled in several ways.
 Thus the Hubble expansion involves energy densities of 
 all sectors, hidden and visible. Further, the strength of the first order phase transition
 in the hidden sector at tunneling time $T_{h,n}$ 
 depends on $\alpha(T_{h,n})= \epsilon/\rho_{rad}$ where $\epsilon$ is the latent heat and
  $\rho_{rad}= \rho^h_{rad}(T_{h,n})+ \rho^v_{rad}(T)$ where 
  $T=\zeta(T_{h,n}) T_{h,n}$ and involves the evolution function $\zeta(T_h)$.
  The same evolution function enters when we impose the $\Delta N_{eff}$ constraint
  at BBN time. Thus imposition of $\Delta N_{eff}$ at BBN requires
 a knowledge of the hidden sector temperature at BBN time which in the visible sector
 is $\sim 1$ MeV. Again one needs the evolution function to deduce 
 the effective degrees of freedom in the hidden sector at the temperature
 synchronous to $\sim 1$ MeV in the visible sector.
In brief since the visible  and the hidden sectors reside in different heat baths, a consistent analysis requires 
that one takes into account the 
dependence of the effective potential on two temperatures: one for the visible 
and the other for the hidden. In this work we have presented
an analysis of the gravitational wave power spectrum which takes into account 
the synchronous evolution of the visible and the hidden sectors. 
Within this framework we discuss nucleation which
involves bubble dynamics in two sectors. The analysis involves solution to the 
evolution function $\xi(T)= T_h/T$ along with solution to yield equations for the 
hidden sector particles. 
  Thus the formalism discussed in this work allows one to 
correlate physics at nucleation time and at BBN time, 
and allows for precision computation of $\Delta N_{eff}$ at BBN and of relic density. 
The formalism presents an improvement over current analyses where
 synchronous evolution of the visible and the hidden sectors is not utilized.  

Several aspects of the gravitational wave power spectrum are analyzed 
within the two temperature evolution formalism. 
Thus we analyze the sensitivity of the gravitational wave power spectrum to sound speed
for  symmetric and broken phases. The analysis includes nucleation 
involving two fields, one from the hidden and the other from the visible. 
 Here it is shown that for the case  two-field nucleation, models that 
pass all the constraints are those where the tunneling in the visible sector precedes 
tunneling in the hidden sector. Further, we discuss the possible imprint of the nucleation modes, i.e., detonation, deflagration and hybrid on the characteristics of the 
gravitational power spectrum. We show that a part of the parameter space of the specific gauged $U(1)$ extension of the standard model discussed here is testable at the proposed gravitational wave detectors.
%%%%%%%%%%%%%%%%%%%%%%%%%%%%%%%%%%%%%%%%%

   { Finally we mention below the novel material contained in the paper.

 All the published works on gravitational waves production  in hidden sector models
 thus far, that we are aware of, do not qualify as cosmologically consistent models since there is no synchronous thermal evolution of the visible and the hidden sectors
 in these works,  and our work is the first one which has accomplishes that.  Details of how a synchronous evolution of the visible  and the hidden sectors  is achieved in a two sector/ two temperature universe is discussed in detail in sections 2,3,4 and in the Appendices A-E. 
 Thus currently this paper is the only cosmologically consistent model and no comparable analysis exists in the literature. This is reflected in the first three words of the title of this paper which read ``Cosmologically consistent analysis". 
 
   This paper is the first work  where the  imprint of different nucleation modes, i.e., detonation, deflagration and hybrid, on the gravitational wave power spectrum is analyzed
   (see Fig. (9)). This aspect of the paper is of great significance since it tells us that experimental data  on gravitational waves  can be used to probe the very early history of the universe when the current universe was in the process of creation via bubble formation.
   No comparable analysis exists in any of  the previous works.
   
   Among other novel things, in this work we discussed sound speeds involving two sectors (see section 4), i.e., the visible and the hidden, which  have a very significant effect on the gravitational power spectrum when both sound speeds are taken into account
  as seen in Fig. 6. Here the very large effect that inclusion of sound speeds of 
   both the visible and of the hidden sector can generate on the power spectrum
   is exhibited. This type of analysis has not been discussed in the existing
  literature to our knowledge, and thus our analysis is more complete than what appears in the previous works for the two sector case.   \\
  }
  
%%%%%%%%%%%%%%%%%%%%%%%%%%%%%%%%%%%%%%%%
%%%%%%%%%%%%%%%%%%%%%%%%%%%%%%%%%%%%%%%
\noindent
Acknowledgements: The research of
WZF was supported in part by the National Natural Science Foundation of China under Grant No. 11935009. The research of PN and JL was supported in part by the NSF Grant PHY-2209903.
%%%%%%%%%%%%%%%%%%%%%%%%%%%%%%%%%%%%%%%%%%%%%
\section{Appendix A: Thermal mass calculation for a general $U(1)$ theory 
 \label{sec:8}}
  We discuss here the calculation of thermal masses for the hidden sector Lagrangian 
  given by Eq.(\ref{bsm}) and Eq.(\ref{pot-hid})
  \footnote{{While this paper was in the refereeing process, the eprint arXiv: 2205.08815
appeared which uses dimensionally reduced 3D thermal field theory to minimize the
uncertainty of the gravitational wave signal. This work along with those referenced in it
are a useful tool in making the thermal analysis  more precise. It is of interest to extend the analysis of this work to two sector/ two temperature case so as to be applicable to gravitational power spectrum involving the standard model and the hidden sector discussed in this work.}}.   
The  calculation is done in the high temperature regime,
where the temperature is much higher than the energy scale of the particles' masses.
We also take all the external momenta to zero.
In thermal
field theory, at some nonzero temperature $T$, the 1PI graphs are
defined in the Euclidean space ($t=i\tau$) with a periodicity in $\tau$. The computations
are governed by the conventional Feynman rules, while replacing the $k^{0}$
integral by a sum over Matsubara frequencies~\cite{Matsubara:1955ws} so that 
\begin{equation}
\int\frac{{\rm d}^{4}k}{(2\pi)^{4}}f(k^{0},\mathbf{k})\to T\sum_{n}\int\frac{{\rm d}^{3}\mathbf{k}}{(2\pi)^{3}}f(k^{0}={\rm i}\omega_{n},\mathbf{k})\,,
\qquad\omega^{\rm{ b}}_{n}=2n\pi T\,, {\qquad\omega_{n}^{{\rm f}}=(2n+1)\pi T\,.}
\label{matsubara}
\end{equation}
where $\omega^{\rm{ b}}_{n}$ is for bosonic modes and 
$\omega_{n}^{{\rm f}}$ for fermionic modes.
For the rest of the calculation it is useful to define a function $q(T)$ so that
\begin{equation}
q(z)=1+\frac{2\varsigma}{{\rm e}^{z/T}-\varsigma}
\end{equation}
with $\varsigma=+1$ for bosons and $\varsigma=-1$ for fermions.
Some of the integrals that appear in the thermal masses can then be given in terms of $\xi(z)$.
Thus we have
\begin{align}
T\sum_{n}\int\frac{{\rm d}^{3}\mathbf{k}}{(2\pi)^{3}}\frac{1}{\omega_{n}^{2}+|\mathbf{k}|^{2}} &=\int_{0}^{\infty}\frac{{\rm dk}\,{\rm k}^{2}}{2\pi^{2}}\frac{q({\rm k})}{2{\rm k}}
  = \left\{ \begin{aligned} & T^{2}/12\qquad\qquad\qquad{\rm bosons}\\
 &  -T^{2}/24\qquad\qquad{\rm fermions}
\end{aligned}
\right.\\
T\sum_{n}\int\frac{{\rm d}^{3}\mathbf{k}}{(2\pi)^{3}}\frac{\omega_{n}^{2}}{\big(\omega_{n}^{2}+|\mathbf{k}|^{2}\big)^{2}} & =\int_{0}^{\infty}\frac{{\rm dk}\,{\rm k}^{2}}{2\pi^{2}}\frac{{\rm k}+2Tq({\rm k})-{\rm k}q^{2}({\rm k})}{8{\rm k}T}
=  \left\{ \begin{aligned} & -T^{2}/24\qquad\quad{\rm bosons}\\
& T^{2}/48\qquad\qquad{\rm fermions}
\end{aligned}
\right.\\
T\sum_{n}\int\frac{{\rm d}^{3}\mathbf{k}}{(2\pi)^{3}}\frac{|\mathbf{k}|^{2}}{\big(\omega_{n}^{2}+|\mathbf{k}|^{2}\big)^{2}} &=\int_{0}^{\infty}\frac{{\rm dk}\,{\rm k}^{2}}{2\pi^{2}}\frac{-{\rm k}+2Tq({\rm k})+{\rm k}q^{2}({\rm k})}{8{\rm k}T}
= \left\{ \begin{aligned} & T^{2}/8\qquad{\rm \qquad bosons}\\
 & -T^{2}/16\qquad{\rm fermions}
\end{aligned}
\right.
\end{align}
where we dropped the non-thermal contribution in the integral which
is UV-divergent and is removed by the counter terms.
There are no thermal corrections to the fermion masses, and only the scalar boson and the
longitudinal components of the gauge boson gain thermal corrections.

\subsection{A1: Thermal mass correction to  scalar $\chi$}

 We discuss the thermal corrections to the scalar boson first. Here the thermal mass corrections come from the scalar loops, from the
 neutral Goldstone loop, and from the gauge
boson loop as shown in Fig.(\ref{fig:STM}). 
\begin{figure}[h]
\begin{center}
\includegraphics[scale=0.03]{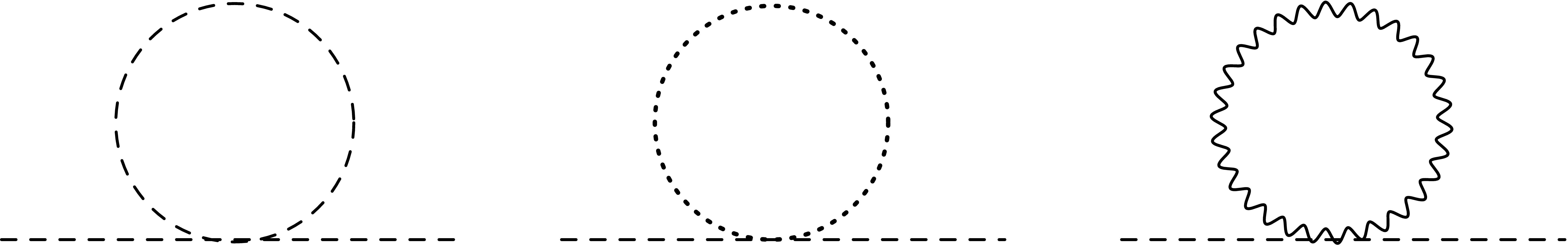}
\caption{Left figure: thermal mass correction to the scalar field $\chi$  complex scalar
loop exchange indicated by the double dashed line.
(dashed line).
Middle figure: same as the left figure except that the thermal loop correction is from the Goldstone loop (small dashed line).
Right figure: same as the left figures except that the thermal loop correction is from the $U(1)$ gauge field loop (wavy line).}
\label{fig:STM}
\end{center}
\end{figure}

The scalar loop contribution from $\chi^4$ term is given by
\begin{equation}
{\rm 
Scalar ~loop\,from\,}\chi^{4}  =3\lambda_h {\rm i}\int\frac{{\rm d}^{4}k}{(2\pi)^{4}}\frac{1}{k^{2}}
\to3\lambda_h T\sum_{n}\int\frac{{\rm d}^{3}\mathbf{k}}{(2\pi)^{3}}\frac{1}{\omega_{n}^{2}+|\mathbf{k}|^{2}}=\frac{\lambda_h}{4}T^{2}\,.
\end{equation}
The pre factors can be understood as the following:
The contraction of 
$\langle\chi(x)|\frac{\lambda_h}{4}\chi\chi\chi\chi(z)|\chi(y)\rangle$
has in total $4\times3$ ways and  give rise to a factor $3\lambda$, and
the additional ${\rm i}$ in the front is from the computation of the amplitudes, i.e., ${\rm i}\mathcal{M}$.
The scalar loop contribution from $\chi^2 (G_h^0)^2$ term is given by
\begin{align}
{\rm Goldstone\, loop\,from\,}\chi^2 (G_h^0)^2  =\lambda_h {\rm i} \int\frac{{\rm d}^{4}k}{(2\pi)^{4}}\frac{1}{k^{2}}
\to\lambda_h T\sum_{n}\int\frac{{\rm d}^{3}\mathbf{k}}{(2\pi)^{3}}\frac{1}{\omega_{n}^{2}+|\mathbf{k}|^{2}}=\frac{\lambda_h}{12}T^{2}\,.
\end{align}
where the contraction of $\langle\chi(x)|\frac{\lambda_h}{2}\chi\chi(z)G^0_hG^0_h(z)|\chi(y)\rangle$
has 2 ways and thus gives rise to a factor $2\times(-{\rm i}\frac{\lambda_h}{2})=-{\rm i}\lambda_h$.
Thus the total scalar thermal contribution is the sum of the two above results:
\begin{equation}
{\rm Scalar\,and\,Goldstone\,Loops}=\frac{\lambda_h}{4}T^{2}+\frac{\lambda_h}{12}T^{2}= \frac{\lambda_h}{3}T^{2}\,,
\end{equation}
which is different from the SM Higgs thermal mass $\frac{\lambda}{2}T^{2}$, 
owing to the fact that there are also contributions from the two charged Goldstone bosons.
The gauge boson loop contributions to the scalar mass is given by
\begin{equation}
{\rm Gauge\,boson\,loop}  ={\rm i} ({\rm i}g_x^{2}) \int\frac{{\rm d}^{4}k}{(2\pi)^{4}}{\rm Tr}\big[\Delta_{\mu\nu}(k)\big]
  \to3g_x^{2}T\sum_{n}\int\frac{{\rm d}^{3}\mathbf{k}}{(2\pi)^{3}}\frac{1}{\omega_{n}^{2}+|\mathbf{k}|^{2}} 
=\frac{g_x^{2}}{4}T^{2}\,,
\end{equation}
where $\Delta_{\mu\nu}(k)$ is the gauge boson propagator in the Landau
gauge given by $\Delta_{\mu\nu}(k)=\frac{-{\rm i}}{k^{2}}\left(g^{\mu\nu}-\frac{k^{\mu}k^{\nu}}{k^{2}}\right)$.
The contraction of $\langle\chi(x)|\frac{1}{2}gx^{2}\chi\chi(z) AA(z)|\chi(y)\rangle$
gives rise to a total front factor $2\times({\rm i}\frac{1}{2}g_x^{2})={\rm i}g_x^{2}$.

Thus in this case the total thermal mass for the dark scalar field $\Pi_{\chi}(T)$ is given by
\begin{equation}
\Pi_{\chi} = \frac{1}{3} \lambda_h T^2 +\frac{1}{4}g_x^2 T^2\,.
\end{equation}

\subsection{Thermal mass for the $U(1)$ gauge boson}
Next we compute the thermal mass for the longitudinal contribution to the $U(1)$ gauge boson mass $\gamma^\prime$.
Here the polarization tensors of vector bosons can split into components of longitudinal (L)
and transverse (T) polarization so that
\begin{equation}
\Pi^{\mu \nu } = \Pi^T T^{\mu \nu } + \Pi^L L^{\mu \nu }
\end{equation}
with projection operators $T^{\mu \nu } = {\rm diag}\{0,2,2,2\}$ and
$L^{\mu \nu } =  {\rm diag}\{-1,0,0,0\}$ in the IR limit~\cite{Carrington:1991hz}.
The gauge boson thermal mass corrections come from scalar and fermion contributions:
\begin{figure}[h]
\begin{center}
\includegraphics[scale=0.03]{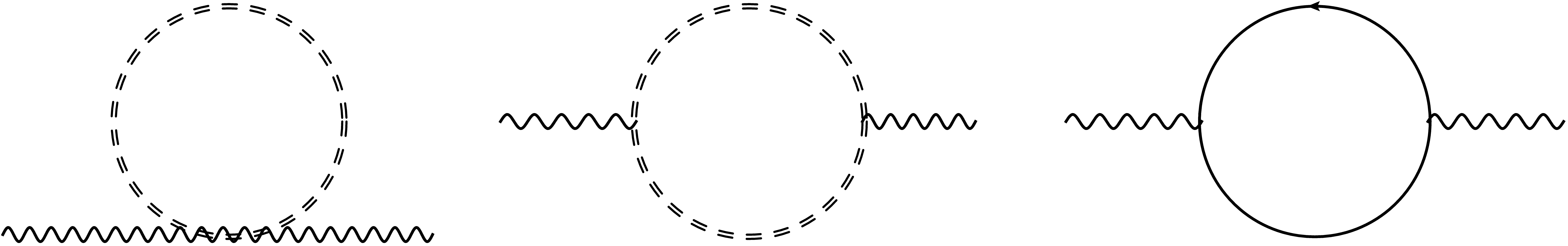}
\caption{Thermal mass corrections to the gauge boson mass from the complex
scalar loop with a four point vertex  (left figure) and with a three point vertex 
(middle figure), and
from the Dirac fermion loop (right figure).}
\label{fig:GBTM}
\end{center}
\end{figure}
In this case the thermal mass corrections to the $\gamma^\prime$ mass come from scalar and fermion loop contributions
as shown in Fig.(\ref{fig:GBTM}). The calculation of the scalar loop contribution is easier to be performed
considering the complex $U(1)_X$ field $\Phi$ which has 2 degrees of freedom and it represented by double dashed line in Fig.(\ref{fig:GBTM}).
The corresponding Lagrangian reads
\begin{equation}
\mathcal{L}\supset|D_{\mu}\Phi|^{2}\to g_x^{2}A^{2}\Phi\Phi^* + {\rm i}g_x A_\mu(\Phi^* \partial^\mu \Phi - \Phi \partial^\mu \Phi^*)\,,
\end{equation}
which gives  ${\rm i}2g_x^{2}$ for the four-point vertex $AA\Phi\Phi^*$
and $-2g_xk^\mu $ for the three-point vertex $A^\mu \Phi\Phi^*$.
The scalar loop contribution is
\begin{equation}
{\rm Scalar\,loop}=2\times\frac{{\rm i}}{2}\int\frac{{\rm d}^{4}k}{(2\pi)^{4}}
\left[({\rm i}2g_x^{2}g^{\mu\nu})\frac{{\rm i}}{k^{2}}
+ \frac{(2gk^\mu)(-2g_xk^\nu)({\rm i})^2}{(k^2)^2}
\right]
\,,
\end{equation}
where the pre factor $2$ is from $\Phi$ being a complex scalar, $1/2$
is the symmetric factor due to the two external gauge boson legs.
One still needs to multiply the $2$ from the complex $\Phi$.
For the non-zero contribution to the longitudinal part we get
\begin{align}
\Pi^{L}  =-\Pi^{00}&={\rm i}2g_x^{2}\int\frac{{\rm d}^{4}k}{(2\pi)^{4}}\left[\frac{1}{k^{2}}
-\frac{2 k_0^2}{(k^2)^2}\right]
  \to 2g_x^{2}T\sum_{n}\int\frac{{\rm d}^{3}
\mathbf{k}}{(2\pi)^{3}}\Bigg[\frac{1}{\omega_{n}^{2}+|\mathbf{k}|^{2}}
-\frac{2 \omega_n^2}{\big(\omega_{n}^{2}+|\mathbf{k}|^{2}\big)^2} \Bigg]\nonumber\\
&=\frac{g_x^{2}}{3}T^{2}\,.
\end{align}
The $U(1)_X$ charged  fermion loop contribution given by the 
right diagram in Fig.(\ref{fig:GBTM}) is given by

\begin{equation}
{\rm Fermion\, loop}  =(-){\rm i}\int\frac{{\rm d}^{4}k}{(2\pi)^{4}}({\rm i}g_x)^{2}\frac{({\rm i})^{2}{\rm Tr}(\gamma^{\mu}\slashed k\gamma^{\nu}\slashed k)}{\big(k^{2}\big)^{2}}
 =-{\rm i}4g_x^{2}\int\frac{{\rm d}^{4}k}{(2\pi)^{4}}\frac{2k^{\mu}k^{\nu}-k^{2}g^{\mu\nu}}{\big(k^{2}\big)^{2}}\,,
\end{equation}
which gives contribution to $\Pi_L$ so that
\begin{equation}
\Pi^{L}  =-\Pi^{00}={\rm i}4g_x^{2}\int\frac{{\rm d}^{4}k}{(2\pi)^{4}} \frac{k_{0}^{2}+|\mathbf{k}|^{2}}{\big(k^{2}\big)^{2}} 
\to-4g_x^{2}T\sum_{n}\int\frac{{\rm d}^{3}\mathbf{k}}{(2\pi)^{3}}\frac{\omega_{n}^{2}+|\mathbf{k}|^{2}}{\big(\omega_{n}^{2}+|\mathbf{k}|^{2}\big)^{2}} =\frac{g_x^{2}}{3}T^{2}\,.
\end{equation}
This is the contribution from a Dirac fermion exchange. For a chiral fermion exchange, either  left-handed or right-handed, 
the contribution to the thermal mass is $\frac{g_x^{2}}{6}T^{2}$.
For an abelian gauge theory there is no gauge boson loop contribution.

From the above analysis we deduce that if in addition to the complex
scalar field $\Phi$, there are $n$ numbers of dark {\it chiral} fermion $X_i$ 
(either left-handed or right-handed) with the $U(1)_X$ charge $Q_i$,
then the thermal mass for the dark sector gauge boson $\gamma^\prime$ is given by
\begin{equation}
\Pi_{\gamma^\prime} = \frac{1}{3} g_x^2 T^2 +\sum_{i=1}^n \frac{1}{6} g_x^2 Q_i^2 T^2\,,
\end{equation}
where the first term on the right hand side arises from a complex 
scalar loop and the second term from $N$ chiral fermion loops.
%%%%%%%%%%%%%%%%%%%%%%%%%%%%%%%%%%%%%%%%%%%%
\subsection{Daisy resummation \label{sec:8.1}}
\begin{figure}[t]
\begin{center}
\includegraphics[scale=0.2]{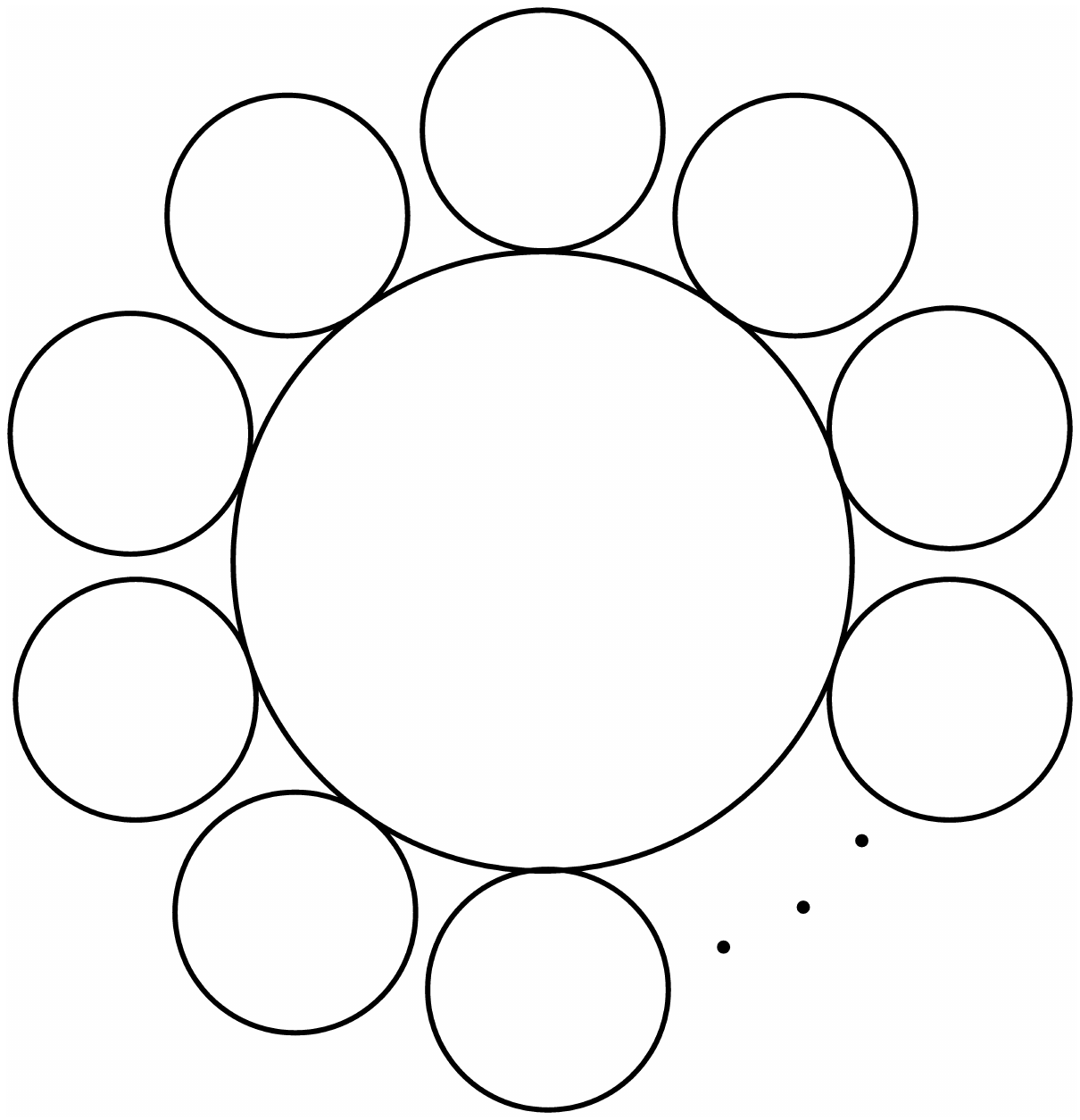}
\caption{A daisy or ring diagram which contributes to thermal potential.}
\label{fig:daisy}
\end{center}
\end{figure}
As discussed in a number of works in  temperature dependent perturbation theory
 the summation over higher loops can produce the same size correction as the one loop and should be taken into account~\cite{Carrington:1991hz,Comelli:1996vm,Quiros:1999jp,Katz:2014bha}.
Thus one finds that at the $n$-th order, the
$n$-loop daisy diagram with
$n-1$ petals  (see Fig.(\ref{fig:daisy})), also called the ring diagram, gives the dominant contribution.
The daisy diagrams can be resumed by adding up propagators with increasing
number of attached loops. Each loop can contribute a thermal mass
correction $\Pi(T_h)=\Pi_{1}(T_h)$, where $\Pi_{1}(T_h)$ is the one-loop
thermal mass correction, derived above. The sum of all the
propagators can be written as
\begin{align*}
 & \frac{1}{p^{2}-m^{2}}+\frac{\Pi(T_h)}{(p^{2}-m^{2})^{2}}+\frac{\Pi^{2}(T_h)}{(p^{2}-m^{2})^{3}}+\cdots
=\frac{1}{p^{2}-m^{2}-\Pi(T_h)}\,.
\end{align*}
which is equivalent to adding a thermal  contribution to the 
mass in the propagator, i.e., 
\begin{equation}
m^{2}(\chi_c)\to m^{2}(\chi_c)+\Pi(T_h)\,.\label{eq:MShiftT}
\end{equation}
Now the one loop contribution at 
zero temperature is given by 
\begin{equation}
V_{{1h}}^{(0)}(\chi_c)=\sum_{i}\frac{N_{i}}{2}\int\frac{{\rm d}^{4}k_{E}}{(2\pi)^{4}}\ln\big[k_{E}^{2}+m_{i}^{2}(\chi_c)\big]\,.
\label{0+1}
\end{equation}
where $i$ runs over all the particles that enter the loop and $N_i$ are the degrees of
freedom for particle $i$.
The regularized and renormalized one loop potential as given by  the
right hand side under
$\overline{MS}$ scheme is the familiar Coleman-Weinberg potential. 
From here on we follow the procedure in the preceding analysis 
  and using the Imaginary Time Formalism we replace the 
integral over  $k_{E}^{0}$ by a summation over the Matsubara frequencies as 
given by Eq.(\ref{matsubara}) 
where $\omega_{n}^{{\rm b}}$ are for bosons and $\omega_{n}^{{\rm f}}$
for fermions and 
$\Delta V_{1h}^{T_h}(\chi_c,T_h))$ at finite temperature is given by
\begin{equation}
\Delta V_{1h}^{T_h}(\chi_c,T_h)=\sum_{i}\frac{N_{i}T_h}{2}\sum_{n=-\infty}^{+\infty}\int\frac{{\rm d}^{3}\mathbf{k}}{(2\pi)^{3}}\ln\big[\mathbf{k}^{2}+\omega_{n}^{2}+m_{i}^{2}(\chi_c)\big]\,.
\end{equation}
After the replacement Eq. (\ref{eq:MShiftT}), the thermal
one-loop 
potential $\Delta V_{1h}^{(T_h)}(\chi_c,T_h)$  reads
\begin{align*}
\Delta V_{1h}^{(T_h)}(\chi_c,T_h) &
 \to\sum_{i}\frac{N_{i}T_h}{2}\left\{ \sum_{n\neq0}\int\frac{{\rm d}^{3}\mathbf{k}}{(2\pi)^{3}}\ln\big[\mathbf{k}^{2}+\omega_{n}^{2}+m_{i}^{2}(\chi_c)\big]+
 \int\frac{{\rm d}^{3}\mathbf{k}}{(2\pi)^{3}}\ln\big[\mathbf{k}^{2}+m_{i}^{2}(\chi_c)+\Pi(T_h)\big]\right\} \\
 & =\sum_{i}\frac{N_{i}T_h}{2}\Bigg\{\sum_{n\neq0}\int\frac{{\rm d}^{3}\mathbf{k}}{(2\pi)^{3}}\ln\big[\mathbf{k}^{2}+\omega_{n}^{2}+m_{i}^{2}(\chi_c)\big]+\int\frac{{\rm d}^{3}\mathbf{k}}{(2\pi)^{3}}\ln\big[\mathbf{k}^{2}+m_{i}^{2}(\chi_c)\big]\\
 & \qquad\qquad\quad+\int\frac{{\rm d}^{3}\mathbf{k}}{(2\pi)^{3}}\ln\Big[1+\frac{\Pi(T_h)}{\mathbf{k}^{2}+m_{i}^{2}(\chi_c)}\Big]\Bigg\}\\
 & =\sum_{i}\frac{N_{i}T_h}{2}\sum_{n=-\infty}^{+\infty}\int\frac{{\rm d}^{3}\mathbf{k}}{(2\pi)^{3}}\ln\big[\mathbf{k}^{2}+\omega_{n}^{2}+m_{i}^{2}(\chi_c)\big]\\
 &~~~+\sum_{i}\frac{\bar{N}_{i}T_h}{2}\int\frac{{\rm d}^{3}\mathbf{k}}{(2\pi)^{3}}\ln\Big[1+\frac{\Pi(T_h)}{\mathbf{k}^{2}+m_{i}^{2}(\chi_c)}\Big]\\
 & =\Delta V_{1h}^{(T_h)}(\chi_c,T_h)+V^{\rm daisy}_{h}(\chi_c,T_h),
\end{align*}
where $\bar{N}_{i}$ are the bosonic degrees of freedom which incur the
mass shift. 
The daisy diagram contribution to the effective potential
from one particle is computed to be
\begin{align*}
V^{\rm daisy}_{h}(\chi_c,T_h)
 & =\frac{T_h}{2}\int\frac{{\rm d}^{3}\mathbf{k}}{(2\pi)^{3}}\ln\Big[1+\frac{\Pi(T_h)}{\mathbf{k}^{2}+m^{2}(\chi_c)}\Big]=\lim_{\Lambda\to+\infty}\frac{T_h}{4\pi^{2}}\int_{0}^{\Lambda}{\rm d}{\rm k}\,{\rm k}^{2}\ln\Big[1+\frac{\Pi(T_h)}{{\rm k}^{2}+m^{2}(\chi_c)}\Big]\\
 & =\lim_{\Lambda\to+\infty}\frac{T_h}{4\pi^{2}}\times\frac{1}{3}\Big\{2\Lambda\Pi(T_h)+\Lambda^{3}\ln\Big[1+\frac{\Pi(T_h)}{m^{2}(\chi_c)+\Lambda^{2}}\Big]\\
 & \quad\qquad+2m^{3}(\chi_c)\tan^{-1}\left(\frac{\Lambda}{m}\right)-2\big[m^{2}(\chi_c)+\Pi(T_h)\big]^{3/2}\tan^{-1}\Big[\frac{\Lambda}{\sqrt{m^{2}(\chi_c)+\Pi(T_h)}}\Big]\Big\}\\
 & \to-\frac{T_h}{12\pi}\Big\{\big[m^{2}(\chi_c)+\Pi(T_h)\big]^{3/2}-m^{3}(\chi_c)\Big\}\,,
\end{align*}
where on the last line we drop the divergent pieces which are
canceled by counter terms, and  $\tan^{-1}\left(\frac{\Lambda}{m_\times}\right)\to\frac{\pi}{2}$
when taking $\Lambda\to+\infty$ where $m_\times$ is a mass taken positive.

%%%%%%%%%%%%%%%%%%%%%%%%%%%%%%%%%%%%%%%%%%
%%%%%%%%%%%%%%%%%%%%%%%%%%%%%%%%%%%%%%%%%
\section{Appendix B: Effective thermal potential of the visible sector \label{sec:9}}
 The  effective 
 Higgs potential in the standard model, including the temperature dependent part, 
  is well known. It is given by the sum of the zero temperature tree and zero temperature 
  Coleman-Weinberg one loop potential\cite{Coleman:1973jx}, temperature
  dependent one loop correction and ``daisy diagrams''\cite{Sher:1988mj,Anderson:1991zb,Dine:1992wr,Quiros:1999jp,Morrissey:2012db,Comelli:1996vm,Matsedonskyi:2020mlz,Katz:2014bha}. 
 We give a brief discussion of it here for completeness.
Thus consider the tree level potential for the standard model with the complex 
Higgs doublet field $H$ so that 
\begin{align}
V(H,H^\dagger)= -\mu^2 H^\dagger H + \lambda (H^\dagger H)^2.
\end{align}

We write the doublet of the Higgs field $H$ so that
\begin{align}
H&= \left(\begin{matrix} G^{+}\\ \frac{(\phi_c+ \phi+ i G_3)}{\sqrt 2}\end{matrix}\right)
\end{align}
where $\phi_c$ is the background fields, $\phi$ is the Higgs field and 
$G^{+}= (G_1+ i G_2)/\sqrt 2$ where $G_{1,2,3}$ 
are the three Goldstone bosons. The tree level potential given by 
\begin{align}
V_0(\phi_c)= -\frac{\mu^2}{2} \phi_c^2 + \frac{\lambda}{4} \phi_c^4.
\end{align}
To one loop order, the
effective potential of the standard model including temperature dependent 
contributions is given by 
\begin{equation}
V_{{\rm eff}}(\phi_c,T)=V_{0}(\phi_c)+V_{1}^{(0)}(\phi_c)
+\Delta V_{1}^{(T)}(\phi_c,T)+V^{{\rm daisy}}(\phi_c,T)
+\delta V^{(T)}(\phi_c,T)\,,
\end{equation}
where $V^{0)}_1$ is the zero temperature one loop potential, 
$\Delta V^{(T)}_1$ is the temperature dependent one loop contribution, $V^{{\rm daisy}}$ is the daisy loop contribution, and $\delta V^{(T)}$ are the counter terms to remove divergent terms.
Thus $V_{1}^{(0)}(\phi_c)$ is given by  
\begin{equation}
V_{1}^{(0)}(\phi_c)=\sum_{i}\frac{N_{i}(-1)^{2{\rm s}_{i}}}{64\pi^{2}}m_{i}^{4}(\phi_c)\left[\ln\left(\frac{m_{i}^{2}(\phi_c)}{\Lambda^{2}}\right)-\mathcal{C}_{i}\right]\,,
\end{equation}
where the sum $i$ runs over all particles in the theory with $N_{i}$
degrees of freedom for particle $i$ with mass $m_{i}(\phi_c)$ and spin ${\rm s}_{i}$, $\Lambda$
is the renormalization scale and $\mathcal{C}_{i}$ equals 5/6
for gauge bosons and 3/2 for fermions and scalars in $\overline{\text{MS}}$ renormalization.
 The relevant contribution arises 
 from the gauge bosons $Z$ and $W^{\pm}$, the top quark, the Higgs boson, and the 
 Goldstone bosons.
Thus for the SM $i$ runs through $\{Z,W,t,H, G_3, G^{\pm}\}$ and the corresponding
front factors are $N_{i}=\{3,6,12,1,1,1,1\}$.
The field dependent masses $m_{i}^{2}(\phi_{c})$ are given by
\begin{align}
m_{h}^{2}(\phi_{c})&= -\mu^2
+3\lambda\phi_{c}^{2}\,,\qquad m_{t}^{2}(\phi_{c})=\frac{1}{2}y_{t}^{2}\phi_{c}^{2}\,,\\
m_{W}^{2}(\phi_{c})&=\frac{1}{4}g_{2}^{2}\phi_{c}^{2}\,,\qquad M_{Z}^{2}(\phi_{c})=\frac{1}{4}(g_{2}^{2}+g_{Y}^{2})\phi_{c}^{2}\,,\\
 m_{G_3}&= m_{G^{\pm}}=-\mu^2+ \lambda \phi_{c}^{2}\,,\qquad
 \end{align} 
The thermal correction in one-loop order arise from bosons and fermions which couple to the Higgs field
and is given by
\begin{align}
\Delta V_{1}^{(T)}(\phi,T)=  \frac{T^4}{2\pi^2}
\Big[ &6J_{B}\left(\frac{m_{W}}{T}\right)+  3J_{B}\left(\frac{m_{Z}}{T}\right)
+   J_{B}\left(\frac{m_{h}}{T}\right)
+ 3  J_{B}\left(\frac{m_{G}}{T}\right)
- 12 J_{F}\left(\frac{m_{t}}{T}\right)\,.\Big]
\end{align}
where  the function $J_{B}$ and $J_{F}$ are defined as in Eq.(\ref{JBF}).
Further, as noted earlier one needs to include the daisy resummation contribution to the potential
which in this case is given by
\begin{equation}
V_{1}^{{\rm daisy}}(\phi,T)=\frac{T}{12\pi}\sum_{B^{\prime}=Z,W,H}g_{B^{\prime}}\big\{ m_{B^{\prime}}^{3}(\phi)-\big[m_{B^{\prime}}^{2}(\phi)+\Pi_{B^{\prime}}(T)\big]^{3/2}\big\}\,,\label{eq:DaisyP}
\end{equation}
where the sum runs only over scalars and longitudinal vectors. Here
  $g_{B^{\prime}}=\{1,2,1\}$ for $B^{\prime}=\{Z,W,H\}$,  
   and there are no
contributions to the transverse modes and to the fermion masses. 
Thus the thermal contributions to the masses $\Pi_{B^{\prime}}(T)$ are given by~\cite{Carrington:1991hz}
\begin{gather}
\Pi_{H}(T)=\left[\frac{1}{6}(3g_{2}^{2}+g_{Y}^{2})+\frac{1}{4}y_{t}^{2}+\frac{1}{2}\lambda\right]T^{2}\,,\\
\Pi_{W}(T)=\Pi_{Z}(T)=\frac{11}{6}g_{2}^{2}T^{2}\,,
\end{gather}
at the leading order in $T^{2}$ where $y_t$ is defined so that $m_t=\frac{1}{\sqrt 2} y_t v$ and
$v\simeq 246$ GeV.

%%%%%%%%%%%%%%%%%%%%%%%%%%%%%%%%%%%%%%%%%%
\section{Appendix C: Further details of visible and hidden sector interactions\label{sec:10}}
As noted in section \ref{sec:2} the analysis of synchronous evolution is very general and 
 applicable to a wide array of portals connecting the hidden and the visible sectors.  In this work for the specific hidden sector with a $U(1)$ gauge invariance broken by the Higgs mechanism, we used
  the kinetic mixing between the hidden and the visible sectors as noted in section \ref{sec:2.2}.
Here one includes  a mixing term 
$-\frac{\delta}{2} A^{\mu\nu}B_{\mu\nu}$ in the Lagrangian, where $A^{\mu\nu}$ is the field strength 
of the hidden sector $U(1)$ field $A^\mu$ and
 $B_{\mu\nu}$ is the field strength of the $U(1)_Y$ hypercharge field $B_\mu$ of the visible sector.
  Since the standard model is based on 
the group $SU(2)\times U(1)_Y$ we will have a coupling of three gauge fields
$A_3^{\mu},B^{\mu}, A^{\mu}$, where $A_3^{\mu}$ is the third component of the $SU(2)_L$ gauge field $A^{\mu}_a (a = 1, 2, 3)$ of the standard model. 
After electroweak symmetry breaking and in the canonical basis where the 
  kinetic energies of the gauge fields are diagonalized and normalized,  
  the physical fields are
$Z^{\mu},A_{\gamma}^{\mu}, A_{\gamma'}^{\mu}$, where $Z$ is the Z-boson
of the standard model, $A_\gamma$ is the photon, and $A_{\gamma'}$ is the
dark photon. Thus the couplings governing the dark sector and the 
feeble interactions of the dark sector with the visible sector are given by
\begin{align}
\label{D-darkphoton}
&\Delta \mathcal{L}^{\rm int}=\bar D\gamma^\mu(g_{\gamma'} A_{\mu}^{\gamma'}+g_{Z} Z_{\mu}+g_{\gamma} A_{\mu}^{\gamma})D
   + \frac{g_2}{2\cos\theta}\bar\psi_f\gamma^{\mu}\Big[(v'_f-\gamma_5 a'_f)A^{\gamma'}_{\mu}\Big]\psi_f -\Delta V_h  \non
&\Delta V_h= \frac{1}{2} m^2_\chi \chi^2 + \frac{1}{2} m^2_{\gamma'} A_\mu^{\gamma'} A^{\gamma' \mu}  + g^2_x v_h \chi  A_\mu^{\gamma'} A^{\gamma' \mu} + \frac{1}{2} g_x^2 \chi^2 A_\mu^{\gamma'} A^{\gamma' \mu}\\
v'_f&=-\cos\psi[(\tan\psi-s_\delta\sin\theta)T_{3f}-2\sin^2\theta(-s_{\delta} \csc\theta+\tan\psi)Q_f],\\
a'_f&=-\cos\psi(\tan\psi-s_{\delta} \sin\theta)T_{3f}.
\end{align}
Here $s_{\delta}=\sinh\delta$ and $c_{\delta} = \cosh\delta$,
 and $f$ runs over all SM fermions, $m_{\gamma'}= g_x v_h$ and $m_{\chi} =\sqrt{2 \lambda_h} v_h$. Further,  
 $T_{3f}$ is the third component of isospin, $Q_f$ is the electric charge for the fermion.
 The  couplings $g_{Z} , g_{\gamma}$ and $g_{\gamma'}$ that appear above 
are given by 
\begin{align}
 g_{\gamma'}= g_X Q_X (\mathcal{R}_{11}- s_{\delta} \mathcal{R}_{21}),
  ~g_{\gamma} = g_X Q_X (\mathcal{R}_{13}- s_{\delta} \mathcal{R}_{23}), 
 ~g_Z &= g_X Q_X (\mathcal{R}_{12}- s_{\delta} \mathcal{R}_{22}). 
  \end{align}
Here the matrix $\mathcal{R}$ is given  by Eq.~(23) of~\cite{Feldman:2007wj} and it involves three Euler angles $(\theta, \phi, \psi)$ which are given by  
\begin{equation}
 \tan\phi=-s_{\delta}, ~~~ \tan\theta=\frac{g_Y}{g_2}c_{\delta}\cos\phi,
~\tan2\psi=\frac{2\bar\delta m^2_Z\sin\theta}{m^2_{\gamma'}-m^2_Z+(m^2_{\gamma'}+m^2_Z-m^2_W)\bar\delta^2},
 \label{hid-exact}
\end{equation}
where $\bar \delta=-\delta/\sqrt{1-\delta^2}$.
In addition to the above, there is also a modification of the  standard model couplings. 
Thus in the canonically diagonalized basis the couplings of  $Z_\mu$ and $A^\gamma_\mu$ are given by~{\cite{Cheung:2007ut,Feldman:2007wj}
\begin{equation}
\Delta \mathcal{L}'_{\rm SM}=\frac{g_2}{2\cos\theta}\bar\psi_f\gamma^{\mu}\Big[(v_f-\gamma_5 a_f)Z_{\mu}\Big]\psi_f+e\bar\psi_f\gamma^{\mu}Q_f A^\gamma_{\mu}\psi_f\,.
\label{SMLag}
\end{equation}
Modifications to the visible sector interactions appear in the vector and axial vector couplings so that (see, ~\cite{Aboubrahim:2020lnr,Aboubrahim:2020afx,Aboubrahim:2021ycj,Aboubrahim:2021ohe})
\begin{equation}
\begin{aligned}
v_f&=\cos\psi[(1+ s_\delta \tan\psi\sin\theta)T_{3f}-2\sin^2\theta(1+ s_\delta \csc\theta\tan\psi)Q_f],\\
a_f&=\cos\psi(1 + s_\delta \tan\psi\sin\theta)T_{3f}. 
\end{aligned}
\label{eqn:v-a}
\end{equation}
\begin{figure}
    \centering
        \includegraphics[width=0.6\textwidth]{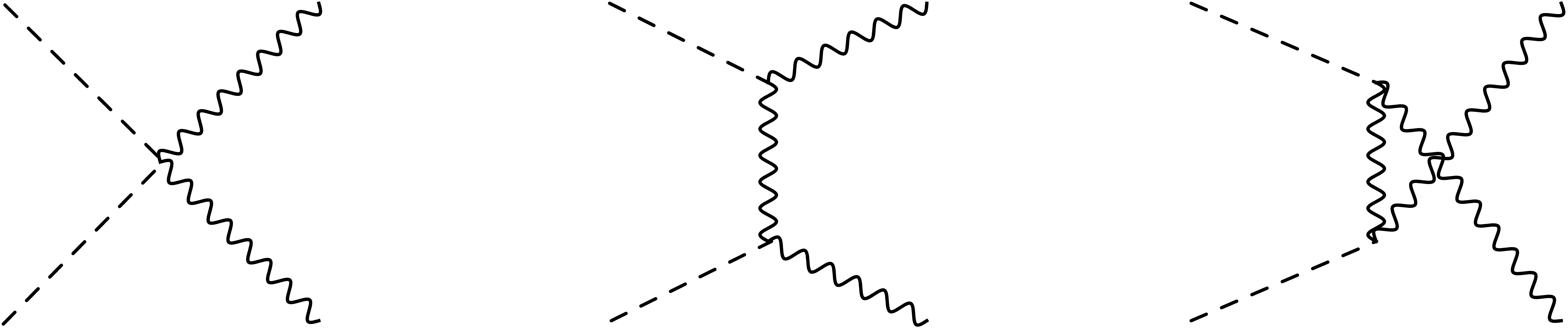}    
    \caption{The Feynman diagram for the annihilation process $\chi\chi\rightarrow \gamma'\gamma'$ }
    \label{fig:Feynman}
\end{figure}
%%%%%%%%%%%%%%%%%%%%%%%%%%%%%%%%%%%%%%%%
\section{Appendix D: Scattering cross sections for $\xi(T)$ and yield equations 
for the hidden sector fields.
\label{sec:11}}

The  analysis of yields in Eq.(\ref{DE2})-Eq.(\ref{DE4}) requires several cross sections. 
The cross sections $\sigma_{D\bar D\to \gamma'\gamma'}$, $\sigma_{D\bar D\to i\bar i}$,
$\sigma_{i\bar i\to \gamma'\gamma'}$ and $\Gamma_{\gamma'\to i \bar i}$ are given in  
 \cite{Li:2023nez,Aboubrahim:2021ohe}. 
 The additional cross section needed is $\sigma_{\chi\chi\to \gamma'\gamma'}$.
  The Feynman diagrams for it are in Figure \ref{fig:Feynman}. This cross section is given by
\begin{align}
    \sigma^{\chi\chi\rightarrow\gamma'\gamma'}(s,T_h) &=\frac{g_x^4(12m_{\gamma'}^4-4m_{\gamma'}^2s+s^2)}{512\pi m_{\gamma'}s^2}\left(\frac{\sqrt{(s-4m_\chi^2)(s-4m_{\gamma'}^2)}}{(m_\chi^2-s)^2(m_\chi^4-4m\chi^2m_{\gamma'}^2+m_{\gamma'}^2s)}(8g_x^4v_h^4(m_\chi^2-s)^2\right.\non
    &+(m_\chi^2-s)^2(m_\chi^4-m_\chi^2m_{\gamma'}^2\non
    &+m_{\gamma'}^2s))+\left.\frac{8g_x^2v_h^2}{2m_\chi^4-3m_\chi^2s+s^2}\left(\log{A}(m_\chi^2(2v_h^2g_x^2-3s)+s(2v_h^2g_x^2+s)+2m_\chi^4)\right)
    \right)\\
    A &= \frac{\sqrt{(s-4m_\chi^2)(s-m_{\gamma'}^2)}-2m_\chi^2+s}{-\sqrt{(s-4m_\chi^2)(s-m_{\gamma'}^2)}-2m_\chi^2+s}
\end{align}
where $s$ is the Mandelstam variable.
The cross section  for the reverse process is then given by
\begin{align}
    2\sqrt{s-4m_\chi^2}\sigma^{\chi\chi\rightarrow\gamma'\gamma'}(s,T_h) = 9\sqrt{s-4m_{\gamma'}^2}\sigma^{\gamma'\gamma'\rightarrow\chi\chi}(s,T_h)
\end{align}
Additionally we also need the decay width for the process
 $\chi\rightarrow\gamma'\gamma'$. This is given by
\begin{align}
    \Gamma_{\chi\rightarrow\gamma'\gamma'}(s) = \frac{g_x^4v_h^2}{128\pi m_\chi m_{\gamma'}^4}\sqrt{1-\frac{4m_{\gamma'}^2}{m_\chi^2}}(-4m_\chi^2m_{\gamma'}^2+m_\chi^4+12m_{\gamma'}^4) 
\end{align}

   We also define here $j_h$ that enters Eq.(\ref{DE1}).
\begin{align}
\label{y6}
j_h=&\sum_i \Big[2Y^{\rm eq}_i(T)^2 J(i\bar{i}\to D\bar{D})(T)
+Y^{\rm eq}_i(T)^2 J(i\bar{i}\to \gamma')(T)\Big]\mathbb{s}^2
-Y_{\gamma'}J(\gamma'\to i\bar i)(T_h)\mathbb{s}, \\
\label{y7}
Y^{\rm eq}_i=&\frac{n_i^{\rm eq}}{\bold{\mathbb{s}}}=\frac{g_i}{2\pi^2 \mathbb{s}}m_i^2 T K_2(m_i/T).
\end{align}
where $K_2$ is the modified Bessel function of the second kind and degree two.
Further,  $g_i$ is the number of degrees of freedom of particle $i$ and mass $m_i$ and the source functions $J$ are defined so that 
The $J$-functions that appear in Eq.~(\ref{y6}) are defined as 
\begin{align}
&n^{\rm eq}_i(T)^2 J(i~\bar{i}\to D\bar{D})(T)
=\frac{T}{32\pi^4}\int_{s_0}^{\infty}ds~\sigma_{D\bar{D}\to i\bar{i}}s(s-s_0)K_2(\sqrt{s}/T), \\
&n^{\rm eq}_i(T)^2 J(i~\bar{i}\to \gamma')(T)
=\frac{T}{32\pi^4}\int_{s_0}^{\infty}ds~\sigma_{i\bar{i}\to \gamma'}s(s-s_0)K_2(\sqrt{s}/T), \\
&n_{\gamma'}J(\gamma'\to i\bar i)(T_h)=n_{\gamma'}m_{\gamma'}\Gamma_{\gamma'\to i\bar i},\\
&n_i^{\rm eq}(T)^2\langle\sigma v\rangle_{i\bar{i}\to\gamma'}(T)
= \frac{T}{32\pi^4}\int_{s_0}^{\infty} ds ~\sigma(s) \sqrt{s}\, (s-s_0)K_1(\sqrt{s}/T),
\end{align}
where $K_1$ is the modified Bessel function of the second kind and degree one and $s_0$ is the minimum of the Mandelstam variable $s$.
We note that  there are additional contributions one can include in the analysis, i.e.,
$i\bar{i}\to \gamma' \gamma,
\gamma'Z, \gamma'\gamma'$. Their contributions are relatively small compared to $i~\bar{i}\to \gamma'$ and are neglected.
%%%%%%%%%%%%%%%%%%%%%%%%%%%%%%%%%%%%
\section{Appendix E: Energy and pressure densities away from equilibrium
\label{sec:12}}
If one assumes that the hidden sector was in thermal equilibrium at all times, then the
particle distributions will follow the Fermi-Dirac or Bose-Einstein statistics as appropriate.
In this case the energy density $\rho_h$ and the  pressure density $p_h$ in the hidden sector   
  are given by \cite{Hindmarsh:2005ix,Husdal:2016haj}
\begin{align}
    \rho_{ h} &= \sum_i \rho_{i}=\sum_i\frac{g_{i}T_h^4}{2\pi^2}\int^{\infty}_{x_i}\frac{\sqrt{x^2-x_{i}^2}}{e^x\pm1}x^2dx, i\in\{\gamma',D,\chi\} \non
    p_{ h} &= \sum_i p_{i}=\sum_i\frac{g_{i}T_h^3}{6\pi^2}\int^{\infty}_{x_i}\frac{(x^2-x_{i}^2)^{\frac{3}{2}}}{e^x\pm1}dx, i\in\{\gamma',D,\chi\}
\end{align}
where $g_{\gamma'} = 3$, $g_D = 4$, $g_\chi = 3$, $x_{i}=\frac{m_i}{T_h}$ and
plus is 
 for fermions while minus is for bosons.
  If a massive particle remained in thermal equilibrium until today, its energy density, $\rho_i\sim (m_i/T)^{3/2}\exp{(-m_i/T)}$, would be negligible because of the exponential factor. 
  However, as pointed out in  \cite{Kolb:1990vq}  if the interactions of the particles freeze out before complete annihilation, the particles may have a significant relic abundance today. 
    Often in the discussion of freeze out, it is generally assumed that $\rho=\rho_{eq}$ where
    $\rho_{eq}$ refers to the equilibrium density. However, the more precise way to compute
    the energy density in a freeze-out situation is to take
\begin{align}
    \rho_{h} = \rho_{h,eq} + \rho_{h,relic}
\end{align}
As suggested in  \cite{Kolb:1990vq} $ \rho_{h,relic}$ could be computed using the yield equation 
to obtain the number density
\begin{align}
    Y_{h,relic} = \frac{n_{h,relic}}{\mathbb{s}} \Rightarrow n_{h,relic} = Y_{h,relic}\mathbb{s}.
\end{align}
which allows a computation of the number density $n_{h,relic}$ 
from where we can compute the $g^h_{n,relic}$ so that 
\begin{align}
    n_{h,relic} &= \frac{\zeta(3)}{\pi^2}g^h_{n,relic}T_h^3. 
 \end{align}
Next we set the effective energy degrees of freedom from the relic density so that 
\begin{align}
 g^h_{\rho,relic} = g^h_{n,relic}   
\end{align}
and use the relation
\begin{align}
    \rho_{h,relic} &= \frac{\pi^2}{30}g^h_{\rho,relic}T_h^4
\end{align}
to  find $\rho_{h,relic}$ and $\rho_{h}$. 
In most cases, this analysis is not necessary since  $\rho_{h,relic} \ll \rho_{h,eq}$.  However, such an analysis becomes relevant when we are dealing with the decoupling of the entire hidden sector since in this situation we have $\rho_{h,relic} \gg \rho_{h,eq}$. A similar analysis holds for  the pressure density $p_h$.
{Aside from  the correction to the density discussed above, the density of the hidden sector should freeze-out when the two sectors are fully decoupled. This analysis will be similar to the analysis in cannibalism dark matter\cite{Pappadopulo:2016pkp}.
For the current model, the decoupling happens when: (i) all interactions between the hidden sector and the visible sector decouple and (ii)  the dark photon decays out. 
}
%%%%%%%%%%%%%%%%%%%%%%%%%%%%%%%%%%%

%%%%%%%%%%%%%%%%%%%%%%%%%%%%%

\end{document}